\documentclass[10pt]{article}

\usepackage{amsmath}
\usepackage{amsfonts}
\usepackage{amssymb}
\usepackage{graphicx}
\usepackage{amsthm}
\usepackage{rotating}

\usepackage{ifthen}
\usepackage[dvipsnames]{xcolor}
\usepackage{colortbl}
\usepackage{array}
\usepackage{pifont}
\usepackage{hyperref}
\usepackage[totoc]{idxlayout}
\usepackage{dirtree}
\usepackage{listings}
\usepackage{makecell}
\usepackage{booktabs}
\usepackage{tikz}

\renewcommand{\hline}{\Xhline{2\arrayrulewidth}}

\DeclareRobustCommand\sampleline[1]{%
\raisebox{-0.55em}{
  \tikz\draw[#1] (-0.8em,0)
  -- (0.8em,0)
  (0,0.8em)
  -- (0,-0.8em);}%
}

\usepackage{datetime}
\usepackage{arydshln}

\definecolor{myorange}{rgb}{0.843, 0.678, 0.000}
\definecolor{mygreen}{rgb}{0,0.6,0}
\definecolor{mygray}{rgb}{0.5,0.5,0.5}
\definecolor{mymauve}{rgb}{0.58,0,0.82}

\renewcommand{\em}{\it}   

\newcolumntype{I}{!{\vrule width 1.5pt}}
\newlength\savedwidth
\newcommand\whline{\noalign{\global\savedwidth\arrayrulewidth
                            \global\arrayrulewidth 1.05pt}%
           \hline
           \noalign{\global\arrayrulewidth\savedwidth}}

\newcommand{\FlaTwoByTwo}[4]{
\left(
\begin{array}{c I c}
#1 & #2 \\ \whline
#3 & #4
\end{array}
\right)
}

\newcommand{\FlaThreeByThreeBR}[9]{
\left(
\begin{array}{c I c | c}
#1 & #2 & #3 \\ \whline 
#4 & #5 & #6 \\ \hline
#7 & #8 & #9 
\end{array}
\right) 
}

\newcommand{\operation}{}

\newcommand{\routinename}{}

\newcommand{\precondition}{~}

\newcommand{\postcondition}{~}

\newcommand{\invariant}{~}

\newcommand{\guard}{~}

\newcommand{\partitionings}{~}

\newcommand{\partitionsizes}{~}

\newcommand{\blocksize}{blank}

\newcommand{\repartitionings}{~}

\newcommand{\repartitionsizes}{~}

\newcommand{\moveboundaries}{~}

\newcommand{\beforeupdate}{~}

\newcommand{\afterupdate}{~}

\newcommand{\update}{~}

\newcommand{\resetsteps}{

\renewcommand{\operation}{\phantom{[A] = op( A )}}

\renewcommand{\routinename}{\operation}

\renewcommand{\precondition}{\phantom{A = \widehat A}}

\renewcommand{\postcondition}{\phantom{A = \widehat A}}

\renewcommand{\invariant}{\phantom{ \FlaTwoByTwo{A_{TL}}{A_{TR}}{A_{BL}}{A_{BR}} =
		\FlaTwoByTwo{A_{TL}}{A_{TR}}{A_{BL}}{A_{BR}}
		\wedge
		\FlaTwoByTwo{A_{TL}}{A_{TR}}{A_{BL}}{A_{BR}} =
		\FlaTwoByTwo{A_{TL}}{A_{TR}}{A_{BL}}{A_{BR}}~~~~~~~
		}}

\renewcommand{\blocksize}{blank}

\renewcommand{\guard}{\phantom{m( A_{BL} ) < m( A )}}

\renewcommand{\partitionings}{
$
\phantom{\FlaTwoByTwo{A_{TL}}{A_{TR}}{A_{BL}}{A_{BR}}
\rightarrow
\FlaThreeByThreeBR
   {A_{00}}{a_{01}}{A_{02}}
   {a_{10}^T}{\alpha_{11}}{a_{12}^T}
   {A_{20}}{a_{21}}{A_{22}}}   
$
}

\renewcommand{\partitionsizes}{$ \phantom{m( A )} $}

\renewcommand{\repartitionings}{
$
\phantom{\FlaTwoByTwo{A_{TL}}{A_{TR}}{A_{BL}}{A_{BR}}
\rightarrow
\FlaThreeByThreeBR
   {A_{00}}{a_{01}}{A_{02}}
   {a_{10}^T}{\alpha_{11}}{a_{12}^T}
   {A_{20}}{a_{21}}{A_{22}}}   
$
}

\renewcommand{\repartitionsizes}{$\phantom{m(A)}$}

\renewcommand{\moveboundaries}{
$
\phantom{\FlaTwoByTwo{A_{TL}}{A_{TR}}{A_{BL}}{A_{BR}}
\rightarrow
\FlaThreeByThreeBR
   {A_{00}}{a_{01}}{A_{02}}
   {a_{10}^T}{\alpha_{11}}{a_{12}^T}
   {A_{20}}{a_{21}}{A_{22}}}   
$
}

\renewcommand{\beforeupdate}{
\phantom{\FlaTwoByTwo{A_{TL}}{A_{TR}}{A_{BL}}{A_{BR}}
\rightarrow
\FlaThreeByThreeBR
   {A_{00}}{a_{01}}{A_{02}}
   {a_{10}^T}{\alpha_{11}}{a_{12}^T}
   {A_{20}}{a_{21}}{A_{22}}}   
}

\renewcommand{\afterupdate}{
\phantom{\FlaTwoByTwo{A_{TL}}{A_{TR}}{A_{BL}}{A_{BR}}
\rightarrow
\FlaThreeByThreeBR
   {A_{00}}{a_{01}}{A_{02}}
   {a_{10}^T}{\alpha_{11}}{a_{12}^T}
   {A_{20}}{a_{21}}{A_{22}}}   
}

\renewcommand{\update}{
\phantom{$
\begin{array}{l}
\\
\\
\\
\end{array}
$}
}
}

\newcommand{\NoShow}[1]{}

\newcommand{\moreinitialize}{}

\newcommand{\FlaAlgorithm}{
\begin{tabular}{|p{0.95\textwidth}|} \hline
$\mbox{\color{blue}Algorithm:~}\routinename$
\\ \hline
\partitionings \\
$\mbox{\color{blue} ~~~where~}$ \partitionsizes 
\moreinitialize 
\\ 
$\mbox{\color{blue}while~} \ShowGuard \mbox{~\color{blue} do}$
\\
\ifthenelse{\equal{\blocksize}{1}}{}%
{%
\ifthenelse{ \equal{\blocksize}{blank} }{}%
{~~~~{\bf Determine block size $ \blocksize $}\\}%
}
\repartitionings 
\NoShow{\\
~$\mbox{\color{blue} ~~~where~}$ \repartitionsizes
}
\\ \hline  \update 
\\ \hline
\moveboundaries 
\\
$\mbox{\color{blue} endwhile} $
\\ \hline 
\end{tabular}
}

\newcounter{WSStep}
\newcommand{\ShowPrecondition}{\ifthenelse{\value{WSStep}<1}%
   {{\color{white} \precondition}}
   {\ifthenelse{\value{WSStep}=1}%
    {\color{red} \precondition}
    {\color{black} \precondition}}}
\newcommand{\ShowPostcondition}{\ifthenelse{\value{WSStep}<1}%
   {{\color{white} \postcondition}}
   {\ifthenelse{\value{WSStep}=1}%
    {\color{red} \postcondition}
    {\color{black} \postcondition}}}
\newcommand{\ShowInvariant}{\ifthenelse{\value{WSStep}<2}%
   {{\color{white} \invariant}}
   {\ifthenelse{\value{WSStep}=2}%
    {\color{red} \invariant}
    {\color{black} \invariant}}}
\newcommand{\ShowGuard}{\ifthenelse{\value{WSStep}<3}%
   {{\color{lightgray!25} \guard}}
   {\ifthenelse{\value{WSStep}=3}%
    {\color{red} \guard}
    {\color{black} \guard}}}
\newcommand{\ShowGuardTwo}{\ifthenelse{\value{WSStep}<3}%
   {{\color{white} \guard}}
   {\ifthenelse{\value{WSStep}=3}%
    {\color{red} \guard}
    {\color{black} \guard}}}
\newcommand{\ShowPartitionings}{\ifthenelse{\value{WSStep}<4}%
   {{\color{lightgray!25} \partitionings}}%
   {\ifthenelse{\value{WSStep}=4}%
    {\color{red} \partitionings}%
    {\color{black} \partitionings}}}
\newcommand{\ShowPartitionSizes}{\ifthenelse{\value{WSStep}<4}%
   {{\color{lightgray!25} \partitionsizes}}
   {\ifthenelse{\value{WSStep}=4}%
    {\color{red} \partitionsizes}
    {\color{black} \partitionsizes}}}

\newcommand{\ShowRepartitionings}{\ifthenelse{\value{WSStep}<5}%
   {{\color{lightgray!25} \repartitionings}}
   {\ifthenelse{\value{WSStep}=5}%
    {\color{red} \repartitionings}
    {\color{black} \repartitionings}}}
    
\newcommand{\ShowRepartitionSizes}{\ifthenelse{\value{WSStep}<5}%
   {{\color{lightgray!25} \repartitionsizes}}
   {\ifthenelse{\value{WSStep}=5}%
    {\color{red} \repartitionsizes}
    {\color{black} \repartitionsizes}}}
    
\newcommand{\ShowMoveBoundaries}{\ifthenelse{\value{WSStep}<5}%
   {{\color{lightgray!25} \moveboundaries}}
   {\ifthenelse{\value{WSStep}=5}%
    {\color{red} \moveboundaries}
    {\color{black} \moveboundaries}}}

\newcommand{\ShowBeforeUpdate}{\ifthenelse{\value{WSStep}<6}%
   {{\color{white} \beforeupdate}}
   {\ifthenelse{\value{WSStep}=6}%
    {\color{red} \beforeupdate}
    {\color{black} \beforeupdate}}}
\newcommand{\ShowAfterUpdate}{\ifthenelse{\value{WSStep}<7}%
   {{\color{white} \afterupdate}}
   {\ifthenelse{\value{WSStep}=7}%
    {\color{red} \afterupdate}
    {\color{black} \afterupdate}}}
\newcommand{\ShowUpdate}{\ifthenelse{\value{WSStep}<8}%
   {{\color{lightgray!25} \update}}
   {\ifthenelse{\value{WSStep}=8}%
    {\color{red} \update}
    {\color{black} \update}}}

\newcommand{\FlaWorksheetNine}{
\begin{tabular}{| c | p{0.9\textwidth} |}\hline
{\color{white}Step} & $\mbox{\color{blue}Algorithm:~}\routinename$
\\ \hline
 &%
$ \phantom{\left\{ 
\begin{minipage}{0.88\textwidth} 
$\ShowPrecondition$  
\end{minipage}
\right\}}
$%
\\ \hline
\rowcolor{lightgray!25}   
& %
\begin{minipage}{0.88\textwidth}%
\vspace{0.05in}
\ShowPartitionings~ \\
\mbox{\color{blue} ~~~where~} \ShowPartitionSizes
\end{minipage}
\\ \hline
& 
$ \phantom{\left\{ 
\begin{minipage}{0.88\textwidth} 
$\ShowInvariant $
\end{minipage}
\right\}} $ 
\\ \hline
\rowcolor{lightgray!25}   
&$\mbox{\color{blue}while~} \ShowGuard \mbox{~\color{blue} do}$
\\ \hline 
 &  
$
\phantom{\left\{
\begin{minipage}[t]{0.88\textwidth}%
~~~~$
\ShowInvariant 
\wedge \ShowGuardTwo$
\end{minipage}
\right\}}
$ 
\\ \hline
\rowcolor{lightgray!25}   
 & ~~~~ \begin{minipage}{0.85\textwidth}%
\vspace{0.05in}
\ifthenelse{\equal{\blocksize}{1}}{}%
{%
\ifthenelse{ \equal{\blocksize}{blank} }{}%
{{\bf Determine block size $ \blocksize $}\\}%
}
\ShowRepartitionings~ \\
$\mbox{\color{blue} ~~~where~}$ \ShowRepartitionSizes
\end{minipage}
\\ \hline
& 
$ \phantom{\left\{ 
\begin{minipage}{0.88\textwidth} 
~~~~ \ShowBeforeUpdate 
\end{minipage}
\right\}}
$
\\ \hline
\rowcolor{lightgray!25}  
 & ~~~~  \ShowUpdate 
\\ \hline 
& 
$ \phantom{\left\{ 
\begin{minipage}{0.88\textwidth} 
~~~~ \ShowAfterUpdate 
\end{minipage}
\right\}}
$
\\ \hline
\rowcolor{lightgray!25}   
 & ~~~~ \begin{minipage}{0.85\textwidth}%
\vspace{0.05in}
\ShowMoveBoundaries~
\end{minipage}
\\ \hline
& 
$ \phantom{\left\{ 
\begin{minipage}{0.88\textwidth} 
~~~~ $ \ShowInvariant  $ 
\end{minipage}
\right\}}
$
\\ \hline
\rowcolor{lightgray!25}  
 &$\mbox{\color{blue} endwhile} $
\\ \hline 
& 
$ \phantom{\left\{ 
\begin{minipage}{0.88\textwidth} 
$ \ShowInvariant \wedge \neg( \ShowGuardTwo )$ 
\end{minipage}
\right\}}
$
\\ \hline
& 
$ \phantom{\left\{ 
\begin{minipage}{0.88\textwidth} 
$ \ShowPostcondition $ 
\end{minipage}
\right\}}
$
\\ \hline
\end{tabular}
}

\newcommand{\FlaCostWorksheet}{
\begin{tabular}{| c | p{0.45\textwidth}
p{0.45\textwidth}|}\hline
Step & $\mbox{\color{blue}Algorithm:~}\routinename $ &
\\ \hline
1a &%
$ \left\{ 
\begin{minipage}{0.44\textwidth} 
$\ShowPrecondition$  
\end{minipage}
\right\}
$
&
\\ \hline
\rowcolor{lightgray!25}   
4 & %
\begin{minipage}{0.88\textwidth}%
\vspace{0.05in}
\ShowPartitionings~ \\
\mbox{\color{blue} ~~~where~} \ShowPartitionSizes
\end{minipage}
& 
\begin{minipage}{0.44\textwidth}
\hfill \CostInit
\end{minipage}
\\ \hline
2 & 
$ \left\{ 
\begin{minipage}{0.44\textwidth} 
$\ShowInvariant $
\end{minipage}
\right\} $ 
&
\begin{minipage}{0.44\textwidth}
$ \{ $ \hfill \CostInvariant
$ \} $ 
\end{minipage}
\\ \hline
\rowcolor{lightgray!25}   
3 &$\mbox{\color{blue}while~} \ShowGuard \mbox{~\color{blue} do}$
&
\\ \hline 
2,3 &  
$
\left\{
\begin{minipage}[t]{0.41\textwidth}%
$
~~~~ \ShowInvariant 
\wedge \ShowGuardTwo$
\end{minipage}
\right\}
$ 
&
\begin{minipage}{0.44\textwidth}
$ \{ $ \hfill \CostInvariant
$ \} $
\end{minipage}
\\ \hline
\rowcolor{lightgray!25}   
5a & ~~~~ \begin{minipage}{0.41\textwidth}%
\vspace{0.05in}
\ifthenelse{\equal{\blocksize}{1}}{}%
{%
\ifthenelse{ \equal{\blocksize}{blank} }{}%
{{\bf Determine block size $ \blocksize $}\\}%
}
\ShowRepartitionings~ 
\NoShow{\\
$\mbox{\color{blue} ~~~where~}$ \ShowRepartitionSizes}
\end{minipage}
&
\\ \hline
6 & 
$ \left\{ 
\begin{minipage}{0.41\textwidth} 
~~~~ \ShowBeforeUpdate 
\end{minipage}
\right\}
$
&
$ \{ $ \hfill 
\CostBefore
$ \} $
\\ \hline
\rowcolor{lightgray!25}  
8 & 
\begin{minipage}{0.41\textwidth} 
~~~~  \ShowUpdate 
\end{minipage}
&
\hfill 
$
C := C + 2 
$
\\ \hline 
7 & 
$ \left\{ 
\begin{minipage}{0.41\textwidth} 
~~~~ \ShowAfterUpdate 
\end{minipage}
\right\}
$
&
$ \{ $ \hfill 
\CostAfter
$ \} $
\\ \hline
\rowcolor{lightgray!25}   
5b & ~~~~ \begin{minipage}{0.41\textwidth}%
\vspace{0.05in}
\ShowMoveBoundaries~
\end{minipage}
&
\\ \hline
2 & 
$ \left\{ 
\begin{minipage}{0.44\textwidth} 
~~~~ $ \ShowInvariant  $ 
\end{minipage}
\right\}
$
&
\begin{minipage}{0.44\textwidth}
$ \{ $ \hfill \CostInvariant
$ \} $
\end{minipage}
\\ \hline
\rowcolor{lightgray!25}  
 &$\mbox{\color{blue} endwhile} $
 &
\\ \hline 
2,3 & 
$ \left\{ 
\begin{minipage}{0.44\textwidth} 
$ \ShowInvariant \wedge \neg( \ShowGuardTwo )$ 
\end{minipage}
\right\}
$
&
\begin{minipage}{0.44\textwidth}
$ \{ $ \hfill \CostInvariant
$ \} $
\end{minipage}
\\ \hline
1b & 
$ \left\{ 
\begin{minipage}{0.44\textwidth} 
$ \ShowPostcondition $ 
\end{minipage}
\right\}
$
&
\begin{minipage}{0.44\textwidth}
$ \{ $ \hfill \CostPostCond
$ \} $
\end{minipage}
\\ \hline
\end{tabular}
}

\usepackage{microtype}

\usepackage{multirow}

\newtheorem{theorem}
{Theorem}[section]
\newtheorem{lemma}[theorem]
{Lemma}
\newtheorem{corollary}[theorem]
{Corollary}
\newtheorem{definition}[theorem]
{Definition}

\newboolean{VersionTOMS}
\setboolean{VersionTOMS}{true}

\title{Deriving Algorithms for Triangular Tridiagonalization of a Skew-symmetric Matrix}

\usepackage{authblk}

\author{Robert  van de Geijn}
\author{Maggie  Myers}
\affil{
	Oden Institute for Computational Engineering \& Sciences 
 and 
 Department of Computer Science,
The University of Texas at Austin,
\tt \{rvdg,myers\}@cs.utexas.edu}

\author{
RuQing G. Xu }
\affil{
Department of Physics,
The University of Tokyo,
\tt r-xu@g.ecc.u-tokyo.ac.jp
}

\author{
Devin Matthews}
\affil{
Department of Chemistry,
Southern Methodist University,
\tt damatthews@smu.edu}

\date{\today}

\begin{document}

\maketitle

\begin{abstract}
    This paper provides technical details regarding the application of the FLAME methodology to derive algorithms hand in hand with their proofs of correctness for the computation of the $ L T L^T $ decomposition (with and without pivoting) of a skew-symmetric matrix.  The approach yields known as well as new algorithms, presented using the FLAME notation, enabling comparing and contrasting.  
A number of BLAS-like primitives are exposed at the core of the resulting unblocked and blocked algorithms.

\end{abstract}

\allowdisplaybreaks
\section{Introduction}

Under well-understood conditions, a skew-symmetric indefinite matrix $ X $ can be factored as $ P X P^T = L T L^T $, where $ P $ is a permutation matrix, $ L $ is a unit lower-triangular matrix and $ T $ is a \mbox{skew-symmetric} tridiagonal matrix.
This is sometimes referred to as {\em triangular tridiagonalization~\cite{Miroslav2011}.}
One may recognize this as a variation on the Cholesky%
\NoShow{($ X = L L^T $)} and $ L D L^T $, where $ D $ is diagonal, factorizations for symmetric positive definite and indefinite matrices, respectively.
We are motivated by the computation of the Pfaffian $ {\rm Pf}(X) $, defined as $ {\rm Pf}( X ) = \frac{1}{ 2^n n! } \sum_{ \sigma \in S_{ 2n } } {\rm sgn}( \sigma ) \prod_i^n x_{ \sigma( 2i-1 ), \sigma( 2i ) } $  for skew-symmetric $ X $ of size $ 2n \times 2n $.  Here, $ S_{ 2n } $ represents the $ 2n $-element permutation set. 
It can be shown that $ {\rm Pf}( X )^2 = \det( X ) $. 
Also, if $ P X P^T = L T L^T $, where 
{
\small
\[
\label{eqn:T}
T = \left( \begin{array}{c c c c c}
0 & -\tau_{1,0} & 0 & \cdots & 0 \\
\tau_{1,0} & 0 & - \tau_{2,1} & \cdots & 0 \\
0 & \tau_{2,1} & 0 & \ddots & 0 \\
\vdots & \vdots & \ddots & \ddots & \vdots \\
0 & 0 & 0 & \cdots & 0
\end{array}
\right),
\]
}%
then $ {\rm Pf}( X ) = {\rm Pf}( T ) = \tau_{1,0}
\times \tau_{3,2}  \times \cdots
\times \tau_{2n-1,2n-2} $.
This quantity arises frequently in physics studies where pairs of Fermions are involved, such as the 2-dimensional Ising spin glass~\cite{thomas2009} and electronic structure quantum Monte Carlo~\cite{bajdich2009}.
At this writing, LAPACK~\cite{LAPACK3} does not provide a routine to factorize skew-symmetric matrices.
This paper provides technical details of the  derivations for algorithms discussed in
our paper ``Performant Tridiagonalization of Skew-symmetric Matrices''~\cite{LTLt_SISC_ArXiv}.

\section{Background}

We gather a number of results related to skew-symmetric matrices and  Gauss transforms.

\subsection{Notation}

\begin{table}
\caption{A summary of the notational conventions used in this work. The symbols $A$, $a$, and $\alpha$ are used to denote arbitrary matrices, vectors, and scalars.}
\label{table:notation}
\begin{tabular}{ >{\centering}m{0.15\columnwidth} m{0.75\columnwidth} }
\toprule
$A$ & Matrix  \\
$a$ & (Column) vector  \\
$\alpha$ & Scalar  \\
$e_f$, $e_l$ & Standard basis vector with a $1$ in the \underline{$f$}irst or \underline{$l$}ast position \\
$\widehat{A}$, $\widehat{a}$, $\widehat{\alpha}$ & Original contents of a matrix, vector, or scalar \\
${A}$, ${a}$, ${\alpha}$ & Current contents of a matrix, vector, or scalar \\
${A}^+$, ${a}^+$, ${\alpha}^+$ & Updated contents of a matrix, vector, or scalar, typically at the bottom of a loop body \\
$\widetilde{A}$, $\widetilde{a}$, $\widetilde{\alpha}$ & Final contents of a matrix, vector, or scalar at the end of the algorithm \\
\textcolor{blue}{${A}$, ${a}$, ${\alpha}$} & Matrix sub-partitions which have already been computed at the current step \\
$\Big(\!\sampleline{black}\Big)$ & Partitioned matrix---the size of each sub-partition is implicit \\
$\Big(\!\sampleline{black,line width=1.5pt}\Big)$ & Partitioned matrix---a thick line typically separates regions of the matrix according to the current progress of a loop-based algorithm
\\
$TL, TM, \ldots $ &
Identify 
Top-Left, Top-Middle, etc. subparts of matrices
\\
$ \star $ &
Implicit (skew-)symmetric part of matrix, assuming only the lower triangular part is stored \\
\bottomrule
\end{tabular}
\end{table}

We adopt {\em Householder notation} where, as a general rule, matrices, (column) vectors, and scalars are denoted with upper-case Roman, lower-case Roman, and lower-case Greek letters, respectively.
As is  customary in computer science, indexing starts at $ 0 $.
We let $ e_i $, $ 0 \leq i < m $, 
denote the standard basis vectors so that the $ m \times m $ identity matrix, $ I $, can be partitioned by columns as
$ I = \left( \begin{array}{c | c  | c | c}
e_0 & e_1 & \cdots & e_{m-1} \end{array} \right) $.
Vectors $ e_f $ and $ e_l $ denote the standard basis vectors with a 1 in the \underline{$f$}irst and \underline{$l$}ast position, respectively. The size of the vectors is determined by context, e.g., in the example above $ e_f = e_0 $ and $ e_l = e_{m-1} $.  
The zero matrix ``of appropriate size'' is denoted by $ 0 $, which means it can also stand for a scalar $ 0 $, the $ 0 $ vector, or even a $ 0 \times 0 $ matrix. These and additional notations applying to matrices and matrix sub-partitions (which can be matrices, vectors, or scalars) are summarized in Table~\ref{table:notation}.

\subsection{Skew-symmetric (antisymmetric) matrices}

\begin{definition}
    Matrix $ X \in \mathbb{R}^{m \times m} $ is said to be 
    \emph{skew symmetric} if
    $ X = -X^T $.
\end{definition}

The diagonal elements of a skew-symmetric matrix equal zero and $ \chi_{i,j} = - \chi_{j,i} $.

\begin{theorem}
    Let matrix $ X \in \mathbb{R}^{m \times m} $ be partitioned as
    $
    X = 
    \left( \begin{array}{c I c}
    X_{TL} & X_{TR} \\ \whline
    X_{BL} & X_{BR}
    \end{array} \right)$,
where $ X_{TL} $ is square.  Then $ X $ is skew symmetric iff $
\left( \begin{array}{c I c}
    X_{TL} & X_{TR} \\ \whline
    X_{BL} & X_{BR}
    \end{array} \right)
    =
    \left( \begin{array}{c I c}
    -X_{TL}^T & -X_{BL}^T \\ \whline
    -X_{TR} & -X_{BR}^T
    \end{array} \right)$.
\end{theorem}

\NoShow{
\begin{theorem}
\label{thm:yTXy}
Let $ X\in \mathbb{R}^{m \times m} $ be a skew symmetric matrix and $ y $ a vector of appropriate size.  Then $ y^T X y = 0 $.
\end{theorem}

\begin{proof}
    $ y^T X y
    = ( y^T X y )^T =
    y^T X^T y
    = y^T (-X) y = - y^T X y $.  Hence $ y^T X y = 0 $.
\end{proof}

\begin{corollary}
Let $ X $ be an $ m \times m $ skew-symmetric matrix and $ \chi_{i,j} $ its entries.  Then $ \chi_{i,i} = 0 $ for all $ 0 \leq i < m $.
\end{corollary}

\begin{proof}
    $ \chi_{i,i} = e_i^T X e_i = 0 $ by Theorem~\ref{thm:yTXy}.
\end{proof}
}

\begin{theorem}
Let $  X $ be $ n \times n $ and $ B $ be $ m \times n $.  If $ X $ is skew symmetric, then so is $ C = B X B^T $. 
\end{theorem}

\begin{proof}
$ C^T = ( B X B^T)^T = B X^T B^T = - B X B^T = - C $.
\end{proof}

The following theorem will become key to understanding the  connection between a simple blocked right-looking algorithm and blocked versions of the two-step (Wimmer's)  algorithms:

\begin{theorem}
\label{thm:TandS}
    Let $ X $ be  a skew-symmetric matrix and assume there exist a (square) matrix $ B $ and tridiagonal skew-symmetric matrix $ T $ such that $ X = B T B^T $.
    Then
    \begin{enumerate}
        \item 
        $ T = S - S^T$ where 
        \begin{equation}
            \label{eqn:TvsS}
        T = \left( \begin{array}{c c c c c}
        0 & - \tau_{10} & 0 & 0 & \cdots \\ 
        \tau_{10}  & 0 & - \tau_{21} & 0 &  \cdots \\
        0 &  \tau_{21} &  0 & - \tau_{32} & \cdots  \\
        0 &  0 & \tau_{31} &  0 & \ddots \\
        \vdots & \vdots &  \vdots & \ddots & \ddots 
        \end{array}
        \right)
        \quad
        \mbox{and}
        \quad
        S = \left( \begin{array}{c c c c c}
        0 & 0
        & 0 & 0 & \cdots \\ 
        \tau_{10}
        & 0 & 
        - \tau_{21}
        & 0 &  \cdots \\
        0 & 0
        &  0 & 0
        & \cdots  \\
        0 &  0 & 
        \tau_{31}
        &  0 & \ddots \\
        \vdots & \vdots &  \vdots & \ddots & \ddots 
        \end{array}
        \right).
        \end{equation}
        \item 
        $ X = B T B^T = B ( S - S^T ) B^T = 
        (B S)
        B^T - B
        (B S)^T
        =  W B^T - B W^T $,
        where $ W = B S $ has every other column equal to zero, starting with the second column.
    \end{enumerate}
\end{theorem}

\subsection{Gauss transforms}

The LU factorization of a matrix $ A \in \mathbb{R}^{m \times m} $
is given by  $ A = LU $, where $ L $ and $ U $ are unit lower triangular 
and upper triangular matrices, respectively.

The computation of the LU factorization
can be organized as the   application of a sequence of {\em Gauss transforms}:
If one
partitions 
\[
A = 
\left( \begin{array}{c | c} 
\alpha_{11} & a_{12}^T \\ \hline
a_{21} & A_{22}
\end{array} \right),
L =
\left( \begin{array}{c | c} 
1 & 0 \\ \hline
l_{21} & L_{22}
\end{array} \right),
\mbox{~and~}
U =
\left( \begin{array}{c | c} 
\upsilon_{11} & u_{12}^T \\ \hline
0 & U_{22}
\end{array} \right).
\]
then $ A = L U $ implies that
\[
\left( \begin{array}{c | c} 
\alpha_{11} & a_{12}^T \\ \hline
a_{21} & A_{22}
\end{array} \right)
=
\left( \begin{array}{c | c} 
1 & 0 \\ \hline
l_{21} & L_{22}
\end{array} \right)
\left( \begin{array}{c | c} 
\upsilon_{11} & u_{12}^T \\ \hline
0 & U_{22}
\end{array} \right)
=
\left( \begin{array}{c | c} 
\upsilon_{11} & u_{12}^T \\ \hline
\upsilon_{11} l_{21} & l_{21} u_{12}^T + L_{22} U_{22}
\end{array} \right).
\]
If we choose $ l_{21} = a_{21} / \alpha_{11} $, then
one updates
\[
\left( \begin{array}{c | c} 
\alpha_{11} & a_{12}^T \\ \hline
a_{21} & A_{22}
\end{array} \right)
:=
\left( \begin{array}{c | c} 
1 & 0 \\ \hline
-l_{21} & I
\end{array} \right)
\left( \begin{array}{c | c} 
\alpha_{11} & a_{12}^T \\ \hline
a_{21} & A_{22}
\end{array} \right)
=
\left( \begin{array}{c | c} 
\alpha_{11} & a_{12}^T \\ \hline
0 & A_{22} - l_{21} a_{12}^T
\end{array} \right).
\]
  Continuing this process with the updated $ A_{22} $ will ultimately overwrite $ A $ with $ U $ (provided $ A $ has nonsingular leading principle submatrices).

\begin{definition}
    A matrix $ L_i $ of form
    $
    L_{i} = 
    \left( \begin{array}{c I c | c}
    I_{i \times i} & 0 & 0 \\ \whline
    0 & 1 & 0 \\ \hline
    0 & l_{21}^{(i)} & I 
    \end{array} \right)
    $
    is called a \emph{Gauss transform}.
\end{definition} 
The inverse of a Gauss transform is also a Gauss transform:
\begin{lemma}
    $
    \left( \begin{array}{c I c | c}
    I_{i \times i} & 0 & 0 \\ \whline
    0 & 1 & 0 \\ \hline
    0 & l_{21}^{(i)} & I 
    \end{array} \right)^{-1}
    =
    \left( \begin{array}{c I c | c}
    I_{i \times i} & 0 & 0 \\ \whline
    0 & 1 & 0 \\ \hline
    0 & - l_{21}^{(i)} & I 
    \end{array} \right)$. 
\end{lemma}

The described process for computing the LU factorization can be summarized as
\[
L_{n-1}^{-1} \cdots L_1^{-1} L_0^{-1} A = U
\mbox{~or, equivalently,~}
A = 
L_0 L_1 \cdots L_{n-1}
U = L U,
\]
where each $ L_i 
$ is a Gauss transform with appropriately chosen $ l_{21}^{(i)}  $.  
The following results tell us that the product of Gauss transforms, $ L_0 L_1 \cdots L_{n-1} $, is a unit lower-triangular matrix $ L $ that simply consists of the identity in which the $ l_{21}^{(i)} $ of $ L_i $ is inserted in the column indexed with $ i $:
\begin{theorem}
If the matrices in the following expression are conformally partitioned, then
\[
\begin{array}[t]{c}
\underbrace{
\left( \begin{array}{c I c | c}
L_{00} & 0 & 0 \\ \whline
l_{10}^T & 1 & 0 \\ \hline
L_{20} & 0 & I 
\end{array} \right)
}
\\
L_0 \cdots L_{i-1}
\end{array}
\begin{array}[t]{c}
\underbrace{
\left( \begin{array}{c I c | c}
I & 0 & 0 \\ \whline
0 & 1 & 0 \\ \hline
0 & l_{21}^{(i)} & I 
\end{array} \right)
} \\
L_{i}
\end{array}
=
\begin{array}[t]{c}
\underbrace{
\left( \begin{array}{c I c | c}
L_{00} & 0 & 0 \\ \whline
l_{10}^T & 1 & 0 \\ \hline
L_{20} & l_{21}^{(i)} & I 
\end{array} \right)
}
\\
L_0 \cdots L_{i-1} L_{i}
\end{array}.
\]
\end{theorem}

\begin{corollary}
$
L_0 L_1 \cdots L_{n-1}
=
\left( \begin{array}{c | c | c | c}
1 & 0 & 0 & 0 \\ \hline
\multirow{3}{*}{
$l_{21}^{(0)}$} & 1 & 0 & \cdots \\ \cline{2-4}
{} &  \multirow{2}{*}{
$l_{21}^{(1)}$} & 1 & \cdots  \\ \cline{3-4}
{} & {} & l_{21}^{(2)}& \ddots 
\end{array}
\right).
$
\end{corollary}

A matrix of the form
$
\left( \begin{array}{ c | c}
L_{TL} & 0 \\ \hline
L_{BL} &  I 
\end{array} \right)
$,
where $ L_{TL} $ is unit lower triangular,
represents an accumulation of Gauss transforms or \emph{block Gauss transform}.
This, and the following corollary,  will  play a critical role in the development of so-called blocked algorithms that cast most computation in terms of matrix-matrix multiplication.
\begin{corollary}
\label{cor:inv}
$
    \left( \begin{array}{ c | c}
    L_{TL} & 0 \\ \hline
    L_{BL}  &  I
    \end{array} \right)^{-1}
    \left( \begin{array}{ c | c}
    L_{TL} & 0 \\ \hline
    L_{BL}  &  L_{BR} 
    \end{array} \right)
    =
    \left( \begin{array}{ c | c}
    I & 0 \\ \hline
    0  &  L_{BR} 
    \end{array} \right)
$.
\end{corollary}

\begin{proof}
    The result follows immediately from the observation that
    \[
     \left( \begin{array}{ c | c}
    L_{TL} & 0 \\ \hline
    L_{BL}  &  I
    \end{array} \right)^{-1}
    =
     \left( \begin{array}{ c | c}
    L_{TL}^{-1} & 0 \\ \hline
    - L_{BL} L_{TL}^{-1} &  I
    \end{array} \right).
    \]
\end{proof}

\NoShow{ 
\begin{corollary}
    If 
    $ L_0 L_1 \cdots L_{b-1} =
    \left( \begin{array}{ c | c}
    L_{11} & 0 \\ \hline
    L_{21} &  I 
    \end{array} \right)
    $ then
    \[
    L_{b-1}^{-1} \cdots L_1^{-1} L_0^{-1} =
    \left( \begin{array}{ c | c}
    L_{11} & 0 \\ \hline
    L_{21} &  I 
    \end{array} \right)^{-1}
    =
    \left( \begin{array}{ c | c}
    L_{11}^{-1} & 0 \\ \hline
    - L_{21} L_{11}^{-1} &  I 
    \end{array} \right)
    =
    \left( \begin{array}{ c | c}
    I & 0 \\ \hline
    - L_{21}  &  I 
    \end{array} \right)
    \left( \begin{array}{ c | c}
    L_{11}^{-1} & 0 \\ \hline
    0 &  I 
    \end{array} \right)
    .
    \]
\end{corollary}
} 

\NoShow{
\begin{theorem}
\label{thm:right-update}
  Let $ C = \left( 
    \begin{array}{c | c}
    C_{TL} & - C_{BL}^T \\ \hline
    C_{BL} & C_{BR} 
    \end{array}
    \right) $ be skew symmetric and $ B = \left( 
    \begin{array}{c | c}
    B_{TL} & 0 \\ \hline
    B_{BL} & I 
    \end{array}
    \right) $ be nonsingular (e.g., a unit lower-triangular matrix).  Consider the equality  $ C^{\rm +} = B^{-1}  C B^{-T} $ in terms of quadrants.  Then
    \[
    \left( 
    \begin{array}{c | c}
    C_{TL}^{\rm +} & - C_{BL}^{T \! \rm +} \\ \hline
    C_{BL}^{\rm +} & C_{BR}^{\rm +} 
    \end{array}
    \right)
    =
    \left( 
    \begin{array}{c | c}
    B_{TL} & 0 \\ \hline
    B_{BL} & I 
    \end{array}
    \right)^{-1}
    \left( 
    \begin{array}{c | c}
    C_{TL} & - C_{BL}^T \\ \hline
    C_{BL} & C_{BR} 
    \end{array}
    \right)
    \left( 
    \begin{array}{c | c}
    B_{TL} & 0 \\ \hline
    B_{BL} & I 
    \end{array}
    \right)^{-T}  
    \]
    implies that
    \begin{itemize}
        \item 
        $ C_{TL}^{\rm +} = B_{TL}^{-1} C_{TL} B_{TL}^{-T} $.
        \item 
        $ 
        C_{BL}^{\rm +} = - B_{BL} C_{TL}^{\rm +} 
        + C_{BL} B_{TL}^{-T}$.
        \item 
        $  C_{BR}^{\rm +}
    =
    C_{BR} - 
    B_{BL} C_{TL}^{\rm +} B_{TR}^T + ( B_{BL}  C_{BL}^{{\rm +} \!T}
    -
    C_{BL}^{\rm +} B_{TR}^T )  
        $.

    \end{itemize}
\end{theorem}

\begin{proof}
\[
\begin{array}{l}
    \left( 
    \begin{array}{c | c}
    C_{TL} & - C_{BL}^T \\ \hline
    C_{BL} & C_{BR} 
    \end{array}
    \right)
    =
\left( 
    \begin{array}{c | c}
    B_{TL} & 0 \\ \hline
    B_{BL} & I 
    \end{array}
    \right)
\left( 
    \begin{array}{c | c}
    C_{TL}^{\rm +} & - C_{BL}^{{\rm +} \!T } \\ \hline
    C_{BL}^{\rm +} & C_{BR}^{\rm +} 
    \end{array}
    \right)
\left( 
    \begin{array}{c | c}
    B_{TL}^T & B_{BL}^T \\ \hline
    0 & I 
    \end{array}
    \right) \\
    \NoShow{
~~ =
\left( 
    \begin{array}{c | c}
    B_{TL} & 0 \\ \hline
    B_{BL} & I 
    \end{array}
    \right)
 \left( 
    \begin{array}{c | c}
    C_{TL}^{\rm +} B_{TL}^T & 
    C_{TL}^{\rm +} B_{TR}^T - C_{BL}^{{\rm +} \!T } \\ \hline
    C_{BL}^{\rm +} B_{TL}^T  & C_{BL}^{\rm +} B_{TR}^T + C_{BR}^{\rm +} 
    \end{array}
    \right)   \\
    } 
~~ =
 \left( 
    \begin{array}{c | c}
    B_{TL} C_{TL}^{\rm +} B_{TL}^T & 
      \star \\ \hline
    B_{BL} C_{TL}^{\rm +} B_{TL}^T +
    C_{BL}^{\rm +} B_{TL}^T  & 
    B_{BL} C_{TL}^{\rm +} B_{TR}^T - B_{BL}  C_{BL}^{{\rm +} \!T}
    +
    C_{BL}^{\rm +} B_{TR}^T + C_{BR}^{\rm +} 
    \end{array}
    \right)   
    \end{array}
\]
from  which the desired results follow via straight-forward manipulation.
\begin{itemize}
    \item 
    $ C_{TL}^{\rm +} = B_{TL}^{-1} C_{TL} B_{TL}^{-T} $.
    \item 
    $ C_{BL}^{\rm +} = ( C_{BL} - B_{BL} C_{TL}^{\rm +} B_{TL}^T )  B_{TL}^{-T}=
    C_{BL}B_{TL}^{-T} - B_{BL} C_{TL}^{\rm +}
    $
    \item 
    $ C_{BR}^{\rm +}
    =
    C_{BR} - 
    B_{BL} C_{TL}^{\rm +} B_{TR}^T + B_{BL}  C_{BL}^{{\rm +} \!T}
    -
    C_{BL}^{\rm +} B_{TR}^T   
    $
\end{itemize}
\end{proof}
} 

\begin{figure}[tb!]
    \input LTLt_piv_unb.tex
\centering
    
    \footnotesize
    \FlaAlgorithm    
    \caption{The unblocked right-looking (modified Parlett-Reid) and left-looking (modified Aasen) algorithms.}
    \label{fig:LTLt_unb}

\end{figure}

\subsection{The (modified) Parlett-Reid algorithm}
\label{sec:simple}

With these tools, we describe 
an algorithm for the skew-symmetric problem given in~\cite{Wimmer2012} that is modified from one  first proposed by Parlett and Reid for the symmetric problem~\cite{ParlettReid}.

Partition
\[
X \rightarrow
\left( \begin{array}{c | c | c}
0 & - \chi_{21} & - x_{31}^T \\ \hline
\chi_{21} &  0 & - x_{32}^T \\ \hline
x_{31} & x_{32} & X_{33}
\end{array} \right)
\]
The purpose of the game is to find a Gauss transform to introduce zeroes in $ x_{31} $:
\begin{equation}
\label{eqn:1}
\left( \begin{array}{c | c | c}
0 & - \chi_{21} & 0 \\ \hline
\chi_{21} &  0 & -  x_{32}^{{\rm +} \! T} \\ \hline
0 & x_{32}^{\rm +} &  X_{33}^{\rm +}
\end{array} \right)
:=
\left( \begin{array}{c | c | c}
1 & 0 & 0 \\ \hline
0 & 1 & 0 \\ \hline
0 & -l_{32} & I
\end{array} \right)
\left( \begin{array}{c | c | c}
0 & - \chi_{21} & - x_{31}^T \\ \hline
\chi_{21} &  0 & - x_{32}^T \\ \hline
x_{31} & x_{32} & X_{33}
\end{array} \right)
\left( \begin{array}{c | c | c}
1 & 0 & 0 \\ \hline
0 & 1 & -l_{32}^T \\ \hline
0 & 0 & I
\end{array} \right)
.
\end{equation}
Here, the $ {}^+ $ submatrices equal the contents of  the indicated parts of the matrix after the update.
Equation~(\ref{eqn:1}) suggests
updating
\begin{itemize}
    \item 
    $ l_{32} := x_{31} / \chi_{21} $. 
    \item 
    $ x_{31}:= 0 $.  
    \item
    \setlength{\arraycolsep}{3pt}
    $ 
    \left( \begin{array}{c | c}
    \chi_{22} & x_{32}^T \\ \hline
    x_{32} & X_{22} 
    \end{array} \right)
    :=
    \left( \begin{array}{c | c}
    1 & 0 \\ \hline
    - l_{32} & I 
    \end{array} \right)
    \left( \begin{array}{c | c}
    0 & - x_{32}^T \\ \hline
    x_{32} & X_{22} 
    \end{array} \right)
     \left( \begin{array}{c | c}
    1 & - l_{32}^T \\ \hline
    0  & I 
    \end{array} \right) 
    \NoShow{\\
    &=&
    \left( \begin{array}{c | c}
    0 & x_{32}^T \\ \hline
    x_{32} & X_{22} - l_{32} x_{32}^T 
    \end{array} \right)
    \left( \begin{array}{c | c}
    1 & - l_{32}^T \\ \hline
    0  & I 
    \end{array} \right)}
    =
    \left( \begin{array}{c | c}
    0 & - x_{32}^T \\ \hline
    x_{32} & X_{22} + ( l_{32} x_{32}^T
    - x_{32} l_{32}^T )
    \end{array} \right)
    $.

    \item
    Continue the factorization with the updated
    $
    \left( \begin{array}{c | c}
    \chi_{22} & - x_{32}^T \\ \hline
    x_{32} & X_{22}  
    \end{array} \right)$.

\end{itemize}
    In practice, $ X_{22} $ is updated by a skew-symmetric rank-2 update (meaning only the lower-triangular part is affected).
The resulting algorithm, in FLAME notation, is given in Figure~\ref{fig:LTLt_unb}.
The partitioning and repartitioning in that algorithm is consistent with the use of the thick lines and the choice of subscripting  earlier in this section.  
\NoShow{
Matlab code is given in Figure~\ref{fig:LTLt_unb_right_matlab}.
}

\NoShow{
Equation (\ref{eqn:1}) can be rearranged as
\[
\left( \begin{array}{c | c | c}
0 & - \chi_{21} & - x_{31}^T \\ \hline
\chi_{21} &  0 & - x_{32}^T \\ \hline
x_{31} & x_{32} & X_{33}
\end{array} \right)
=
\begin{array}[t]{c}
\underbrace{
\left( \begin{array}{c | c | c}
1 & 0 & 0 \\ \hline
0 & 1 & 0 \\ \hline
0 & l_{31} & I
\end{array} \right)
} \\
L_0 
\end{array}
\left( \begin{array}{c | c | c}
0 & - \chi_{21} & 0 \\ \hline
\chi_{21} &  0 & - \widehat x_{32}^T \\ \hline
0 & \widehat x_{32} & \widehat X_{33}
\end{array} \right)
\begin{array}[t]{c}
\underbrace{
\left( \begin{array}{c | c | c}
1 & 0 & 0 \\ \hline
0 & 1 & l_{31}^T \\ \hline
0 & 0 & I
\end{array} \right)
}
\\
L_0^T
\end{array}.
\]
After $ n-1 $ iterations, we find that
\[
X
=
L_0 
\cdots
L_{n-2} 
T
L_{n-2}^T 
\cdots
L_0^T,
\]
where $ T $ is a tridiagonal skew-symmetric matrix.

The observation that
\begin{equation}
    \label{eqn1:L1}
\left( \begin{array}{c I c | c}
L_{00} & 0 & 0 \\ \whline
l_{10} & 1 & 0 \\ \hline
L_{20} & 0 & I 
\end{array} \right)
\left( \begin{array}{c I c | c}
I & 0 & 0 \\ \whline
0 & 1 & 0 \\ \hline
0 & l_{21} & I 
\end{array} \right)
=
\left( \begin{array}{c I c | c}
L_{00} & 0 & 0 \\ \whline
l_{10} & 1 & 0 \\ \hline
L_{20} & l_{21} & I 
\end{array} \right)
\end{equation}
tells us each encountered
$ \left( \begin{array}{c}
0 \\ \hline
1 \\ \hline
l_{31}= 
\end{array} \right) $ becomes 
the corresponding column of a unit lower-triangular matrix $ L $ so that 
\[
X =
L X L^T =
\left( \begin{array}{c | c}
1 & 0 \\ \hline
0 & \widetilde L
\end{array} \right)
T
\left( \begin{array}{c | c}
1 & 0 \\ \hline
0 & \widetilde L^T
\end{array} \right),
\]
where $ L $ and $ \widetilde L $ are unit lower triangular matrices.
The strictly lower triangular part of $ L $ overwrites $ X $ below the first subdiagonal.  In other words, in Matlab notation,
\begin{center}
    {\tt Ltilde = tril( X( 2:n, 1:n-1 ), -1 ) + eye( n-1, n-1 )}.
\end{center}
}

\section{Systematic derivation of a family of algorithms}

We now turn to how multiple  algorithms can be systematically derived from specifications.

\subsection{The FLAME workflow}
\label{sec:FLAME}

\begin{figure}[tb!]

\includegraphics[width=\textwidth]{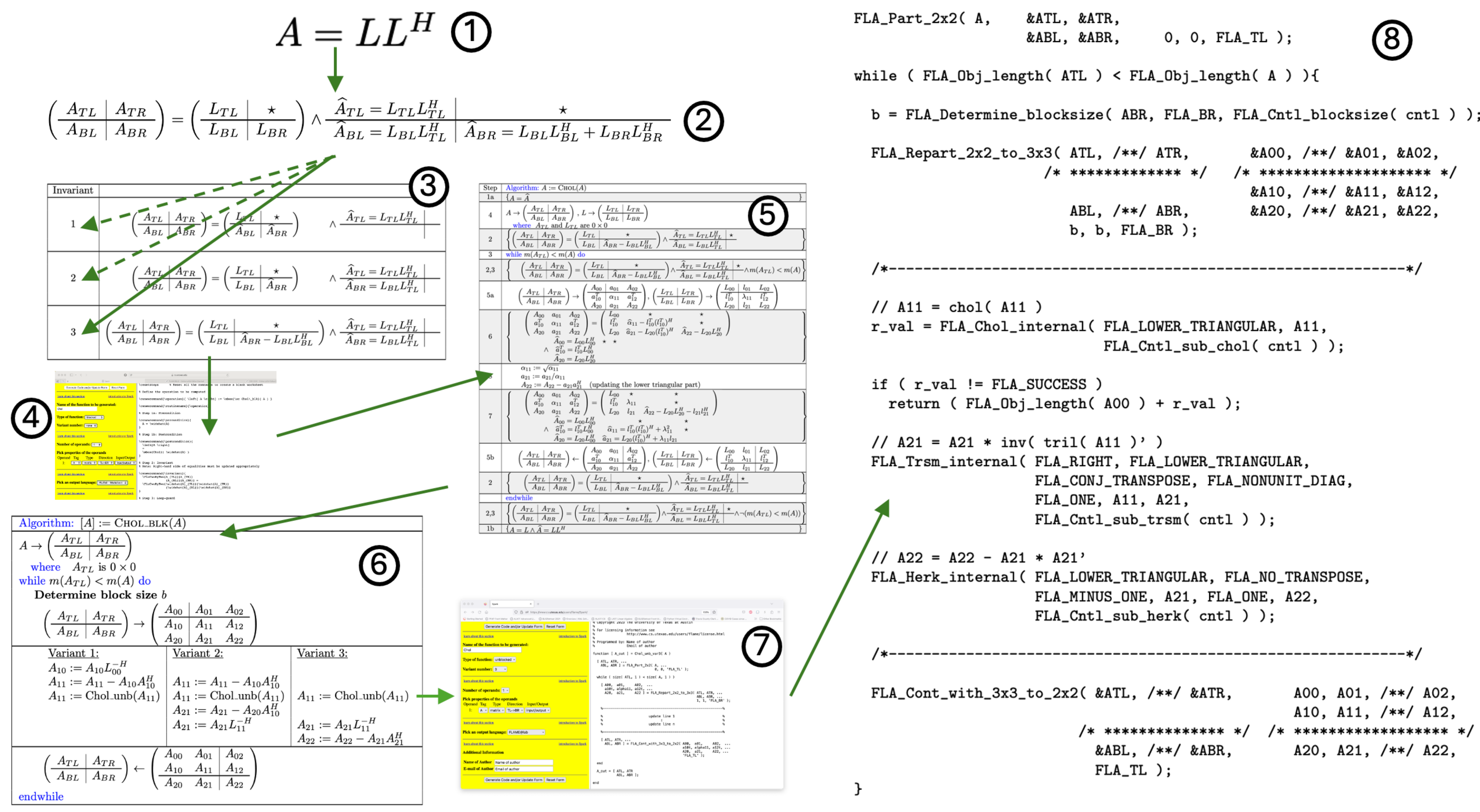}

\caption{The FLAME methodology workflow.}
\label{fig:workflow}
\end{figure}

   We briefly review how the FLAME methodology supports the systematic discovery of families of algorithms, using the Cholesky factorization as an example.  Together with the translation of those algorithms into code using a FLAME API is what we now call the {\em FLAME methodology workflow} (Fig.~\ref{fig:workflow}).
   
A key advance that enables rapid discovery of algorithms was the presentation of algorithms without explicit indexing, what we now call the {\em FLAME notation}~\cite{FLAME_WoCo,FLAME,inverse-siam}.  This is illustrated by \textcircled{6} in Fig.~\ref{fig:workflow} for the three blocked algorithmic variants for Cholesky factorizing.
    
    Embracing the FLAME notation has enabled the application of formal derivation techniques to this domain~\cite{FLAME_WoCo,FLAME_TR,Recipe,TSoPMC}.
   Using the Cholesky factorization in Fig.~\ref{fig:workflow}, one starts with \textcircled{1} the definition of the operation from which the \textcircled{2} Partitioned Matrix Expression (PME)  (a recursive definition of the operation) is derived.  From this, \textcircled{3} a complete set of loop invariants (logical conditions that captures the state of variables before and after each iteration) can be systematically deduced.  \textcircled{4} A menu generates a \textcircled{5} worksheet outline, which is used to derive (hand in hand  with their proofs of correctness) \textcircled{6} algorithmic variants, presented using the FLAME notation.  
    Whole families of algorithms for a broad range of DLA operations (within and beyond LAPACK) have been systematically derived~\cite{FLAME,TSoPMC,Recipe,Bientinesi:2008:FAR:1377603.1377606,10.1145/3544585.3544597}. 
    By adopting APIs that mirror the FLAME notation, correct algorithms can be translated (for example using an automated system \textcircled{7}) to correct code \textcircled{8},
    e.g. the FLAMEC API used by the  libflame DLA library~\cite{libflame_github,libflamebook,CiSE09}.

    We now turn to applying this process to the problem of skew-symmetric triangular tridiagionalization.

\subsection{Specification}
\label{sec:specification}

Given a skew-symmetric matrix $ X $,
the goal is to compute a unit lower triangular matrix $ L $ and tridiagonal matrix $ T $ such that $ X = L T L^T $, overwriting $ X $ with $ T $, provided such a factorization exists. We omit pivoting for now---it will be addressed in Section~\ref{sec:pivoting}.
We formalize this goal as a \emph{precondition} 
$
X = \widehat X \wedge ( \exists L, T ~ \vert ~ \widehat X = L T L^T ) 
$
and \emph{postcondition} 
$ X = T \wedge \widehat X = L T L^T $, where $\widehat X $ equals the original contents of $ X $ and the special structures of the various matrices are implicit.
Since in practice the strictly lower triangular part of $ L $ typically overwrites the entries below the first subdiagonal of $ T $, the first column of $ L $ equals $ e_0 $.  However, as was pointed out in~\cite{Miroslav2011}, this is only one choice for the first column of $ L $.
Indeed, if
\[
\begin{array}[t]{c}
\underbrace{
\left( \begin{array}{ c | c}
0 & - \widehat  x_{21}^T \\ \hline
\widehat x_{21} & \widehat X_{22}
\end{array} \right)
} \\
\widehat X
\end{array}
=
\begin{array}[t]{c}
\underbrace{
\left( \begin{array}{c | c}
1 & 0 \\ \hline
l_{21} & L_{22}
\end{array}
\right)
} \\
L
\end{array}
\begin{array}[t]{c}
\underbrace{
\left( \begin{array}{c | c}
0 & - \tau_{21} e_f^T \\ \hline
\tau_{21} e_f & T_{22}
\end{array}
\right)
} \\
T
\end{array}
\begin{array}[t]{c}
\underbrace{
\left( \begin{array}{c | c}
1 & 0 \\ \hline
l_{21} & L_{22}
\end{array}
\right)^T
} \\
L^T
\end{array}
\]
for some choice of $ l_{21} $, 
then 
\[
\left( \begin{array}{c | c}
1 & 0 \\ \hline
- l_{21} & I
\end{array}
\right)
\left( \begin{array}{ c | c}
0 & - \widehat x_{21}^T \\ \hline
\widehat x_{21} & \widehat X_{22}
\end{array} \right)
\left( \begin{array}{c | c}
1 & 0 \\ \hline
- l_{21} & I
\end{array}
\right)^T
=
\left( \begin{array}{c | c}
1 & 0 \\ \hline
0 & L_{22}
\end{array}
\right)
\left( \begin{array}{c | c}
0 & - \tau_{21} e_f^T \\ \hline
\tau_{21} e_f & T_{22}
\end{array}
\right)
\left( \begin{array}{c | c}
1 & 0 \\ \hline
0 & L_{22}
\end{array}
\right)^T,
\]
which means the original matrix $ X $ can always be updated by applying the first Gauss transform, defined by $ l_{21} $, from the left and right or, equivalently, 
$
X_{22} := 
X_{22} + ( l_{21} x_{21}^T - x_{21} l_{21}^T )
$,
before executing the algorithm given in Section~\ref{sec:simple}.

\subsection{Deriving the Partitioned Matrix Expession}

In the FLAME methodology, the Partitioned Matrix Expression  (PME) is a recursive definition of the operation to be computed.
One derives it from the specification of the operation by substituting the partitioned matrices into the postcondition.
For most dense linear algebra factorization algorithms that were previously  derived using the FLAME methodology, matrices were partitioned into quadrants.  When the methodology was applied to derive Krylov subspace methods~\cite{Eijkhout20101805}, where upper Hessenberg and tridiagonal matrices are encountered, $ 3 \times 3 $ partitionings were necessary.  Not surprisingly, especially given the algorithm presented in Figure~\ref{fig:LTLt_unb}, this is also found to be the case when deriving algorithms for the $ L T L^T $ factorization.

\NoShow{
Still, let's start by trying to  derive a PME using a more traditional $ 2 \times 2 $ partitioning.
In the postcondition, substitute  the various partitioned matrices:
\begin{eqnarray}
    \nonumber
\lefteqn{
\FlaTwoByTwo
  {X_{TL}}{\star}
  {X_{BL}}{X_{BR}}
=
\FlaTwoByTwo
  { T_{TL}}{ \star}
  { \tau_{BL} e_l e_f^T }{ T_{BR}} \wedge}  \\
  \label{eqn:constraint_0}
  && 
\FlaTwoByTwo
  {\widehat X_{TL}}{\star}
  {\widehat X_{BL}}{\widehat X_{BR}} 
  =
\FlaTwoByTwo
  {L_{TL}}{0}
  {L_{BL}}{L_{BR}}
\FlaTwoByTwo
  { T_{TL}}{ \star}
  { \tau_{BL} e_f e_l^T }{ T_{BR}}
\FlaTwoByTwo
  {L_{TL}^T}{L_{BL}^T}
  {0}{L_{BR}^T}.
\end{eqnarray}
\NoShow{
which can be manipulated to
\[
\begin{array}{l}
\FlaTwoByTwo
  {X_{TL}}{\star}
  {X_{BL}}{X_{BR}}
=
\FlaTwoByTwo
  { T_{TL}}{ \star}
  { \tau_{BL} e_f e_l^T }{ T_{BR}} \\
  ~~~~ \wedge 
\FlaTwoByTwo
  {\widehat X_{TL}}{\star}
  {\widehat X_{BL}}{\widehat X_{BR}} 
  =
\FlaTwoByTwo
  {L_{TL}}{0}
  {L_{BL}}{I}
\FlaTwoByTwo
  { T_{TL}}{ \star}
  { \tau_{BL} L_{BR} e_l e_f^T }{ L_{BR} T_{BR}L_{BR}^T }
\FlaTwoByTwo
  {L_{TL}^T}{L_{BL}^T}
  {0}{I},
  \end{array}
\]
}
From experience with other operations 
(like LU factorization),
two loop invariants are  of particular interest because they 
can be extended to support pivoting: 
\begin{itemize}
    \item
An invariant corresponding to the right-looking algorithm in Section~\ref{sec:simple}:
\[
\FlaTwoByTwo
  {X_{TL}}{\star}
  {X_{BL}}{X_{BR}}
=
\FlaTwoByTwo
  { T_{TL}}{ \star}
  {\tau_{BL} e_l e_f^T }{ \widehat X - \mbox{updated}}
  \wedge
  ~
\mbox{(\ref{eqn:constraint_0})}
  ~
  \wedge
  ~
  \mbox{\begin{minipage}{2.5in}
  what part of $ L $ has been computed.
  \end{minipage}
  }
\]
\item
An invariant that corresponds to what is often called a left-looking algorithm:
\begin{equation}
    \label{eqn:left}
\FlaTwoByTwo
  {X_{TL}}{\star}
  {X_{BL}}{X_{BR}}
=
\FlaTwoByTwo
  { T_{TL}}{ \star}
  {\tau_{BL} e_l e_f^T }{ \widehat X_{BR}}
    \wedge
  ~
\mbox{(\ref{eqn:constraint_0})}
  ~
  \wedge
  ~
  \mbox{\begin{minipage}{2.5in}
  what part of $ L $ has been computed.
  \end{minipage}
  }
\end{equation}
\end{itemize}
The problem lies with capturing what part of $ L $ has been computed, because it involves the first column of $ L_{BR} $.
What we glean from this is that in reasoning about the PME and loop invariant, it pays to expose an additional row and column of the matrices that are involved, just like those that showed up in the discussion in Section~\ref{sec:simple}.
}  

For the PME we find
\begin{eqnarray}
\nonumber
\lefteqn{
\left( \begin{array}{c I c | c}
X_{TL} &  \star
& \star \\ \whline
x_{ML}^T & \chi_{MM} & \star \\ \hline
X_{BL} & x_{BM} & X_{BR}
\end{array} \right)
=
\left( \begin{array}{c I c | c}
T_{TL} & \star & \star \\ \whline
\tau_{ML} e_l^T & 0 &  \star \\ \hline
0  & \tau_{BM} e_f & T_{BR}
\end{array} \right) \wedge 
\left( \begin{array}{c I c | c}
\widehat X_{TL} & 
- \widehat x_{ML} & 
- \widehat X_{BL} \\ \whline
\widehat x_{ML}^T & 0 & 
- \widehat x_{BM}^T \\ \hline
\widehat X_{BL} & \widehat x_{BM} & \widehat X_{BR}
\end{array} \right)
}
\\ 
\label{eqn:PR-1}
& 
= &
\left( \begin{array}{c I c | c}
L_{TL} & 0 & 0 \\ \whline
l_{ML}^T & 1 & 0 \\ \hline
L_{BL} & l_{BM} &  L_{BR}
\end{array}\right)
\left( \begin{array}{c I c | c}
T_{TL} & - \tau_{ML} e_l  & 0 \\ \whline
\tau_{ML} e_l^T  & 0 & -\tau_{BM} e_f^T \\ \hline
0 & \tau_{BM} e_f & T_{BR}
\end{array} \right)
\left( \begin{array}{c I c | c}
L_{TL}^T & l_{ML} & L_{BL}^T \\ \whline
0 & 1 & l_{BM}^T \\ \hline
0 & 0 &  L_{BR}^T
\end{array}\right).
\end{eqnarray}
The $ \star $s capture that those expressions are not stored.  The right hand side of the second condition can be rewritten as
\begin{eqnarray*}
\NoShow{\lefteqn{
\left( \begin{array}{c I c | c}
X_{TL} & \star & \star \\ \whline
x_{ML}^T & \chi_{MM} & \star \\ \hline
X_{BL} & x_{BM} & X_{BR}
\end{array} \right)
=
\left( \begin{array}{c I c | c}
T_{TL} & \star & \star \\ \whline
\tau_{ML} e_l^T & 0 &  \star \\ \hline
0  & \tau_{BM} e_f & T_{BR}
\end{array} \right) \wedge 
\left( \begin{array}{c I c | c}
\widehat X_{TL} & \star & \star \\ \whline
\widehat x_{ML}^T & 0 & \star \\ \hline
\widehat X_{BL} & \widehat x_{BM} & \widehat X_{BR}
\end{array} \right)} \\
} 
\left( \begin{array}{c I c | c}
L_{TL} & 0 & 0 \\ \whline
l_{ML}^T & 1 & 0 \\ \hline
L_{BL} & l_{BM} &  I
\end{array}\right)
\left( \begin{array}{c I c | c}
T_{TL} & - \tau_{ML} e_l  & 0 \\ \whline
\tau_{ML} e_l^T  & 0 & -\tau_{BM} (L_{BR} e_f)^T \\ \hline
0 & \tau_{BM} L_{BR} e_f & L_{BR} T_{BR} L_{BR}^T
\end{array} \right)
\left( \begin{array}{c I c | c}
L_{TL}^T & l_{ML} & L_{BL}^T \\ \whline
0 & 1 & l_{BM}^T \\ \hline
0 & 0 &  I
\end{array}\right).
\end{eqnarray*}
Here
\[
\left( \begin{array}{c | c}
0 & - \tau_{BM} (L_{BR}  e_f)^T \\ \hline
\tau_{BM} L_{BR}  e_f & L_{BR} T_{BR} L_{BR}^T 
\end{array} \right)
=
\begin{array}[t]{c}
\underbrace{
\left( \begin{array}{c | c}
1 & 0 \\ \hline
0 & L_{BR} 
\end{array} \right)
} \\
L_{k} \cdots L_{m-2}
\end{array}
\left( \begin{array}{c | c}
0 & - \tau_{BM} e_f^T \\ \hline
\tau_{BM} e_f & T_{BR} 
\end{array} \right)
\begin{array}[t]{c}
\underbrace{
\left( \begin{array}{c | c}
1 & 0 \\ \hline
0 & L_{BR} 
\end{array} \right)^T
}, \\
L_{m-2}^T \cdots L_{k}^T
\end{array}
\]
which captures that it represents the result at a particular intermediate stage of the calculation
expressed as  the final result but with the yet-to-be-computed transformations not yet applied.%
\footnote{The exact number of Gauss transforms applied at a given step is tricky to account for due to the offset in $L$, leading to infamous ``off by one'' errors.  This becomes inconsequential since we avoid indices in our subsequent reasoning.}
This insight will play an important role in our derivation and deviates from how the FLAME methodology has been traditionally deployed. 

\subsection{Loop invariants}

A loop invariant is a predicate that captures the state of the variables before and after each iteration of the loop.
The strength of the FLAME methodology is that this condition is derived {\em a priori} from the PME so that it can guide the derivation of the loop.
From the PME, taking into account that we eventually wish to add pivoting, we find  the following 
loop invariants%
\footnote{There may be other loop invariants.}%
:
\begin{itemize}
 \item 
Invariant~1 (for the right-looking variant from Section~\ref{sec:simple}):
    {
\setlength{\arraycolsep}{2pt}
  \begin{eqnarray}
  \label{eqn:inv_unb_right_1}
  \lefteqn{
\left( \begin{array}{c I c | c}
X_{TL} & \star & \star \\ \whline
x_{ML}^T & \chi_{MM} & \star \\ \hline
X_{BL} & x_{BM} & X_{BR}
\end{array} \right)
=
\left( \begin{array}{c I c | c}
T_{TL} & \star & \star \\ \whline
 \tau_{ML} e_l^T & 0 &  \star \\ \hline
0  &  \tau_{BM} 
L_{BR} e_f
& L_{BR} T_{BR} L_{BR}^T
\end{array} \right)  \wedge 
\left( \begin{array}{c I c | c}
\widehat X_{TL} & 
- \widehat x_{ML}
& 
- \widehat X_{BL}^T
\\ \whline
\widehat x_{ML}^T & 0 & 
- \widehat x_{BM}^T \\ \hline
\widehat X_{BL} & \widehat x_{BM} & \widehat X_{BR}
\end{array} \right) } \\
\label{eqn:inv_unb_right_2}
&& ~~~~~ =
\left( \begin{array}{c I c | c}
\color{blue} L_{TL} & 0 & 0 \\ \whline
\color{blue} l_{ML}^T & 1 & 0 \\ \hline
\color{blue} L_{BL} & \color{blue} l_{BM} &  I
\end{array}\right)
\left( \begin{array}{c I c | c}
T_{TL} & - \tau_{ML} e_l  & 0 \\ \whline
\tau_{ML} e_l^T  & 0 & -\tau_{BM} (L_{BR} e_f)^T \\ \hline
0 & \tau_{BM} L_{BR} e_f & L_{BR} T_{BR} L_{BR}^T
\end{array} \right)
\left( \begin{array}{c I c | c}
\color{blue} L_{TL}^T & \color{blue} l_{ML} & \color{blue} L_{BL}^T \\ \whline
0 & 1 & \color{blue} l_{BM}^T \\ \hline
0 & 0 &  I
\end{array}\right).
\end{eqnarray}
    }
    \NoShow{
    \item 
    Invariant for  a variation on the right-looking algorithm, which we will call lazy right-looking, that in a current iteration computes the next Gauss transform but does not yet apply it: 
{
\setlength{\arraycolsep}{2pt}
\[
\begin{array}{l}
\left( \begin{array}{c I c | c}
X_{TL} & \star & \star \\ \whline
x_{ML}^T & \chi_{MM} & \star \\ \hline
X_{BL} & x_{BM} & X_{BR}
\end{array} \right)
=
\left( \begin{array}{c I c | c}
T_{TL} & ~~~~~~~\star ~~~~~~~& \star \\ \whline
 \tau_{ML} e_l^T & 
 \multicolumn{2}{c}{
 \multirow{2}{*}{
 $
 \left( \begin{array}{c | c}
 1 &  0 \\ \hline
l_{BM} & L_{BR} 
\end{array}
\right)
 \left( \begin{array}{c | c}
 0 &  \star \\ \hline
\tau_{BM} 
e_f
& T_{BR} 
\end{array}
\right)
\left( \begin{array}{c | c}
 1 &  0 \\ \hline
l_{BM} & L_{BR} 
\end{array}
\right)^T
$
 }
 }
  \\ \cline{1-1}
0  &  \multicolumn{2}{c}{}
\end{array} \right)
\wedge \\
\\[-0.15in]
~~ 
\left( \begin{array}{c I c | c}
\widehat X_{TL} & \star & \star \\ \whline
\widehat x_{ML}^T & 0 & \star \\ \hline
\widehat X_{BL} & \widehat x_{BM} & \widehat X_{BR}
\end{array} \right)  =
\left( \begin{array}{c I c | c}
\color{blue} L_{TL} & 0 & 0 \\ \whline
\color{blue} l_{ML}^T & 1 & 0 \\ \hline
\color{blue} L_{BL} & \color{blue} l_{BM} &  L_{BR}
\end{array}\right)
\left( \begin{array}{c I c | c}
T_{TL} & - \tau_{ML} e_l  & 0 \\ \whline
\tau_{ML} e_l^T  & 0 & -\tau_{BM} e_f^T \\ \hline
0 & \tau_{BM} e_f & T_{BR}
\end{array} \right)
\left( \begin{array}{c I c | c}
\color{blue} L_{TL}^T & \color{blue} l_{ML} & \color{blue} L_{BL}^T \\ \whline
0 & 1 & \color{blue} l_{BM}^T \\ \hline
0 & 0 &  L_{BR}^T
\end{array}\right).
\end{array}
\]
}
}
\item 
    Invariant~2a (later used to derive a blocked fused right-looking variant):
What we will see is that Invariant~1 leads to a blocked algorithm that casts most computation in an update that is a matrix-matrix (level-3 BLAS-like) operation, but requires an additional matrix-vector (level-2 BLAS-like) operation that forces data to be brought into memory an additional time.  
The following loop invariant lead to algorithms that avoid this, shifting some computation from one iteration to an adjacent iteration by delaying the application of the most recently computed Gauss transform:
    {
\setlength{\arraycolsep}{2pt}
\begin{eqnarray}
\label{eqn:inv_unb_right_2a-1}
\lefteqn{
\left( \begin{array}{c I c | c}
X_{TL} & \star & \star \\ \whline
x_{ML}^T & \chi_{MM} & \star \\ \hline
X_{BL} & x_{BM} & X_{BR}
\end{array} \right)
=
\left( \begin{array}{c I c | c}
T_{TL} & ~~~~~~~~~~~~~\star ~~~~~~~~~~~~~& \star \\ \whline
 \tau_{ML} e_l^T & 
 \multicolumn{2}{c}{
 \multirow{2}{*}{
 $
 \left( \begin{array}{c | c}
 1 &  0 \\ \hline
\color{blue} l_{BM} & L_{BR} 
\end{array}
\right)
 \left( \begin{array}{c | c}
 0 &  \star \\ \hline
\tau_{BM} 
e_f
& T_{BR} 
\end{array}
\right)
\left( \begin{array}{c | c}
 1 &  \color{blue} l_{BM}^T \\ \hline
0 & L_{BR}^T
\end{array}
\right)
$
 }
 }
  \\ \cline{1-1}
0  &  \multicolumn{2}{c}{}
\end{array} \right)
\wedge }\\
\label{eqn:inv_unb_right_2a-2} 
&&
\left( \begin{array}{c I c | c}
\widehat X_{TL} & \star & \star \\ \whline
\widehat x_{ML}^T & 0 & \star \\ \hline
\widehat X_{BL} & \widehat x_{BM} & \widehat X_{BR}
\end{array} \right)  =
\left( \begin{array}{c I c | c}
\color{blue} L_{TL} & 0 & 0 \\ \whline
\color{blue} l_{ML}^T & 1 & 0 \\ \hline
\color{blue} L_{BL} & \color{blue} l_{BM} &  L_{BR}
\end{array}\right)
\left( \begin{array}{c I c | c}
T_{TL} & - \tau_{ML} e_l  & 0 \\ \whline
\tau_{ML} e_l^T  & 0 & -\tau_{BM} e_f^T \\ \hline
0 & \tau_{BM} e_f & T_{BR}
\end{array} \right)
\left( \begin{array}{c I c | c}
\color{blue} L_{TL}^T & \color{blue} l_{ML} & \color{blue} L_{BL}^T \\ \whline
0 & 1 & \color{blue} l_{BM}^T \\ \hline
0 & 0 &  L_{BR}^T
\end{array}\right).
\end{eqnarray}
}%
 \item 
Invariant~2b (later used to derive an alternative blocked fused right-looking variant):
Alternatively, rather than delaying the application of the most recently computed Gauss transform, one can compute one additional Gauss transform, but not yet apply it:
    {
\setlength{\arraycolsep}{2pt}
  \begin{eqnarray}
  \label{eqn:inv_unb_right_2b-1}
  \lefteqn{
\left( \begin{array}{c I c | c}
X_{TL} & \star & \star \\ \whline
x_{ML}^T & \chi_{MM} & \star \\ \hline
X_{BL} & x_{BM} & X_{BR}
\end{array} \right)
=
\left( \begin{array}{c I c | c}
T_{TL} & \star & \star \\ \whline
 \tau_{ML} e_l^T & 0 &  \star \\ \hline
0  &  \tau_{BM} 
e_f
& L_{BR} T_{BR} L_{BR}^T
\end{array} \right)  \wedge 
\left( \begin{array}{c I c | c}
\widehat X_{TL} & 
- \widehat x_{ML}
& 
- \widehat X_{BL}^T
\\ \whline
\widehat x_{ML}^T & 0 & 
- \widehat x_{BM}^T \\ \hline
\widehat X_{BL} & \widehat x_{BM} & \widehat X_{BR}
\end{array} \right) } \\
\label{eqn:inv_unb_right_2b-2}
&& ~~~~~ =
\left( \begin{array}{c I c | c}
\color{blue} L_{TL} & 0 & 0 \\ \whline
\color{blue} l_{ML}^T & 1 & 0 \\ \hline
\color{blue} L_{BL} & \color{blue} l_{BM} &  I
\end{array}\right)
\left( \begin{array}{c I c | c}
T_{TL} & - \tau_{ML} e_l  & 0 \\ \whline
\tau_{ML} e_l^T  & 0 & -\tau_{BM} ({\color{blue} L_{BR} e_f})^T \\ \hline
0 & \tau_{BM} 
\color{blue} L_{BR} e_f & L_{BR} T_{BR} L_{BR}^T
\end{array} \right)
\left( \begin{array}{c I c | c}
\color{blue} L_{TL}^T & \color{blue} l_{ML} & \color{blue} L_{BL}^T \\ \whline
0 & 1 & \color{blue} l_{BM}^T \\ \hline
0 & 0 &  I
\end{array}\right).
\end{eqnarray}
    }
    \item 
    Invariant~3 (for left-looking variants):
    {
\setlength{\arraycolsep}{2pt}
  \begin{eqnarray}
  \label{eqn:inv_unb_left-1}
  \lefteqn{
\left( \begin{array}{c I c | c}
X_{TL} & \star & \star \\ \whline
x_{ML}^T & \chi_{MM} & \star \\ \hline
X_{BL} & x_{BM} & X_{BR}
\end{array} \right)
=
\left( \begin{array}{c I c | c}
T_{TL} & \star & \star \\ \whline
\tau_{ML} e_l^T & 0 &  \star \\ \hline
0   & \widehat x_{BM}  & \widehat X_{BR}
\end{array} \right) \wedge 
\left( \begin{array}{c I c | c}
\widehat X_{TL} & 
- \widehat x_{ML}  & 
- \widehat x_{BL}^T \\ \whline
\widehat x_{ML}^T & 0 & 
- \widehat x_{BM}^T \\ \hline
\widehat X_{BL} & \widehat x_{BM} & \widehat X_{BR}
\end{array} \right)} \\
\label{eqn:inv_unb_left-2}
& & ~~~~~ =
\left( \begin{array}{c I c | c}
\color{blue} L_{TL} & 0 & 0 \\ \whline
\color{blue} l_{ML}^T & 1 & 0 \\ \hline
\color{blue} L_{BL} & \color{blue} l_{BM} &  L_{BR}
\end{array}\right)
\left( \begin{array}{c I c | c}
T_{TL} & - \tau_{ML} e_l  & 0 \\ \whline
\tau_{ML} e_l^T  & 0 & -\tau_{BM} e_f^T \\ \hline
0 & \tau_{BM} e_f & T_{BR}
\end{array} \right)
\left( \begin{array}{c I c | c}
\color{blue} L_{TL}^T & \color{blue} l_{ML} & \color{blue} L_{BL}^T \\ \whline
0 & 1 & \color{blue} l_{BM}^T \\ \hline
0 & 0 &  L_{BR}^T
\end{array}\right).
\end{eqnarray}
    }
\end{itemize}
Note that
\begin{itemize}
\item
In all cases, only the parts of $ L $ highlighted in blue have been computed.
\item
The constraints in~(\ref{eqn:inv_unb_right_2}),
(\ref{eqn:inv_unb_right_2a-2}),(\ref{eqn:inv_unb_right_2b-2}) and~(\ref{eqn:inv_unb_left-2}) are equivalent but stated somewhat differently.  This is a choice that we found makes deriving algorithms corresponding to the respective invariants slightly easier.
\end{itemize}
We will see that the loop that implements the algorithm is prescribed by the pre- and postconditions, the loop invariant, and how we choose to stride through the operands.

\subsection{Right-looking (modified Parlett-Reid) algorithm}
\label{sec:unb-right}

\resetsteps      


\renewcommand{\operation}{ \left[ X, L \right] := \mbox{\sc LTLt\_unb\_right}( X )}

\renewcommand{\routinename}{\operation}


\renewcommand{\precondition}{
  X = \widehat{X}
  \wedge ( \exists L, T ~ \vert ~ \widehat X = L T L^T ) 
}


\renewcommand{\postcondition}{ 
  X = T \wedge \widehat X =  L T L^T 
}


\renewcommand{\invariant}{
\setlength{\arraycolsep}{2pt}
\left( \begin{array}{c I c | c}
X_{TL} & \star & \star \\ \whline
x_{ML}^T & \chi_{MM} & \star \\ \hline
X_{BL} & x_{BM} & X_{BR}
\end{array} \right)
=
\left( \begin{array}{c I c | c}
T_{TL} & \star & \star \\ \whline
 \tau_{ML} e_l^T & 0 &  \star \\ \hline
0  &  \tau_{BM} 
L_{BR} e_f
& L_{BR} T_{BR} L_{BR}^T
\end{array} \right)  \wedge \cdots
}


\renewcommand{\guard}{
  m( X_{TL} ) < m( X )-1
}


\renewcommand{\partitionings}{
\setlength{\arraycolsep}{2pt}
  \vspace{-0.13in}
  \begin{tabular}{@{}l}
  $L = I$ \\
  $X \rightarrow
  \left( \begin{array}{c I c | c }
  {X_{TL}} & {x_{TM}} & {X_{TR}} \\ \whline
  {x_{ML}^T} & {\chi_{MM}} & {x_{MR}^T} \\ \hline
  {X_{BL}} & {x_{BM}} & {X_{BR}} 
  \end{array} \right)
  $
,
  $
  L \rightarrow 
\left( \begin{array}{c I c | c }
  {L_{TL}} & {l_{TM}} & {L_{TR}} \\ \whline
  {l_{ML}^T} & {\lambda_{MM}} & {l_{MR}^T} \\ \hline
  {L_{BL}} & {l_{BM}} & {L_{BR}} 
  \end{array} \right)
  $
,
  $
  T \rightarrow 
  \left( \begin{array}{c I c | c }
  {T_{TL}} & {t_{TM}} & {T_{TR}} \\ \whline
  {t_{ML}^T} & {\tau_{MM}} & {t_{MR}^T} \\ \hline
  {T_{BL}} & {t_{BM}} & {T_{BR}} 
  \end{array} \right)
  $
  \end{tabular}
}

\renewcommand{\partitionsizes}{
$ X_{TL} $ is $ 0 \times 0 $,
$ L_{TL} $ is $ 0 \times 0 $,
$ T_{TL} $ is $ 0 \times 0 $
}


\renewcommand{\repartitionings}{
\setlength{\arraycolsep}{1pt}
$ 
  \left( \begin{array}{c I c | c }
  {X_{TL}} & {x_{TM}} & {X_{TR}} \\ \whline
  {x_{ML}^T} & {\chi_{MM}} & {x_{MR}^T}  \\ \hline
  {X_{BL}} & {x_{BM}} & {X_{BR}} 
  \end{array} \right)
  \rightarrow
  \left( \begin{array}{c I c | c | c }
  {X_{00}} & {x_{01}} & x_{02} & {X_{03}} \\ \whline
  {x_{10}^T} & {\chi_{11}} & {\chi_{12}} & {x_{13}^T}  \\ \hline
  {x_{20}^T} & {\chi_{21}} & {\chi_{22}} & {x_{23}^T}  \\ \hline
  {X_{30}} & {x_{31}} & x_{32} & {X_{33}}
  \end{array} \right)
  $
  ,
  $
  \left( \begin{array}{c I c | c }
  {L_{TL}} & {l_{TM}} & {L_{TR}} \\ \whline
  {l_{ML}^T} & {\lambda_{MM}} & {l_{MR}^T} \\ \hline
  {L_{BL}} & {l_{BM}} & {L_{BR}} 
  \end{array} \right) \rightarrow \cdots
  $
,
  $
  \left( \begin{array}{c I c | c }
  {T_{TL}} & {t_{TM}} & {T_{TR}} \\ \whline
  {t_{ML}^T} & {\tau_{MM}} & {t_{MR}^T} \\ \hline
  {T_{BL}} & {t_{BM}} & {T_{BR}} 
  \end{array} \right)
  \rightarrow \cdots
  $
}

\renewcommand{\repartitionsizes}{
  $ \chi_{11} $ is $ 1 \times 1 $,
  $ \lambda_{11} $ is $ 1 \times 1 $,
  $ \tau_{11} $ is $ 1 \times 1 $}


\renewcommand{\moveboundaries}{
\setlength{\arraycolsep}{1pt}
$ 
  \left( \begin{array}{c I c | c }
  {X_{TL}} & {x_{TM}} & {X_{TR}} \\ \whline
  {x_{ML}^T} & {\chi_{MM}} & {x_{MR}^T}  \\ \hline
  {X_{BL}} & {x_{BM}} & {X_{BR}} 
  \end{array} \right)
  \leftarrow
  \left( \begin{array}{c | c I c | c }
  {X_{00}} & {x_{01}} & x_{02} & {X_{03}} \\ \hline
  {x_{10}^T} & {\chi_{11}} & {\chi_{12}} & {x_{13}^T}  \\ \whline
  {x_{20}^T} & {\chi_{21}} & {\chi_{22}} & {x_{23}^T}  \\ \hline
  {X_{30}} & {x_{31}} & x_{32} & {X_{33}}
  \end{array} \right)
  $
    ,
  $
  \left( \begin{array}{c I c | c }
  {L_{TL}} & {l_{TM}} & {L_{TR}} \\ \whline
  {l_{ML}^T} & {\lambda_{MM}} & {l_{MR}^T} \\ \hline
  {L_{BL}} & {l_{BM}} & {L_{BR}} 
  \end{array} \right) \leftarrow \cdots
  $
,
  $
  \left( \begin{array}{c I c | c }
  {T_{TL}} & {t_{TM}} & {T_{TR}} \\ \whline
  {t_{ML}^T} & {\tau_{MM}} & {t_{MR}^T} \\ \hline
  {T_{BL}} & {t_{BM}} & {T_{BR}} 
  \end{array} \right)
  \leftarrow \cdots
  $
  }


\renewcommand{\beforeupdate}{
\setlength{\arraycolsep}{3pt}
$
\left( \begin{array}{c I c | c | c}
X_{00} & \star & \star & \star \\ \whline
x_{10}^T & \chi_{11} & 
\star & \star \\ \hline
x_{20}^T & 
\chi _{21} & \chi_{22} &  \star \\ \hline
X_{30} & x_{31}
& x_{32} & X_{33}
\end{array} \right)
=
\left( \begin{array}{c I c | c | c}
 T_{00} & \star & \star & \star \\ \whline
\tau_{10} e_l^T  & 0 & 
~~~\star~~~ & \star \\ \hline
0 & 
\multirow{2}{*}{$
\tau_{21}
\left( \begin{array}{c}
1   \\ \hline
l_{32} 
\end{array}
\right)
$}
& 
\multicolumn{2}{c}
{
\multirow{2}{*}{
$
\left( \begin{array}{c | c}
1 &  0 \\ \hline
l_{32} &
L_{33}
\end{array}
\right)
\left( \begin{array}{c | c}
0 &  \star \\ \hline
\tau_{32} e_f &
T_{33}
\end{array}
\right)
\left( \begin{array}{c | c}
1 &  l_{32}^T \\ \hline
0 &
L_{33}^T
\end{array}
\right)
$ } }
\\
\cline{1-1} 
 0 & 
&  \multicolumn{2}{c}{}
\end{array} \right)
\wedge \cdots 
$
}


\renewcommand{\afterupdate}{
\setlength{\arraycolsep}{2pt}
$
\left( \begin{array}{c | c I c | c}
X_{00} & \star & \star & \star \\ \hline
x_{10}^{T} & \chi_{11} & 
\star & \star \\ \whline
x_{20}^{T} & 
\chi_{21} & \chi_{22} &  \star \\ \hline
X_{30} & x_{31}
& x_{32} & X_{33}
\end{array} \right)
=
\left( \begin{array}{c | c I c | c}
T_{00} & \star & \star & \star \\ \hline
 \tau_{10} e_l^T & 0 & 
~~~\star~~~ & \star \\ \whline
0 & 
\tau_{21} & 0 & \star
\\
\hline
 0 & 
 0 & 
\tau_{32}
L_{33} e_f
 &
L_{33}
T_{33} 
L_{33}^T
\end{array} \right) \wedge \cdots
$
}


\renewcommand{\update}{
$
  \begin{array}{l}
    l_{32} := x_{31} /  \chi_{21} \\ 
    x_{31} := 0 \\
    X_{33} := X_{33} + ( l_{32} x_{32}^T - x_{32} l_{32}^T ) 
    \mbox{~~~~(skew symmetric rank-2 update)}
  \end{array}
$
}



\begin{figure}[tbp]


{
\scriptsize


\begin{tabular}{| c | p{0.9\textwidth} |}\hline
Step & 
$\mbox{\color{blue}Algorithm:~}\routinename$
\\ \hline
\rowcolor{lightgray!25}   1a &%
$ \left\{\begin{minipage}[t]{0.88\textwidth} 
$\ShowPrecondition$  
\end{minipage}
\right\}
$%
\\ \hline
4 & \ShowPartitionings~ \\
& \mbox{\color{blue} ~~~where~} \ShowPartitionSizes
\\ \hline
\rowcolor{lightgray!25}   
2 & 
$ \left\{ 
\begin{minipage}[t]{0.88\textwidth} 
$\ShowInvariant $
\end{minipage}
\right\} $ 
\\ \hline  
3 &
$\mbox{\color{blue}while~} \ShowGuard \mbox{~\color{blue} do}$
\\ \hline 
\rowcolor{lightgray!25} 
2,3 &  
$
\left\{
\begin{minipage}[t]{0.88\textwidth}%
$
~~~~ \ShowInvariant 
\wedge \ShowGuardTwo$
\end{minipage}
\right\}
$ 
\\ \hline 
5a & 
 \begin{minipage}{0.85\textwidth}%
~~~~  \ShowRepartitionings
\end{minipage}
\\ \hline 
\rowcolor{lightgray!25} 6 & 
$ \left\{ 
\begin{minipage}{0.88\textwidth} 
~~~~ \beforeupdate 
\end{minipage}
\right\}
$
\\ \hline 
8 & 
\begin{minipage}{0.85\textwidth}%
~~~~ \ShowUpdate 
\end{minipage}
\\ \hline  
5b & 
\begin{minipage}{0.85\textwidth}%
~~~~  \ShowMoveBoundaries~
\end{minipage}
\\ \hline
\rowcolor{lightgray!25} 7 & 
$ \left\{ 
\begin{minipage}{0.88\textwidth} 
~~~~ \afterupdate
\end{minipage}
\right\}
$
\\ \hline 
\rowcolor{lightgray!25} 2 & 
$ \left\{ 
\begin{minipage}{0.88\textwidth} 
~~~~ $ \ShowInvariant  $ 
\end{minipage}
\right\}
$
\\ \hline
 &
 $\mbox{\color{blue} endwhile} $
\\ \hline 
\rowcolor{lightgray!25} 2,3 & 
$ \left\{ 
\begin{minipage}[t]{0.88\textwidth} 
$ \ShowInvariant \wedge \neg( \ShowGuardTwo )$ 
\end{minipage}
\right\}
$
\\ \hline
\rowcolor{lightgray!25} 1b & 
$ \left\{ 
\begin{minipage}[t]{0.88\textwidth} 
$ \ShowPostcondition $ 
\end{minipage}
\right\}
$
\\ \hline
\end{tabular}

\caption{Worksheet for deriving the unblocked right-looking algorithm.}
\label{fig:LTLt_unb_right_ws}
}

\end{figure}

Let us adopt
Invariant~1 in 
(\ref{eqn:inv_unb_right_1})-(\ref{eqn:inv_unb_right_2})%
\ifthenelse{\boolean{VersionTOMS}}{.}{:
\[
    \begin{array}{l}
\left( \begin{array}{c I c | c}
X_{TL} & \star & \star \\ \whline
x_{ML}^T & \chi_{MM} & \star \\ \hline
X_{BL} & x_{BM} & X_{BR}
\end{array} \right)
=
\left( \begin{array}{c I c | c}
T_{TL} & \star & \star \\ \whline
 \tau_{ML} e_l^T & 0 &  \star \\ \hline
0  &  \tau_{BM} 
L_{BR} e_f
& L_{BR} T_{BR} L_{BR}^T
\end{array} \right)  \wedge
\color{gray}
\left( \begin{array}{c I c | c}
\widehat X_{TL} & \star & \star \\ \whline
\widehat x_{ML}^T & 0 & \star \\ \hline
\widehat X_{BL} & \widehat x_{BM} & \widehat X_{BR}
\end{array} \right) \\
\color{gray}
~~ =
\left( \begin{array}{c I c | c}
\color{blue} L_{TL} & 0 & 0 \\ \whline
\color{blue} l_{ML}^T & 1 & 0 \\ \hline
\color{blue} L_{BL} & \color{blue} l_{BM} &  I
\end{array}\right)
\left( \begin{array}{c I c | c}
T_{TL} & - \tau_{ML} e_l  & 0 \\ \whline
\tau_{ML} e_l^T  & 0 & -\tau_{BM} (L_{BR} e_f)^T \\ \hline
0 & \tau_{BM} L_{BR} e_f & L_{BR} T_{BR} L_{BR}^T
\end{array} \right)
\left( \begin{array}{c I c | c}
\color{blue} L_{TL}^T & \color{blue} l_{ML} & \color{blue} L_{BL}^T \\ \whline
0 & 1 & \color{blue} l_{BM}^T \\ \hline
0 & 0 &  I
\end{array}\right).
\end{array}
    \]%
Here we grayed out most of the constraint since it will not affect our derivation while highlighting in blue what parts of $ L $ have been computed.

    }
As briefly discussed in Section~\ref{sec:FLAME}, the FLAME methodology systematically derives the algorithm by filling out what we call the {\em worksheet}~\cite{Recipe}, given in Figure~\ref{fig:LTLt_unb_right_ws} for the right-looking algorithm.
The column on the left indicates the order in which it is filled with assertions (in the highlighted lines) and commands.  
It starts with entering the precondition and postcondition in Steps~1a and~1b.  Then the invariant is entered in the four places where it must hold (Step~2): before the loop, after the loop, at the top of the loop body, and at the bottom of the loop body.  
This exposes a structure for the inductive proof that guides the derivation of the algorithm.  
The loop guard (Step 3) and initialization (Step 4) are prescribed by the loop invariant, the postcondition, and the precondition.  
Each iteration exposes submatrices and the thick lines highlight how the computation progresses through the matrices (Steps~5a and~5b).  
This brings us to the most important steps: determining the contents of $ X $ and $L $ after the matrix is repartitioned (Step 6) and the contents of the exposed submatrices so that the invariant again holds at the bottom of the loop (Step 7).

After repartitioning (Step 6), we get
\[
\begin{array}{l}
\left( \begin{array}{c I c | c | c}
X_{00} & \star & \star & \star \\ \whline
x_{10}^T & \chi_{11} & 
\star & \star \\ \hline
x_{20}^T & 
\chi _{21} & \chi_{22} &  \star \\ \hline
X_{30} & x_{31}
& x_{32} & X_{33}
\end{array} \right)
=
\setlength{\arraycolsep}{4pt}
\left( 
\begin{array}{c I c | c | c}
 T_{00} & \star & \mbox{\hspace{0.55in}}  \star \mbox{\hspace{0.55in}} & \star \\ \whline
\tau_{10} e_l^T  & 0 & 
\star & \star \\ \hline
0 & 
\multirow{2}{*}{
$
\tau_{21}
\left( \begin{array}{c}
1   \\ \hline
l_{32} 
\end{array}
\right)
$
} & 
\multicolumn{2}{c}
{
\multirow{2}{*}{
$\left( \begin{array}{c | c}
1 &  0 \\ \hline
l_{32} &
L_{33}
\end{array}
\right)
\left( \begin{array}{c | c}
0 &  \star \\ \hline
\tau_{32} e_f &
T_{33}
\end{array}
\right)
\left( \begin{array}{c | c}
1 &  l_{32}^T \\ \hline
0 &
L_{33}^T
\end{array}
\right)
$
} }
\\
\cline{1-1} 
 0 & 
&  \multicolumn{2}{c}{}
\end{array}
\right)
\NoShow{
\wedge \\
~~
\left( \begin{array}{c I c | c | c}
\widehat X_{00} & \star & \star & \star \\ \whline
\widehat x_{10}^T & 0 & 
\star & \star \\ \hline
\widehat x_{20}^T & 
\widehat \chi _{21} &  0 &  \star \\ \hline
\widehat X_{30} & \widehat x_{31}
& \widehat x_{32} & \widehat X_{33}
\end{array} \right) \\
~~~~
=
\left( \begin{array}{c I c | c | c}
L_{00} & 0 & 0 & 0 \\ \whline
l_{10}^T & 1 & 
0 & 0 \\ \hline
l_{20}^T & 
\lambda _{21} & 
1
&  0 \\ \hline
L_{30} & l_{31}
& l_{32} & L_{33}
\end{array} \right)
\left( \begin{array}{c I c | c | c}
T_{00} & 
\tau_{10} e_l & 0 &  \\ \whline
\tau_{10} e_l^T & 0 & 
\tau_{21}  & 0 \\ \hline
0 & 
\tau_{21} & 
0
&  \tau_{32} e_f^T \\ \hline
0 & 
0 &
\tau_{32} e_f & T_{33}
\end{array} \right)
\left( \begin{array}{c I c | c | c}
L_{00}^T & l_{10} & l_{20} & L_{30}^T \\ \whline
0 & 1 & 
\lambda_{21} & l_{31}^T \\ \hline
0
& 
0 & 
1
&  l_{32}^T \\ \hline
0 & 
& 0 & L_{33}^T
\end{array} \right)
}
\end{array}
\]
and at the bottom of the loop (Step 7) we find that
\[
\begin{array}{l}
\left( \begin{array}{c | c I c | c}
X_{00}^{\rm +} & \star & \star & \star \\ \hline
x_{10}^{\rm +\!T} & \chi_{11} & 
\star & \star \\ \whline
x_{20}^{\rm +\!T} & 
\chi_{21}^{\rm +} & \chi_{22}^{\rm +} &  \star \\ \hline
X_{30}^{\rm +} & x_{31}^{\rm +}
& x_{32}^{\rm +} & X_{33}^{\rm +}
\end{array} \right)
=
\left( \begin{array}{c | c I c | c}
T_{00} & \star & \star & \star \\ \hline
 \tau_{10} e_l^T & 0 & 
~~~\star~~~ & \star \\ \whline
0 & 
\tau_{21} & 0 & \star
\\
\hline
 0 & 
 0 & 
\tau_{32}
L_{33} e_f
 &
L_{33}
T_{33} 
L_{33}^T
\end{array} \right)
\NoShow{
\wedge \\
~~
\left( \begin{array}{c I c | c | c}
\widehat X_{00} & \star & \star & \star \\ \whline
\widehat x_{10}^T & 0 & 
\star & \star \\ \hline
\widehat x_{20}^T & 
\widehat \chi _{21} &  0 &  \star \\ \hline
\widehat X_{30} & \widehat x_{31}
& \widehat x_{32} & \widehat X_{33}
\end{array} \right)
=
\left( \begin{array}{c I c | c | c}
L_{00} & 0 & 0 & 0 \\ \whline
l_{10}^T & 1 & 
0 & 0 \\ \hline
l_{20}^T & 
\lambda _{21} & 
1
&  0 \\ \hline
L_{30} & l_{31}
& l_{32} & L_{33}
\end{array} \right)
\left( \begin{array}{c I c | c | c}
T_{00} & 
\tau_{10} e_l & 0 &  \\ \whline
\tau_{10} e_l^T & 0 & 
\tau_{21}  & 0 \\ \hline
0 & 
\tau_{21} & 
0
&  \tau_{32} e_f^T \\ \hline
0 & 
0 &
\tau_{32} e_f & T_{33}
\end{array} \right)
\left( \begin{array}{c I c | c | c}
L_{00}^T & l_{10} & l_{20} & L_{30}^T \\ \whline
0 & 1 & 
\lambda_{21} & l_{31}^T \\ \hline
0
& 
0 & 
1
&  l_{32}^T \\ \hline
0 & 
& 0 & L_{33}^T
\end{array} \right)
}
.
\end{array}
\]
Here the $ {}^{\rm +} $ is used to  distinguish the contents of $ X $ at the bottom of the loop body from those at the top.

The assertions in Steps~6 and~7 prescribe the updates to the various exposed submatrices.  Comparing
\[
\left( \begin{array}{c}
\chi_{21} \\ \hline
x_{31}
\end{array} \right)
=
\tau_{21}
\left( \begin{array}{c}
1 \\ \hline
l_{32}
\end{array} \right)
\quad
\mbox{and}
\quad
\left( \begin{array}{c}
\chi_{21}^{\rm  +} \\ \hline
x_{31}^{\rm  +} 
\end{array} \right)
=
\left( \begin{array}{c}
\tau_{21} \\ \hline
0
\end{array} \right)
\]
prescribes the updates
\begin{eqnarray*}
l_{32} &:=&  x_{31} / \chi_{21} \\
x_{31} &:=& 0. 
\end{eqnarray*}
Next, 
\begin{eqnarray*}
\left( \begin{array}{c | c}
\chi_{22}^{\rm +} & \star \\ \hline
x_{32}^{\rm +} & X_{33}^{\rm +}
\end{array} \right)
&=&
\left( \begin{array}{c | c}
0 & 
- \tau_{32} ( L_{33} e_f )^T \\ \hline
\tau_{32} L_{33} e_f & L_{33}
 T_{33} L_{33}^T 
 \end{array} \right)
~=~
 \left( \begin{array}{c | c}
1 &  0 \\ \hline
0 &
L_{33}
\end{array}
\right)
\left( \begin{array}{c | c}
0 &  
- \tau_{32} e_f^T \\ \hline
\tau_{32} e_f &
T_{33}
\end{array}
\right)
\left( \begin{array}{c | c}
1 &  0 \\ \hline
0 &
L_{33}^T
\end{array}
\right)
\\
 & = &
 \left( \begin{array}{c | c}
1 &  0 \\ \hline
- l_{32} &
I
\end{array}
\right)
\left( \begin{array}{c | c}
1 &  0 \\ \hline
l_{32} &
L_{33}
\end{array}
\right)
\left( \begin{array}{c | c}
0 &  
- \tau_{32} e_f^T \\ \hline
\tau_{32} e_f &
T_{33}
\end{array}
\right)
\left( \begin{array}{c | c}
1 &  l_{32}^T  \\ \hline
0 &
L_{33}^T
\end{array}
\right)
\left( \begin{array}{c | c}
1 &  - l_{32}^T \\ \hline
0 &
I
\end{array}
\right)
\\
 & = &
 \left( \begin{array}{c | c}
1 &  0 \\ \hline
- l_{32} &
I
\end{array}
\right)
\left( \begin{array}{c | c}
0 &  
- x_{32}^T  \\ \hline
x_{32} &
X_{33}
\end{array}
\right)
\left( \begin{array}{c | c}
1 &  -l_{32}^T \\ \hline
0 &
I
\end{array}
\right)
= 
\left( \begin{array}{c | c}
0 &  
- x_{32}^T \\ \hline
x_{32} &
X_{33} + 
( l_{32} x_{32}^T -
x_{32} l_{32}^T )
\end{array}
\right)
\end{eqnarray*}
prescribes the update
\[
X_{33} := 
X_{33} + ( l_{32} x_{32}^T - x_{32} l_{32}^T
).
\]
This completes the formal derivation in Figure~\ref{fig:LTLt_unb_right_ws} from the  invariant.
By removing the various assertions, one is left with the right-looking algorithm in Figure~\ref{fig:LTLt_unb}.

The cost of this algorithm can be analyzed as follows:
The dominant cost term comes from the skew-symmetric rank-2 update.  
 If $ X $ is $ m \times m $ and $ X_{TL} $ is $ k \times k $, then $ X_{BR} $ is $ (m-k-1) \times (m-k-1) $ and updating it requires
$ 2 (m-k-1) \times (m-k-1) $ flops (updating only the lower-triangular part).
The approximate total cost is hence
$
\sum_{k=0}^{m-2}
2 ( m-k-1 )^2 
\approx
2 m^3/3 \mbox{~flops}$.

We do not derive unblocked algorithms corresponding  to Invariants~2a and~2b, instead deriving blocked algorithms corresponding to those invariants in Section~\ref{sec:2a} since that is where fusing will have a benefit.

\NoShow{
\subsection{Lazy right-looking algorithm}
\label{sec:unb-lazy}

A slight variation on the invariant
in (\ref{sec:unb-right})
that yields the right-looking algorithm is given by
{
 \setlength{\arraycolsep}{2pt}
\[
    \begin{array}{l}
\left( \begin{array}{c I c | c}
X_{TL} & \star & \star \\ \whline
x_{ML}^T & \chi_{MM} & \star \\ \hline
X_{BL} & x_{BM} & X_{BR}
\end{array} \right)
=
\left( \begin{array}{c I c | c}
T_{TL} & ~~~~~~~\star ~~~~~~~& \star \\ \whline
 \tau_{ML} e_l^T & 
 \multicolumn{2}{c}{
 \multirow{2}{*}{
 $
 \left( \begin{array}{c | c}
 1 &  0 \\ \hline
l_{BM} & L_{BR} 
\end{array}
\right)
 \left( \begin{array}{c | c}
 0 &  \star \\ \hline
\tau_{BM} 
e_f
& T_{BR} 
\end{array}
\right)
\left( \begin{array}{c | c}
 1 &  0 \\ \hline
l_{BM} & L_{BR} 
\end{array}
\right)^T
$
 }
 }
  \\ \cline{1-1}
0  &  \multicolumn{2}{c}{}
\end{array} \right) \wedge \\
~~ 
\color{gray}
\left( \begin{array}{c I c | c}
\widehat X_{TL} & \star & \star \\ \whline
\widehat x_{ML}^T & 0 & \star \\ \hline
\widehat X_{BL} & \widehat x_{BM} & \widehat X_{BR}
\end{array} \right) 
 =
\left( \begin{array}{c I c | c}
\color{blue} L_{TL} & 0 & 0 \\ \whline
\color{blue} l_{ML}^T & 1 & 0 \\ \hline
\color{blue} L_{BL} & \color{blue} l_{BM} &  I
\end{array}\right)
\left( \begin{array}{c I c | c}
T_{TL} & - \tau_{ML} e_l  & 0 \\ \whline
\tau_{ML} e_l^T  & 0 & -\tau_{BM} (L_{BR} e_f)^T \\ \hline
0 & \tau_{BM} L_{BR} e_f & L_{BR} T_{BR} L_{BR}^T
\end{array} \right)
\left( \begin{array}{c I c | c}
\color{blue} L_{TL}^T & \color{blue} l_{ML} & \color{blue} L_{BL}^T \\ \whline
0 & 1 & \color{blue} l_{BM}^T \\ \hline
0 & 0 &  I
\end{array}\right).
\end{array}
    \]
}%
This invariant captures that the latest Gauss transform, defined by $ l_{BM} $ has been computed, but not yet applied to the remainder of $ X $.
Going through the motions of derivation yields the  algorithm also given in Figure~\ref{fig:LTLt_unb_right}.
}  

\subsection{Two-step right-looking (Wimmer's) algorithm}

\label{sec:unb-right-wimmer}

Observe that  in the right-looking algorithm corresponding to Invariant~1, the application of the current Gauss transform does not change the ``next column,'' 
$
\left( \begin{array}{c}
\chi_{32} \\ \hline
x_{42}
\end{array}
\right) $.
Building on this observation, we next systematically derive an extension of Wimmer's unblocked algorithm~\cite{Wimmer2012} that computes the factorization two Gauss transforms at a time.
Surprisingly, this halves the operation count.

We again start  with Invariant~1 in~(\ref{eqn:inv_unb_right_1})--(\ref{eqn:inv_unb_right_2}).
This time we expose two rows and columns so that after repartitioning (in Step~6) we get
{
\setlength{\arraycolsep}{2pt}
\begin{eqnarray}
    \label{Before-unb-right-2step}
\lefteqn{
\left( \begin{array}{c I c | c | c | c }
X_{00} & \star & \star & \star & \star \\ \whline
x_{10}^T & \chi_{11} & 
\star & \star& \star \\ \hline
x_{20}^T & 
\chi_{21} & \chi_{22} &  \star & \star \\ \hline
x_{30}^T & \chi_{31}
& \chi_{32} & \chi_{33}
& \star \\ \hline
X_{40} & x_{41}
& x_{42} & x_{43} & X_{44}
\end{array} \right) }  \nonumber \\
&&\quad =
\setlength{\arraycolsep}{2pt}
\left( \begin{array}{c I c | c | c | c }
T_{00} & \star & 
\hspace{0.5in}
\star
\hspace{0.5in}~ & \hspace{0.5in}\star \hspace{0.5in}
~ & \hspace{0.5in}\star \hspace{0.5in}~ \\ \whline
\tau_{10} e_l^T & 0 & 
\hspace{0.5in}\star \hspace{0.5in}~  & \hspace{0.5in} \star \hspace{0.5in}~ & \hspace{0.5in} \star \hspace{0.5in} ~ \\ \hline
0 & 
\multirow{3}{*}{$
\tau_{21}
\left( \begin{array}{c}
1  \\ \hline
\lambda_{32} \\ \hline
l_{42} 
\end{array}
\right) 
$} & 
\multicolumn{3}{c}
{
\multirow{2}{*}{
$
\left( \begin{array}{c | c | c}
1 & 0 &  0 \\ \hline
\lambda_{32} &
1 & 0 \\ \hline
l_{42} &
l_{43} &
L_{44}
\end{array}
\right)
\left( \begin{array}{c | c | c}
0 & -\tau_{32}  &  0 \\ \hline
\tau_{32}  &
0 & - \tau_{43} e_f^T \\ \hline
0 &
\tau_{43} e_f &
T_{44}
\end{array}
\right)
\left( \begin{array}{c | c | c}
1 & \lambda_{32}  &  l_{42}^T \\ \hline
0 &
1 & l_{43}^T \\ \hline
0 & 0 &
L_{44}^T
\end{array}
\right)
$ } }
\\
\cline{1-1} 
0 & 
&  \multicolumn{3}{c}{}
\\
\cline{1-1} 
0 & 
&  \multicolumn{3}{c}{}
\end{array} \right)
\end{eqnarray}%
}%
and at the bottom of the loop (in  Step~7) we find
{
\begin{equation}
    \label{After-unb-right-2step}
\begin{array}{l}
\left( \begin{array}{c | c | c I c | c }
X_{00}^{\rm +} & \star & \star & \star & \star \\ \hline
x_{10}^{\rm +\!T} & \chi_{11}^{\rm +} & 
\star & \star& \star \\ \hline
x_{20}^{\rm +\!T} & 
\chi _{21}^{\rm +} & \chi_{22}^{\rm +} &  \star & \star \\ \whline
x_{30}^{\rm +\!T} & \chi_{31}^{\rm +}
& \chi_{32}^{\rm +} & \chi_{33}^{\rm +}
& \star \\ \hline
X_{40}^{\rm +} & x_{41}^{\rm +}
& x_{42}^{\rm +} & x_{43}^{\rm +} & X_{44}^{\rm +}
\end{array} \right)
=
\left( \begin{array}{c | c | c I c | c}
T_{00} & \star & \star & \star 
& \star \\ \hline
\tau_{10} e_l^T & 0 & 
\star & \star & \star\\ \hline
0 &
\tau_{21}
&
0
&
\star & \star \\ \whline
0 & 
0 &
\tau_{32} 
& 0 & \star
\\
\hline
0 & 
0 & 
0 & 
\tau_{43} L_{44} e_f &

L_{44} T_{44} L_{44}^T
\end{array} \right)
.
\end{array}
\end{equation}
}%
From \mbox{(\ref{Before-unb-right-2step})}, second column on each side, we find that
$
\tau_{21} \left( \begin{array}{c}
1 \\ \hline
\lambda_{32} \\ \hline
l_{42}
\end{array}
\right) 
=
 \left( \begin{array}{c}
\chi_{21} \\ \hline
\chi_{31} \\ \hline
x_{41}
\end{array}
\right) 
$
so that
$ \tau_{21} = \chi_{21} $ and 
$ \left( \begin{array}{c}
\lambda_{32} \\ \hline
l_{42}
\end{array}
\right)
:=  
\left( \begin{array}{c}
\chi_{31} \\ \hline
x_{41}
\end{array}
\right)
/
\chi_{21}  $, after which $
\left( \begin{array}{c}
\chi_{31} \\ \hline
x_{41}
\end{array}
\right)
:= \left( \begin{array}{c}
0 \\ \hline
0
\end{array}
\right)  $.
Also from \mbox{(\ref{Before-unb-right-2step})} we see that
\begin{eqnarray}
\nonumber
\lefteqn{
\left( \begin{array}{c I c | c}
\chi_{22} & \star & \star \\ \whline 
\chi_{32}  &
\chi_{33} & \star \\ \hline
x_{42} &
x_{43} & X_{44}
\end{array} \right)
= 
\left( \begin{array}{c | c | c}
1 & 0 &  0 \\ \hline
\lambda_{32} &
1 & 0 \\ \hline
l_{42} &
l_{43} &
L_{44}
\end{array}
\right)
\left( \begin{array}{c | c | c}
0 & -\tau_{32}  &  0 \\ \hline
\tau_{32}  &
0 & - \tau_{43} e_f^T \\ \hline
0 &
\tau_{43} e_f &
T_{44}
\end{array}
\right)
\left( \begin{array}{c | c | c}
1 & \lambda_{32}  &  l_{42}^T \\ \hline
0 &
1 & l_{43}^T \\ \hline
0 & 0 &
L_{44}^T
\end{array}
\right) }
\\
\nonumber
&& = 
\left( \begin{array}{c | c | c}
0 & -\tau_{32} & 0 \\ \hline
\tau_{32} &
-  \lambda_{32} \tau_{32} 
&
- \tau_{43} e_f^T \\ \hline
\tau_{32} l_{43} &
-  \tau_{32} l_{42} +  \tau_{43} L_{44} e_f &
-  \tau_{43} l_{43} e_f^T + L_{44} T_{44}
\end{array}
\right)
\left( \begin{array}{c | c | c}
1 & \lambda_{32}  &  l_{42}^T \\ \hline
0 &
1 & l_{43}^T \\ \hline
0 & 0 &
L_{44}^T
\end{array}
\right)  \\
\label{Before-unb-right-2step-c}
&& =
\left( \begin{array}{c | c | c}
0 & -\tau_{32} & 
- \tau_{32} l_{43}^T \\ \hline
\tau_{32} &
0
&
-\tau_{32} \lambda_{32} l_{43}^T + \tau_{32} l_{42}^T   
- \tau_{43} (L_{44} e_f)^T \\ \hline
\tau_{32} l_{43} &
\tau_{32} \lambda_{32}  l_{43}
-  \tau_{32} l_{42} +  
\tau_{43} 
L_{44} e_f
&
\begin{array}[t]{l}
~~~~~~~~~~(-  \tau_{32} l_{42} +  \tau_{43} L_{44} e_f ) l_{43}^T \\

- l_{43}(- \tau_{32} l_{42} + \tau_{43}  L_{44} e_f)^T  + 
L_{44} T_{44} L_{44}^T 
\end{array}
\end{array}
\right).
\end{eqnarray}
Hence we find that $\tau_{32}=\chi_{32}$ and compute
\begin{eqnarray*}
l_{43} & := & x_{42} / \chi_{32}
\\
x_{42} & := & 0.
\end{eqnarray*}
Finally, (\ref{Before-unb-right-2step-c}) tells us that 
\begin{eqnarray}
    x_{43} &=&
  \tau_{32} \lambda_{32} l_{43} -
  \tau_{32} l_{42} 
  +
  \tau_{43} L_{44} e_f \\
  X_{44} & = & 
  \tau_{32} l_{43} l_{42}^T +
(
-  \tau_{32} l_{42} +  \tau_{43} L_{44} e_f ) l_{43}^T 
-  \tau_{43} l_{43} ( L_{44} e_f)^T  + 
L_{44} T_{44} L_{44}^T \\
& = &
l_{43} ( \tau_{32}  l_{42} -
\tau_{43} L_{44} e_f )^T - 
(
 \tau_{32} l_{42} 
-  \tau_{43}  L_{44} e_f) l_{43}^T + 
L_{44} T_{44} L_{44}^T \\
& = &
l_{43} ( \tau_{32} \lambda_{32}  l_{43} -
x_{43} )^T - 
(
 \tau_{32} \lambda_{32}  l_{43} -
x_{43}) l_{43}^T + 
L_{44} T_{44} L_{44}^T
=
x_{43} l_{43} - l_{43} x_{43}^T + L_{44} T_{44} L_{44}^T .
\end{eqnarray}
Since $\tau_{32}=\chi_{32}$,
$
x_{43}^{\rm +} = \tau_{43} L_{44} e_f $,
and $ X_{44}^{\rm +} = L_{44} T_{44} L_{44}^T $, this prescribes  the updates (in this order)
\begin{eqnarray*}
X_{44} &:=& X_{44} + 
( l_{43} x_{43}^T - x_{43} l_{43}^T ) \\
 x_{43} &:=& 
 x_{43} + \chi_{32} l_{42} - \chi_{32} \lambda_{32} l_{43}.
\end{eqnarray*}
\NoShow{
We deduce that the following commands will change the state from that described in Step~6 to that in Step~7:
\[
\begin{array}{l}
\chi_{21} :=
\tau_{21} =  \chi_{21} ~~\mbox{(no-op)}\\
\left( \begin{array}{c}
\lambda_{32} \\ \hline
l_{42}
\end{array}
\right)
:=  
\left( \begin{array}{c}
\chi_{31} \\ \hline
x_{41}
\end{array}
\right)
/
\tau_{21} 
\\
\chi_{32} :=
\tau_{32} =  \chi_{32} ~~\mbox{(no-op)} \\
\left( \begin{array}{c}
\chi_{31} \\ \hline
x_{41}
\end{array}
\right) :=
\left( \begin{array}{c}
0 \\ \hline
0
\end{array}
\right)
\\
l_{43} := x_{42} / \tau_{32} \\
x_{42} :=  0 
\\
x_{43} := 
\tau_{43} L_{44} e_f
=
x_{43} + \tau_{32} l_{42} - \tau_{32} \lambda_{32} l_{43} \\
\begin{array}{@{}r@{~}c@{~}l}
X_{44}  &:=&
L_{44} T_{44} L_{44}^T
=
X_{44} + l_{43} ( \tau_{43}  L_{44} e_f 
-
\tau_{32}  l_{42} )^T
- ( \tau_{43} L_{44} e_f + \tau_{32} l_{42} ) l_{43}^T \\
& = & X_{44} + l_{43} ( x_{43} + \tau_{32} l_{42} )^T
- ( x_{43} + \tau_{32} l_{42} ) l_{43}^T
\end{array}
\end{array}
\]
The update to $ X_{44} $ can be described as a skew-symmetric rank-2 update, where only the (strictly) lower triangular part of the matrix is updated.  
}
The resulting algorithm is summarized in Figure~\ref{fig:LTLt_unb_Wimmer}.

\begin{figure}[tbp]
    \input LTLt_unb_Wimmer
    
\centering
    \footnotesize
    \FlaAlgorithm    
    \caption{Two-step unblocked  (Wimmer's) algorithm.}
    \label{fig:LTLt_unb_Wimmer}
\end{figure}

It is in the skew-symmetric rank-2 update   that most of the operations are performed, yielding an approximate cost for the algorithm of
$
m^3/3 \mbox{~flops}$,
or half of the cost of the more straight-forward unblocked right-looking 
(Parlett-Reid) algorithm.

Wimmer's original algorithm skips the computation of $ l_{43} $ (which defines the second Gauss transform in a two-step iteration) and $\tau_{32}$,
since only every other subdiagonal element of the tridiagonal matrix was required for his application (the computation of the Pfaffian). 
His implementation (PFAPACK) reverts back to the unblocked right-looking (Parlett-Reid) algorithm when full $ LTL^T $ output is demanded.
Our derivation in FLAME ``completes'' Wimmer's work and is beneficial for situations where the full $ LTL^T $ factorization is needed, for example when fast-updating computed Pfaffians~\cite{blockedvmc}.

\NoShow{
As of this writing, we have not attempted to derive a  two-step algorithm for \LTLt.  We suspect  that the zeroes on  the diagonal  of a skew-symmetric matrix are key to Wimmer's algorithm for \skewLTLt\ and that hence there is no  beneficial equivalent algorithm for \LTLt.
}

\subsection{Left-looking (Aasen's) algorithm}

\label{sec:unb-left}

Next, let us consider Invariant~3 in (\ref{eqn:inv_unb_left-1})--(\ref{eqn:inv_unb_left-2})%
\ifthenelse{\boolean{VersionTOMS}}{.}{
     \[
\begin{array}{l}
\left( \begin{array}{c I c | c}
X_{TL} & \star & \star \\ \whline
x_{ML}^T & \chi_{MM} & \star \\ \hline
X_{BL} & x_{BM} & X_{BR}
\end{array} \right)
=
\left( \begin{array}{c I c | c}
T_{TL} & \star & \star \\ \whline
\tau_{ML} e_l^T & 0 &  \star \\ \hline
0   & \widehat x_{BM}  & \widehat X_{BR}
\end{array} \right)  \wedge
\left( \begin{array}{c I c | c}
\widehat X_{TL} & \star & \star \\ \whline
\widehat x_{ML}^T & 0 & \star \\ \hline
\widehat X_{BL} & \widehat x_{BM} & \widehat X_{BR}
\end{array} \right) \\
~~ =
\left( \begin{array}{c I c | c}
\color{blue} L_{TL} & 0 & 0 \\ \whline
\color{blue} l_{ML}^T & 1 & 0 \\ \hline
\color{blue} L_{BL} & \color{blue} l_{BM} &  L_{BR}
\end{array}\right)
\left( \begin{array}{c I c | c}
T_{TL} & - \tau_{ML} e_l  & 0 \\ \whline
\tau_{ML} e_l^T  & 0 & -\tau_{BM} e_f^T \\ \hline
0 & \tau_{BM} e_f & T_{BR}
\end{array} \right)
\left( \begin{array}{c I c | c}
\color{blue} L_{TL}^T & \color{blue} l_{ML} & \color{blue} L_{BL}^T \\ \whline
0 & 1 & \color{blue} l_{BM}^T \\ \hline
0 & 0 &  L_{BR}^T
\end{array}\right).
\end{array}
    \]%
    }
At the top of the loop, 
we expose one row and column, as in Figure~\ref{fig:LTLt_unb}.
\NoShow{
\[
\left( \begin{array}{c I c | c}
X_{TL} & \star & \star \\ \whline
x_{ML}^T & \chi_{MM} & \star \\ \hline
X_{BL} & x_{BM} & X_{BR}
\end{array} \right)
\rightarrow
\left( \begin{array}{c I c | c | c}
X_{00} & \star & \star & \star \\ \whline
x_{10}^T & \chi_{11} & 
\star & \star \\ \hline
x_{20}^T & 
\chi _{21} & \chi_{22} &  \star \\ \hline
X_{30} & x_{31}
& x_{32} & X_{33}
\end{array} \right)
\] 
and the other matrices conformally.
} 
This means
that at the top of the loop (Step~6)
\[
\begin{array}{l}
\left( \begin{array}{c I c | c | c}
X_{00} & \star & \star & \star \\ \whline
x_{10}^T & \chi_{11} & 
\star & \star \\ \hline
x_{20}^T & 
\chi _{21} & \chi_{22} &  \star \\ \hline
X_{30} & x_{31}
& x_{32} & X_{33}
\end{array} \right)
=

\left( \begin{array}{c I c | c | c}
T_{00} & \star & \star & \star \\ \whline
  \tau_{10} e_l^T  & 0 & 
\star & \star \\ \hline
 0 & 
\widehat \chi _{21} & 0 &  \star \\ \hline
0  & \widehat x_{31}
& \widehat x_{32} & \widehat X_{33}
\end{array} \right)
\wedge 
\left( \begin{array}{c I c | c | c}
\widehat X_{00} &  - \widehat x_{10} &  -\widehat x_{20} &  -\widehat X_{30}^T \\ \whline
\widehat x_{10}^T & 0 & 
 -\widehat \chi_{21}^T &  -\widehat x_{31}^T \\ \hline
\widehat x_{20}^T & 
\widehat \chi _{21} &  0 &   -\widehat x_{32}^T \\ \hline
\widehat X_{30} & \widehat x_{31}
& \widehat x_{32} & \widehat X_{33}
\end{array} \right) \\
~~ =
\left( \begin{array}{c I c | c | c}
\color{blue} L_{00} & 0 & 0 & 0 \\ \whline
\color{blue} l_{10}^T & 1 & 
0 & 0 \\ \hline
\color{blue} l_{20}^T & 
\color{blue} \lambda _{21} & 
1
&  0 \\ \hline
\color{blue} L_{30} & \color{blue} l_{31}
& l_{32} & L_{33}
\end{array} \right)
\left( \begin{array}{c I c | c | c}
T_{00} & 
 -
\tau_{10} e_l & 0 & 0 \\ \whline
\tau_{10} e_l^T & 0 & 
 -
\tau_{21}   & 0 \\ \hline
0 & 
\tau_{21} & 
0
&   -
\tau_{32} e_f^T \\ \hline
0 & 
0 &
\tau_{32} e_f & T_{33}
\end{array} \right)
\left( \begin{array}{c I c | c | c}
\color{blue} L_{00}^T & \color{blue} l_{10} & \color{blue} l_{20} & \color{blue} L_{30}^T \\ \whline
0 & 1 & 
\color{blue} \lambda_{21} & \color{blue} l_{31}^T \\ \hline
0
& 
0 & 
1
&  l_{32}^T \\ \hline
0 & 0
& 0 & L_{33}^T
\end{array} \right)
\end{array}
\]
\NoShow{ 
while redefining
\[
\left( \begin{array}{c I c | c}
X_{TL} & \star & \star \\ \whline
x_{ML}^T & \chi_{MM} & \star \\ \hline
X_{BL} & x_{BM} & X_{BR}
\end{array} \right)
\leftarrow
\left( \begin{array}{c | c I c | c}
X_{00} & \star & \star & \star \\ \hline
x_{10}^T & \chi_{11} & 
\star & \star \\ \whline
x_{20}^T & 
\chi _{21} & \chi_{22} &  \star \\ \hline
X_{30} & x_{31}
& x_{32} & X_{33}
\end{array} \right),
 \mbox{etc.}
\] 
means that
} 
holds,
and at the bottom of the loop (Step~7)
\[
\begin{array}{l}
\left( \begin{array}{c | c I c | c}
X_{00}^{\rm +} & \star & \star & \star \\ \hline
x_{10}^{\rm +\!T} & \chi_{11}^{\rm +} & 
\star & \star \\ \whline
x_{20}^{\rm +\!T} & 
\chi _{21}^{\rm +} & \chi_{22}^{\rm +} &  \star \\ \hline
X_{30}^{\rm +} & x_{31}^{\rm +}
& x_{32}^{\rm +} & X_{33}^{\rm +}
\end{array} \right)
=
\left( \begin{array}{c | c I c | c}
T_{00} & \star & \star & \star \\ \hline
  \tau_{10} e_l^T  & 0 & 
\star & \star \\ \whline
 0 & 
\tau_{21} & 0 &  \star \\ \hline
0  &  0
& \widehat x_{32} & \widehat X_{33}
\end{array} \right)
\wedge 
\setlength{\arraycolsep}{3pt}
\left( \begin{array}{c | c I c | c}
\widehat X_{00} & - \widehat x_{10} & -\widehat x_{20} & -\widehat X_{30}^T \\ \hline
\widehat x_{10}^T & 0 & 
- \widehat \chi_{21} & - \widehat x_{31}^T \\ \whline
\widehat x_{20}^T & 
\widehat \chi _{21} &  0 &  - \widehat x_{32}^T \\ \hline
\widehat X_{30} & \widehat x_{31}
& \widehat x_{32} & \widehat X_{33}
\end{array} \right) \\
\setlength{\arraycolsep}{3pt}
~~
=
\left( \begin{array}{c | c I c | c}
\color{blue} L_{00} & 0 & 0 & 0 \\ \hline
\color{blue} l_{10}^T & 1 & 
0 & 0 \\ \whline
\color{blue} l_{20}^T & 
\color{blue} \lambda _{21} & 
1
&  0 \\ \hline
\color{blue} L_{30} & \color{blue} l_{31}
& \color{blue} l_{32} & L_{33}
\end{array} \right)
\left( \begin{array}{c | c I c | c}
T_{00} & 
 -
\tau_{10} e_l & 0 &  0 \\ \hline
\tau_{10} e_l^T & 0 & 
 -
\tau_{21}   & 0 \\ \whline
0 & 
\tau_{21} & 
0
&   -
\tau_{32} e_f^T \\ \hline
0 & 
0 &
\tau_{32} e_f & T_{33}
\end{array} \right)
\left( \begin{array}{c I c | c | c}
\color{blue} L_{00}^T & \color{blue} l_{10} & \color{blue} l_{20} & \color{blue} L_{30}^T \\ \whline
0 & 1 & 
\color{blue}   \lambda_{21} & \color{blue} l_{31}^T \\ \hline
0
& 
0 & 
1
&  \color{blue} l_{32}^T \\ \hline
0 & 0
& 0 & L_{33}^T
\end{array} \right) .
\end{array}
\]
The first goal is to compute $ \tau_{21} $ and $ l_{32} $.
From the constraint we note that at the top of the loop
\begin{eqnarray*}
\left( \begin{array}{c}
\chi_{21} \\ \hline
x_{31}
\end{array}
\right)
=
\left( \begin{array}{c}
\widehat \chi_{21} \\ \hline
\widehat x_{31}
\end{array}
\right)
&
=
&
\left( \begin{array}{c | c I c | c}
l_{20}^T & \lambda_{21}
& 1 & 0 \\ \hline
L_{30} & l_{31}
& l_{32} & L_{33}
\end{array} \right)
\left( \begin{array}{c | c I c | c}
T_{00} & 
-\tau_{10} e_l & 0 & 0 \\ \hline
\tau_{10} e_l^T & 0 & 
- \tau_{21}  & 0 \\ \whline
0 & 
\tau_{21} & 
0
&  - \tau_{32} e_f^T \\ \hline
0 & 
0 &
\tau_{32} e_f & T_{33}
\end{array} \right)
\left( \begin{array}{c}
l_{10}  \\ \hline
1  \\ \whline
0  \\ \hline
0
\end{array} \right)
\\
\NoShow{
&
=&
\left( \begin{array}{c | c I c }
l_{20}^T & \lambda_{21}
& 1  \\ \hline
L_{30} & l_{31}
& l_{32} 
\end{array} \right)
\left( \begin{array}{c | c  }
T_{00} & 
\tau_{10} e_l   \\ \hline
\tau_{10} e_l^T & 0   \\ \whline
0 & 
\tau_{21} 
\end{array} \right)
\left( \begin{array}{c}
l_{10}  \\ \hline
1  
\end{array} \right)
\\
}
&
=&
\left( \begin{array}{c | c  }
l_{20}^T & \lambda_{21}
  \\ \hline
L_{30} & l_{31} 
\end{array} \right)
\left( \begin{array}{c | c  }
T_{00} & 
- \tau_{10} e_l   \\ \hline
\tau_{10} e_l^T & 0
\end{array}
\right)
\left( \begin{array}{c}
l_{10}  \\ \hline
1  
\end{array} \right) 
+
\tau_{21}
\left( \begin{array}{c }
 1  \\ \hline
 l_{32} 
\end{array} \right).
\end{eqnarray*}
This suggests that
first 
\begin{equation}
\label{eqn:deriv-left-unb-1}
\left( \begin{array}{c}
\chi_{21} \\ \hline
x_{31}
\end{array}
\right)
:=
\left( \begin{array}{c}
 \chi_{21} \\ \hline
 x_{31}
\end{array}
\right)
-
\left( \begin{array}{c | c  }
l_{20}^T & \lambda_{21}
  \\ \hline
L_{30} & l_{31} 
\end{array} \right)
\left[
\left( \begin{array}{c | c  }
X_{00} & 
\star   \\ \hline
x_{10}^T & 0
\end{array}
\right)
\left( \begin{array}{c}
l_{10}  \\ \hline
1  
\end{array} \right) 
\right]
\end{equation}
after which $ \chi_{21} = \tau_{21} $.  Then $ l_{32} $ can be computed  and $ x_{31} $ updated by
\[
\begin{array}{@{}l}
l_{32} := x_{31} / \chi_{21} \\
x_{31} := 0.
\end{array}
\]
The resulting algorithm is given in Figure~\ref{fig:LTLt_unb}.
The described algorithm works whether the elements below the diagonal of the first column of $ L $ equal zero or not.

If $ X $ is initially $ m \times m $, the cost of this algorithm can be analyzed as follows:
The dominant cost term comes from  
(\ref{eqn:deriv-left-unb-1}).
\NoShow{
\[
\left( \begin{array}{c | c  }
l_{20}^T & \lambda_{21}
  \\ \hline
L_{30} & l_{31} 
\end{array} \right)
\left[
\left( \begin{array}{c | c  }
T_{00} & 
\tau_{10} e_l   \\ \hline
\tau_{10} e_l^T & 0
\end{array}
\right)
\left( \begin{array}{c}
l_{10}  \\ \hline
1  
\end{array} \right) 
\right].
\]
}  
Since $ T $ is skew-symmetric and tridiagonal, this incurs roughly the cost of a matrix-vector multiplication.  If $ X_{TL} $ is $ k \times k $, then the matrix is $ (m-k) \times (k+1) $ and multiplying with it requires
approximately
$ 2 (m-k) \times k $ flops%
\footnote{
Not including a lower order term which may be affected by whether the first column equals zero or not.}.
The approximate total cost is hence
\begin{equation}
    \nonumber
\sum_{k=0}^{m-2}
2 k( m-k ) =
2 \left( \sum_{k=0}^{m-2}
k m 
-
\sum_{k=0}^{m-2}k^2  \right)
\approx
2 \left(
m^3/2 - m^3/3 \right) = m^3 / 3  \mbox{~flops}.
\end{equation}
This is half the approximate cost of the unblocked right-looking (modified Parlett-Reid) algorithm and matches the approximate cost of Wimmer's unblocked two-step algorithm.

What we have described is a variation on Aasen's algorithm~\cite{Aasen}.
Aasen recognizes that
$ X = L T L^T = L H $, where $ H = T L^T $ is an upper-Hessenberg matrix.  As noted in his paper, in each iteration only one column of H needs to be computed and used in an iteration and  hence $ H $ needs not be stored.  This column of $ H $ is  
\[
\left( \begin{array}{c | c  }
X_{00} & 
\star   \\ \hline
x_{10}^T & 0
\end{array}
\right)
\left( \begin{array}{c}
l_{10}  \\ \hline
1  
\end{array} \right)
=
\left( \begin{array}{c | c  }
T_{00} & 
- \tau_{10} e_l   \\ \hline
\tau_{10} e_l^T & 0
\end{array}
\right)
\left( \begin{array}{c}
l_{10}  \\ \hline
1  
\end{array} \right)
\]
in our algorithm.

\NoShow{
\begin{figure}[tb!]
    \input LTLt_unb_left_separate_L
\centering
    \FlaAlgorithm    
    \caption{Unblocked left-looking algorithm.  Version that stores $ L $ separately.}
    \label{fig:LTLt_unb_left_separate_L}
\end{figure}

\begin{figure}[tb!]
    \input LTLt_unb_left
\centering
    \FlaAlgorithm    
    \caption{Unblocked left-looking algorithm.  Version that overwrites $ X $ with $ T $ and $ L $.}
    \label{fig:LTLt_unb_left}
    \end{figure}
    \begin{figure}[tb!]
\begin{lstlisting}[belowskip=-1.8 \baselineskip]
function A_out = TriDiag( A )

if size( A, 1 ) <= 1
    Aout = A;
else
    A =  tril( A ) + tril( A, -1 )';   % make A symmetric
    Aout = triu( tril( A, 1 ), -1 );   % extract tridiagonal
end
end

function Xout = LTLt_unb_left( X )

n = size( X, 1 );
for i=1:n-1
    if i > 1  
        % update the current column according to previous computation  
        X( i+1:n, i ) = X( i+1:n, i ) - ...
            [ zeros( n-i, 1 )  X( i+1:n, 1:i-1 ) ] * ...
            TriDiag( X( 1:i, 1:i ) ) * ...
            [ 0 
              X( i, 1:i-2 )'
              1 ];    
    end
    % Compute the multipliers in the current column
    X( i+2:n, i ) = X( i+2:n, i ) / X( i+1, i );
end
Xout = X;
end
\end{lstlisting}
    \caption{Matlab implementatin of unblocked left-looking algorithm.  For simplicity, the upper triangular part of matrix $ X $ is kept consistent with the lower triangular part.  Taking advantage of, for example, the zeroes when computing with the tridiagonal matrix needs to be added.}
    \label{fig:LTLt_unb_left_matlab}
\end{figure}

{\bf We will probably remove this paragraph.}
If instead $ L $ overwrites the zeroes introduces in $ X $, the following modification need to be made: Since $
\left( \begin{array}{c} \chi_{21} \\ \hline
 x_{31}
\end{array}
\right)
$ stores
$
\left( \begin{array}{c}
\widehat \chi_{21} \\ \hline
\widehat x_{31}
\end{array}
\right)
$
and
$
\left( \begin{array}{c | c  }
l_{20}^T & \lambda_{21}
  \\ \hline
L_{30} & l_{31} 
\end{array} \right)
=
\left( \begin{array}{c : c  }
0 & x_{20}^T 
  \\ \hdashline
0 & X_{30} 
\end{array} \right)
$, we deduce that the following commands update $ X $ appropriately:
\begin{itemize}
    \item 
    $ \left( \begin{array}{c} \chi_{21} \\ \hline
 x_{31}
\end{array}
\right) :=
\left( \begin{array}{c} \chi_{21} \\ \hline
 x_{31}
\end{array}
\right)
-
\left( \begin{array}{c : c  }
0 & x_{20}^T 
  \\ \hdashline
0 & X_{30} 
\end{array} \right)
\left( \begin{array}{c | c  }
T_{00} & 
\tau_{10} e_l   \\ \hline
\tau_{10} e_l^T & 0
\end{array}
\right)
\left( \begin{array}{c}
0 \\ \hdashline
\widetilde x_{10}   
\end{array} \right)
$,
where
$ \widetilde x_{10} $
equals the transpose of the row vector stored in $ x_{10} $, except with the last element replaced by a $1$.
Obviously, in computing this the computation takes the special structure into account.
\item 
$ x_{31} := l_{32} = x_{31}/\chi_{21}$.
\end{itemize}
The resulting algorithm is given in Figure~\ref{fig:LTLt_unb_left}, with a Matlab implementation in
Figure~\ref{fig:LTLt_unb_left_matlab}.
}

\section{Deriving blocked algorithms}

It is well known that high performance for dense linear algebra operations like the one discussed in this paper can be attained by casting computation in terms of matrix-matrix operations (level-3 BLAS)~\cite{BLAS3}.  We now discuss how such blocked algorithms can be derived.

\subsection{Right-looking algorithm}
\label{sec:blk-right}

Let us derive a blocked algorithm from the
invariant in~(\ref{eqn:inv_unb_right_1})-(\ref{eqn:inv_unb_right_2}).
The repartitioning now exposes a new block of columns and rows in each iteration.
After repartitioning in  Step~5a, we get for Step~6 that
{
\small
\setlength{\unitlength}{2pt}
\begin{eqnarray*}
\label{Before-blk-right}
\lefteqn{
\left( \begin{array}{c I c | c | c | c }
X_{00} & \cellcolor{lightgray!35}  \star & \cellcolor{lightgray!35}  \star & \star & \star \\ \whline
\cellcolor{lightgray!35} x_{10}^T & \cellcolor{lightgray!35} \chi_{11} & 
\cellcolor{lightgray!35} \star & \cellcolor{lightgray!35} \star& \cellcolor{lightgray!35} \star \\ \hline
\cellcolor{lightgray!35} X_{20} & 
\cellcolor{lightgray!35} x _{21} & \cellcolor{lightgray!35} X_{22} &  \cellcolor{lightgray!35} \star & \cellcolor{lightgray!35} \star \\ \hline
x_{30}^T & \cellcolor{lightgray!35} \chi_{31}
& \cellcolor{lightgray!35} x_{32}^T & \chi_{33}
& \star \\ \hline
X_{40} & \cellcolor{lightgray!35} x_{41}
& \cellcolor{lightgray!35} X_{42} & x_{43} & X_{44}
\end{array} \right) }\\
& 
= &
\setlength{\arraycolsep}{2pt}
\left( \begin{array}{c I c | c | c | c }
T_{00} & \star & 
\mbox{\hspace{0.5in}}
\star
\mbox{\hspace{0.5in}} & \mbox{\hspace{0.5in}}\star \mbox{\hspace{0.5in}}
 & \mbox{\hspace{0.4in}} \star \mbox{\hspace{0.4in}} \\ \whline
\tau_{10} e_l^T & 0 & 
\star   & \star  & \star \\ \hline
0 & 
\multirow{3}{*}{$
\tau_{21}
\left( \begin{array}{c | c | c}
L_{22} & 0 &  0 \\ \hline
l_{32}^T &
1 & 0 \\ \hline
L_{42} &
l_{43} &
L_{44}
\end{array}
\right) e_f
$} & 
\multicolumn{3}{c}
{
\multirow{2}{*}{
$
\left( \begin{array}{c | c | c}
L_{22} & 0 &  0 \\ \hline
l_{32}^T &
1 & 0 \\ \hline
L_{42} &
l_{43} &
L_{44}
\end{array}
\right)
\left( \begin{array}{c | c | c}
T_{22} & -\tau_{32} e_l &  0 \\ \hline
\tau_{32} e_l^T &
0 & - \tau_{43} e_f^T \\ \hline
0 &
\tau_{43} e_f &
T_{44}
\end{array}
\right)
\left( \begin{array}{c | c | c}
L_{22}^T & l_{32}  &  L_{42}^T \\ \hline
0 &
1 & l_{43}^T \\ \hline
0 & 0 &
L_{44}^T
\end{array}
\right)
$ } }
\\
\cline{1-1} 
0 & 
&  \multicolumn{3}{c}{}
\\
\cline{1-1} 
0 & 
&  \multicolumn{3}{c}{}
\end{array} \right),
\end{eqnarray*}%
}%
where the gray highlighting captures the block of rows and columns being exposed in this iteration.
At the bottom of the loop we find
for Step~7 that 
{
\small
\begin{equation}
    \label{eqn:blk:1}
\begin{array}{l}
\left( \begin{array}{c | c | c I c | c }
X_{00}^{\rm +} & \cellcolor{lightgray!35} \star & \cellcolor{lightgray!35} \star & \star & \star \\ \hline
\rowcolor{lightgray!35} x_{10}^{\rm +\!T} & \chi_{11}^{\rm +} & 
\star & \star& \star \\ \hline
\rowcolor{lightgray!35} X_{20}^{\rm +} & 
\cellcolor{lightgray!35} x _{21}^{\rm +} & \cellcolor{lightgray!35} X_{22}^{\rm +} &  \star & \star \\ \whline
x_{30}^{\rm +\!T} & \cellcolor{lightgray!35} \chi_{31}^{\rm +}
& \cellcolor{lightgray!35} x_{32}^{\rm +\!T} & \chi_{33}^{\rm +}
& \star \\ \hline
X_{40}^{\rm +} & \cellcolor{lightgray!35} x_{41}^{\rm +} 
& \cellcolor{lightgray!35} X_{42}^{\rm +} & x_{43}^{\rm +} & X_{44}^{\rm +}
\end{array} \right)
=
\setlength{\arraycolsep}{2pt}
\left( \begin{array}{c | c | c I c | c}
T_{00} & \star & \star & \star 
& \star \\ \hline
\tau_{10} e_l^T & 0 & 
~~~\star~~~ & ~~~~~~~\star ~~~~~~~& \star\\ \hline
0 &
\tau_{21} e_f
&
T_{22}
&
\star & \star \\ \whline
0 & 
0 &
\tau_{32} e_l^T
& 
0 & \star
\\
\hline
0 & 
0 & 
0 & 
\tau_{43} L_{44} e_f &
L_{44} T_{44} L_{44}^T
\end{array} \right)
\NoShow{
\wedge \\
~~
\left( \begin{array}{c I c | c | c}
\widehat X_{00} & \star & \star & \star \\ \whline
\widehat x_{10}^T & 0 & 
\star & \star \\ \hline
\widehat x_{20}^T & 
\widehat \chi _{21} &  0 &  \star \\ \hline
\widehat X_{30} & \widehat x_{31}
& \widehat x_{32} & \widehat X_{33}
\end{array} \right)
=
\left( \begin{array}{c I c | c | c}
L_{00} & 0 & 0 & 0 \\ \whline
l_{10}^T & 1 & 
0 & 0 \\ \hline
l_{20}^T & 
\lambda _{21} & 
1
&  0 \\ \hline
L_{30} & l_{31}
& l_{32} & L_{33}
\end{array} \right)
\left( \begin{array}{c I c | c | c}
T_{00} & 
\tau_{10} e_l & 0 &  \\ \whline
\tau_{10} e_l^T & 0 & 
\tau_{21}  & 0 \\ \hline
0 & 
\tau_{21} & 
0
&  \tau_{32} e_f^T \\ \hline
0 & 
0 &
\tau_{32} e_f & T_{33}
\end{array} \right)
\left( \begin{array}{c I c | c | c}
L_{00}^T & l_{10} & l_{20} & L_{30}^T \\ \whline
0 & 1 & 
\lambda_{21} & l_{31}^T \\ \hline
0
& 
0 & 
1
&  l_{32}^T \\ \hline
0 & 
& 0 & L_{33}^T
\end{array} \right).
}
.
\end{array}
\end{equation}
}
We observe that
\begin{eqnarray*}
  \lefteqn{
 \left( \begin{array}{c | c | c | c }
 \chi_{11} & 
\star & \star& \star \\ \hline 
x _{21} & X_{22} &  \star & \star \\ \hline
 \chi_{31}
& x_{32}^T & \chi_{33}
& \star \\ \hline
 x_{41}
& X_{42} & x_{43} & X_{44}
\end{array} \right) 
}
\\
& = &
\setlength{\arraycolsep}{2pt}
\left( \begin{array}{c | c | c | c }
 0 & 
\hspace{0.55in}\star \mbox{\hspace{0.55in}}  & \hspace{0.55in} \star \mbox{\hspace{0.55in}} & \hspace{0.35in} \star \mbox{\hspace{0.35in}} \\ \hline 
\tau_{21}
\left( \begin{array}{c | c | c}
L_{22} & 0 &  0 \\ \hline
l_{32}^T &
1 & 0 \\ \hline
L_{42} &
l_{43} &
L_{44}
\end{array}
\right) e_f
 & 
\multicolumn{3}{c}
{
\left( \begin{array}{c | c | c}
L_{22} & 0 &  0 \\ \hline
l_{32}^T &
1 & 0 \\ \hline
L_{42} &
l_{43} &
L_{44}
\end{array}
\right)
\left( \begin{array}{c | c | c}
T_{22} & -\tau_{32} e_l &  0 \\ \hline
\tau_{32} e_l^T &
0 & - \tau_{43} e_f^T \\ \hline
0 &
\tau_{43} e_f &
T_{44}
\end{array}
\right)
\left( \begin{array}{c | c | c}
L_{22}^T & l_{32}  &  L_{42}^T \\ \hline
0 &
1 & l_{43}^T \\ \hline
0 & 0 &
L_{44}^T
\end{array}
\right)
 }
\end{array} \right)  
\\
& = & 
\left( \begin{array}{c | c | c | c}
1 & 0 & 0 & 0 \\ \hline
0 & L_{22} & 0 &  0 \\ \hline
0 & l_{32}^T &
1 & 0 \\ \hline
0 & L_{42} &
l_{43} &
L_{44}
\end{array}
\right)
\left( \begin{array}{c | c | c | c}
0 & - \tau_{21} e_f^T & 0 & 0 \\
\hline
\tau_{21} e_f & T_{22} & -\tau_{32} e_f &  0 \\ \hline
0 & \tau_{32} e_l^T &
0 & - \tau_{43} e_f^T \\ \hline
0 & 0 &
\tau_{43} e_f &
T_{44}
\end{array}
\right)
\left( \begin{array}{c | c | c | c}
1 & 0 & 0 & 0 \\ \hline
0 & L_{22}^T & l_{32}  &  L_{42}^T \\ \hline
0 & 0 &
1 & l_{43}^T \\ \hline
0 & 0 & 0 &
L_{44}^T
\end{array}
\right),
\end{eqnarray*}
which implies that
\[
 \left( \begin{array}{c | c}
 \chi_{11} & 
\star  \\ \hline 
x _{21} & X_{22}  \\ \hline
 \chi_{31}
& x_{32}^T \\ \hline
 x_{41}
& X_{42} 
\end{array} \right) 
=
\left( \begin{array}{c | c | c }
1 & 0 & 0  \\ \hline
0 & L_{22} & 0  \\ \hline
0 & l_{32}^T &
1  \\ \hline
0 & L_{42} &
l_{43} 
\end{array}
\right)
\left( \begin{array}{c | c }
0 & - \tau_{21} e_f^T  \\
\hline
\tau_{21} e_f^T & T_{22}   \\ \hline
0 & \tau_{32} e_l^T  
\end{array}
\right)
\left( \begin{array}{c | c}
1 & 0 \\ \hline
0 & L_{22}^T 
\end{array}
\right).
\]
Examining~\mbox{(\ref{eqn:blk:1})} tells us that  
$
\left( \begin{array}{c | c}
\chi_{11}^{\rm +} & 
\star  \\ \hline 
x _{21}^{\rm +} & X_{22}^{\rm +}  \\ \hline
 \chi_{31}{\rm +}
& x_{32}^{\rm +\!T} \\ \hline
 x_{41}^{\rm +}
& X_{42}^{\rm +} 
\end{array} \right) 
~
\mbox{and}
~
\left( \begin{array}{c | c | c }
1 & 0 & 0  \\ \hline
0 & L_{22} & 0  \\ \hline
0 & l_{32}^T &
1  \\ \hline
0 & L_{42} &
l_{43} 
\end{array}
\right)
~
\mbox{are computed from}
~
\left( \begin{array}{c | c}
 \chi_{11} & 
\star  \\ \hline 
x _{21} & X_{22}  \\ \hline
 \chi_{31}
& x_{32}^T \\ \hline
 x_{41}
& X_{42} 
\end{array} \right) 
$
by factoring that panel. The presence of zeroes below the first diagonal element of the $L$ panel used indicates that the recursive sub-problem factorizing that panel should assume implicit zeroes in the first column. In later derivations, we will see that sometimes a non-zero leading column of $L$ must instead be assumed in the sub-problem.

The purpose of the game now becomes to update the remaining part of $ X $ by separating what is known from what is yet to be computed.  Notice that
\begin{eqnarray}
\nonumber
    \lefteqn{
    \left( \begin{array}{c | c}
    \chi_{33} & \star \\ \hline
    x_{43}& X_{44}
    \end{array}
    \right) =
    \left( \begin{array}{c | c | c | c}
0 & l_{32}^T &
1 & 0 \\ \hline
0 & L_{42} &
l_{43} &
L_{44}
\end{array}
\right)
\left( \begin{array}{c | c | c | c}
0 & -\tau_{21} e_f^T & 0 & 0 \\
\hline
\tau_{21} e_f & T_{22} & -\tau_{32} e_l &  0 \\ \hline
0 & \tau_{32} e_l^T &
0 & - \tau_{43} e_f^T \\ \hline
0 & 0 &
\tau_{43} e_f &
T_{44}
\end{array}
\right)
\left( \begin{array}{c | c}
0 & 0 \\ \hline
l_{32}  &  L_{42}^T \\ \hline
1 & l_{43}^T \\ \hline
0 &
L_{44}^T
\end{array}
\right)
    } \\
    \label{eqn:bk-right-1}
    & = &
    \left( \begin{array}{ c | c | c}
l_{32}^T &
1 & 0 \\ \hline
L_{42} &
l_{43} &
L_{44}
\end{array}
\right)
\left[
\left( \begin{array}{c | c | c}
 T_{22} & -\tau_{32} e_l &  0 \\ \hline
 \tau_{32} e_l^T &
0 & 0 \\ \hline
0 &
0 &
0
\end{array}
\right)
+
\left( \begin{array}{c | c | c}
 0 & 0 &  0 \\ \hline
0 &
0 & - \tau_{43} e_f^T \\ \hline
0 &
\tau_{43} e_f &
T_{44}
\end{array}
\right)
\right]
\left( \begin{array}{c | c}
l_{32}  &  L_{42}^T \\ \hline
1 & l_{43}^T \\ \hline
0 &
L_{44}^T
\end{array}
\right)  \\
\nonumber
& = & 
    \left( \begin{array}{ c | c }
l_{32}^T &
1 \\ \hline
L_{42} &
l_{43} 
\end{array}
\right)
\left( \begin{array}{c | c }
 T_{22} & -\tau_{32} e_l  \\ \hline
 \tau_{32} e_l^T &
0 
\end{array}
\right)
\left( \begin{array}{c | c}
l_{32}  &  L_{42}^T \\ \hline
1 & l_{43}^T 
\end{array}
\right)
+
\left( \begin{array}{c | c}
1 & 0 \\ \hline
l_{43} &
L_{44} 
\end{array}
\right)
\left( \begin{array}{c | c}
0 & \star \\ \hline
\tau_{43}  e_f &
 T_{44} 
\end{array}
\right)
\left( \begin{array}{c | c}
1 & l_{43}^T \\ \hline
0 &
L_{44}^T
\end{array}
\right)
\\
\nonumber
& = & 
\begin{array}[t]{@{}c@{}}
\underbrace{
    \left( \begin{array}{ c | c }
l_{32}^T &
1 \\ \hline
L_{42} &
l_{43} 
\end{array}
\right)
\left( \begin{array}{c | c }
 T_{22} & -\tau_{32} e_l  \\ \hline
 \tau_{32} e_l^T &
0 
\end{array}
\right)
\left( \begin{array}{c | c}
l_{32}  &  L_{42}^T \\ \hline
1 & l_{43}^T 
\end{array}
\right)
} \\
\mbox{known}
\end{array}
+
\left( \begin{array}{c | c}
1 & 0 \\ \hline
l_{43} &
I 
\end{array}
\right)
\begin{array}[t]{@{}c@{}}
\underbrace{
\left( \begin{array}{c | c}
0 & \star \\ \hline
\tau_{43} L_{44} e_f &
L_{44} T_{44} L_{44}^T
\end{array}
\right)
}
\\
\mbox{to be computed}
\end{array}
\left( \begin{array}{c | c}
1 & l_{43}^T \\ \hline
0 &
I 
\end{array}
\right)
\\
\nonumber
& = & 
    \left( \begin{array}{ c | c }
l_{32}^T &
1 \\ \hline
L_{42} &
l_{43} 
\end{array}
\right)
\left( \begin{array}{c | c }
 T_{22} & -\tau_{32} e_l  \\ \hline
 \tau_{32} e_l^T &
0 
\end{array}
\right)
\left( \begin{array}{c | c}
l_{32}  &  L_{42}^T \\ \hline
1 & l_{43}^T 
\end{array}
\right)
+
\left( \begin{array}{c | c}
1 & 0 \\ \hline
l_{43} &
I 
\end{array}
\right)
\left( \begin{array}{c | c}
    \chi_{33}^{\rm +} & \star \\ \hline
    x_{43}^{\rm +}& X_{44}^{\rm +}
    \end{array}
    \right)
\left( \begin{array}{c | c}
1 & l_{43}^T \\ \hline
0 &
I 
\end{array}
\right) .
\end{eqnarray}
This prescribes  the updates
\begin{eqnarray}
\label{eqn:blk-right-2}
\left( \begin{array}{c | c}
    \chi_{33} & \star \\ \hline
    x_{43}& X_{44}
    \end{array}
    \right)
    &:=&
    \left( \begin{array}{c | c}
    \chi_{33} & \star \\ \hline
    x_{43}& X_{44}
    \end{array}
    \right)
    -
 \left( \begin{array}{ c | c }
l_{32}^T &
1 \\ \hline
L_{42} &
l_{43} 
\end{array}
\right)
\left( \begin{array}{c | c }
 T_{22} & -\tau_{32} e_l  \\ \hline
 \tau_{32} e_l^T &
0 
\end{array}
\right)
\left( \begin{array}{c | c}
l_{32}  &  L_{42}^T \\ \hline
1 & l_{43}^T 
\end{array}
\right) \\
\nonumber
&=&
    \left( \begin{array}{c | c}
    \chi_{33} & \star \\ \hline
    x_{43}& X_{44}
    \end{array}
    \right)
    -
 \left( \begin{array}{ c | c }
l_{32}^T &
1 \\ \hline
L_{42} &
l_{43} 
\end{array}
\right)
\left( \begin{array}{c | c }
 X_{22} & \star  \\ \hline
 x_{32}^T  &
0 
\end{array}
\right)
\left( \begin{array}{c | c}
l_{32}  &  L_{42}^T \\ \hline
1 & l_{43}^T 
\end{array}
\right) 
\\
\left( \begin{array}{c | c}
    \chi_{33} & \star \\ \hline
    x_{43}& X_{44}
    \end{array}
    \right)
  &  := &
    \left( \begin{array}{c | c}
1 & 0 \\ \hline
- l_{43} &
I 
\end{array}
\right)
\left( \begin{array}{c | c}
    0 & \star \\ \hline
    x_{43}& X_{44}
    \end{array}
    \right)
    \left( \begin{array}{c | c}
1 & -l_{43}^T \\ \hline
0 &
I 
\end{array}
\right) \\
\label{eqn:blk-right-3}
& = &
\left( \begin{array}{c | c}
    0 & \star \\ \hline
    x_{43}& X_{44}
    +
    ( l_{43} x_{43}^T
    - x_{43} l_{43}^T )
    \end{array}
    \right).
    \end{eqnarray}
\NoShow{
Invoking Corollary~\ref{cor:inv} reveals
\begin{eqnarray*}
\lefteqn{
\left( \begin{array}{c I c | c}
T_{22} & \star & 0 \\ \whline 
\tau_{32} e_l^T &
\chi_{33}^{\rm +} & \star \\ \hline
0 &
x_{43}^{\rm +} & X_{44}^{\rm +}
\end{array} \right)
 = 
\left( \begin{array}{c I c | c}
T_{22} & \star & 0 \\ \whline 
\tau_{32} e_l^T & 
0 & \star \\ \hline
0 & 
\tau_{43} L_{44} e_f & L_{44}
 T_{44} L_{44}^T 
 \end{array} \right) }
 \\
& & =
 \left( \begin{array}{c I c | c}
 I & 0 & 0 \\ \whline
0 & 1 &  0 \\ \hline
0 & 0 &
L_{44}
\end{array}
\right)
\left( \begin{array}{c I c | c}
T_{22} & \star & 0 \\ \whline 
\tau_{32} e_l^T & 
0 & \star \\ \hline
0 & 
\tau_{43}  e_f & 
 T_{44}  
 \end{array} \right)
\left( \begin{array}{c I c | c}
 I & 0 & 0 \\ \whline
0 & 1 &  0 \\ \hline
0 & 0 &
L_{44}
\end{array}
\right)^T
\\
& & =
 \left( \begin{array}{c I c | c}
 I & 0 & 0 \\ \whline
0 & 1 &  0 \\ \hline
0 & 0 &
L_{44}
\end{array}
\right)
\left( \begin{array}{c | c | c}
L_{22} & 0 &  0 \\ \hline
l_{32}^T &
1 & 0 \\ \hline
L_{42} &
l_{43} &
L_{44}
\end{array}
\right)^{-1} \\
& & ~~~~~~
\left( \begin{array}{c | c | c}
L_{22} & 0 &  0 \\ \hline
l_{32}^T &
1 & 0 \\ \hline
L_{42} &
l_{43} &
L_{44}
\end{array}
\right) 
\left( \begin{array}{c I c | c}
T_{22} & \star & 0 \\ \whline 
\tau_{32} e_l^T & 
0 & \star \\ \hline
0 & 
\tau_{43}  e_f & 
 T_{44}  
 \end{array} \right)
\left( \begin{array}{c | c | c}
L_{22} & 0 &  0 \\ \hline
l_{32}^T &
1 & 0 \\ \hline
L_{42} &
l_{43} &
L_{44}
\end{array}
\right)^{T}
 \\
& &  ~~~~~~ ~~~~~~
\left( \begin{array}{c | c | c}
L_{22} & 0 &  0 \\ \hline
l_{32}^T &
1 & 0 \\ \hline
L_{42} &
l_{43} &
L_{44}
\end{array}
\right)^{-T}
\left( \begin{array}{c I c | c}
 I & 0 & 0 \\ \whline
0 & 1 &  0 \\ \hline
0 & 0 &
L_{44}
\end{array}
\right)^T
\\
& &  =
 \left( \begin{array}{c I c | c}
 L_{22} & 0 & 0 \\ \whline
l_{32}^T & 1 &  0 \\ \hline
L_{42} & l_{43} &
I
\end{array}
\right)^{-1}
\left( \begin{array}{c I c | c}
X_{22} & \star & \star \\ \whline 
x_{32}^T & 
0 & \star \\ \hline
X_{42} & 
x_{43} & 
 X_{44}  
 \end{array} \right) 
 \left( \begin{array}{c I c | c}
 L_{22} & 0 & 0 \\ \whline
l_{32}^T & 1 &  0 \\ \hline
L_{42} & l_{43} &
I
\end{array}
\right)^{-T}.
\end{eqnarray*}
This allows us to employ Theorem~\ref{thm:right-update} by noting that in that theorem
\begin{itemize}
    \item 
    $ C_{TL}^{\rm +} = 
    \left( \begin{array}{c I c}
    T_{22} & \star \\ \whline
    \tau_{32} e_l^T & 0 
    \end{array} \right) $.
    \item 
    $ 
    \begin{array}[t]{@{}r @{~} c @{~} l}
    \begin{array}[t]{c}
    \underbrace{
    \left( \begin{array}{c I c}
    0 & x_{43}^{\rm +} 
    \end{array} \right) 
    } \\
    C_{BL}^{\rm +}
    \end{array}
    & = &
    - 
    \begin{array}[t]{c}
    \underbrace{
    \left( 
    \begin{array}{c I c }
    L_{42} & l_{43}
    \end{array} \right)
    } \\
    B_{BL}
    \end{array}
    \begin{array}[t]{c}
    \underbrace{
    \left( \begin{array}{c I c}
    T_{22} & - \tau_{32} e_l \\ \whline
    \tau_{32} e_l^T & 0 
    \end{array} \right) 
    } \\
    C_{TL}^{\rm +}
    \end{array}
    + 
    \begin{array}[t]{c}
    \underbrace{
    \left( \begin{array}{c I c}
    X_{42} & x_{43}
    \end{array} \right) 
    } \\
    C_{BL}
    \end{array}
    \begin{array}[t]{c}
    \underbrace{
    \left( \begin{array}{c I c}
    L_{22} & 0 \\ \whline
    l_{32}^T& 1 
    \end{array} \right)^{-T} 
    }
    \\
    B_{TL}^{-T}
    \end{array} \\
    & = & 
    \left( \begin{array}{ c I c}
    \star\star &
    \tau_{32} L_{42} e_l + 
    \end{array}
    \right)
    \end{array}
    $.
    \item
    $
    \begin{array}[t]{r@{~}c@{~}l@{~}c@{~}l}
    \begin{array}[t]{c}
    \underbrace{
    X_{44}^{\rm +}
    } \\
    C_{BR}^{\rm +}
    \end{array}
    &= &
    \begin{array}[t]{c}
    \underbrace{
    X_{44}
    } \\
    C_{BR}
    \end{array}
    &-& 
    \begin{array}[t]{c}
    \underbrace{
    \left( \begin{array}{c | c}
    L_{42} &
    l_{43}
    \end{array} \right)
    } \\
    B_{BL}
    \end{array}
    \begin{array}[t]{c}
    \underbrace{
    \left( \begin{array}{c I c}
    T_{22} & \star \\ \whline
    \tau_{32} e_l^T & 0 
    \end{array} \right) 
    }
    \\
    C_{TL}^{\rm +}
    \end{array}
    \begin{array}[t]{c}
    \underbrace{
    \left( \begin{array}{c | c}
    L_{42} &
    l_{43}
    \end{array} \right)^T}
    \\
    B_{BL}^T
    \end{array}
    \\
    && & 
    + &
    \begin{array}[t]{c}
    \underbrace{    \left( \begin{array}{c | c }
    L_{42} & 
    l_{43}
    \end{array} \right)
    } \\
    B_{BL}
    \end{array}
    \begin{array}[t]{c}
    \underbrace{
     \left( \begin{array}{c I c}
    0 & x_{43}^{\rm +} 
    \end{array} \right)^T
    } \\
    C_{BL}^T
    \end{array}
    -
    \begin{array}[t]{c}
    \underbrace{
    \left( \begin{array}{c I c}
    0 & x_{43}^{\rm +} 
    \end{array} \right)
    } \\
    C_{BL}^{\rm +}
    \end{array}
    \begin{array}[t]{c}
    \underbrace{
    \left( \begin{array}{c | c }
    L_{42} & 
    l_{43}
\end{array} \right)^T
} \\
B_{BL}^T
\end{array} \\
    &= &
    X_{44} &-& 
    \left( \begin{array}{c | c}
    L_{42} &
    l_{43}
    \end{array} \right)
    \left( \begin{array}{c I c}
    T_{22} & \star \\ \whline
    \tau_{32} e_l^T & 0 
    \end{array} \right)
    \left( \begin{array}{c | c}
    L_{42} &
    l_{43}
    \end{array} \right)^T +
    ( l_{43} x_{43}^{\rm +\!T} -
    x_{43}^{\rm +} l_{43}^T ).
    \end{array}
    $.
\end{itemize}
}
\NoShow{
We can do slightly better, in  the process linking the blocked right-looking algorithm to Wimmer's unblocked algorithm.
\begin{eqnarray*}
\lefteqn{(\ref{eqn:bk-right-1}) = 
\left( \begin{array}{ c | c | c}
l_{32}^T &
1 & 0 \\ \hline
L_{42} &
l_{43} &
L_{44}
\end{array}
\right)
\left[
\left( \begin{array}{c | c | c}
 T_{22} & 0 &  0 \\ \hline
 0 &
0 & 0 \\ \hline
0 &
0 &
0
\end{array}
\right)
+
\left( \begin{array}{c | c | c}
 0 & -\tau_{32} e_l &  0 \\ \hline
 \tau_{32} e_l^T &
0 & 0 \\ \hline
0 &
0 &
0
\end{array}
\right)
\right.
}
\\
&&
\hspace{2in}
\left.
+
\left( \begin{array}{c | c | c}
 0 & 0 &  0 \\ \hline
0 &
0 & - \tau_{43} e_f^T \\ \hline
0 &
\tau_{43} e_f &
T_{44}
\end{array}
\right)
\right]
\left( \begin{array}{c | c}
l_{32}  &  L_{42}^T \\ \hline
1 & l_{34}^T \\ \hline
0 &
L_{44}^T
\end{array}
\right)   \\
& = & 
\left( \begin{array}{ c }
l_{32}^T  \\ \hline
L_{42}
\end{array}
\right)
T_{22}
\left( \begin{array}{c | c}
l_{32}  &  L_{42}^T 
\end{array}
\right)
+
\left( \begin{array}{ c | c }
l_{32}^T &
1  \\ \hline
L_{42} &
l_{43} 
\end{array}
\right)
\left[
\left( \begin{array}{c | c }
 0 & -\tau_{32} e_l \\ \hline
 \tau_{32} e_l^T &
0 
\end{array}
\right)
\right]
\left( \begin{array}{c | c}
l_{32}  &  L_{42}^T \\ \hline
1 & l_{34}^T 
\end{array}
\right) \\
& & \hspace{2in}
+
\left( \begin{array}{c | c}
1 & 0 \\ \hline
l_{43} &
I 
\end{array}
\right)
\left( \begin{array}{c | c}
    \chi_{33}^{\rm +} & \star \\ \hline
    x_{43}^{\rm +}& X_{44}^{\rm +}
    \end{array}
    \right)
\left( \begin{array}{c | c}
1 & l_{43}^T \\ \hline
0 &
I 
\end{array}
\right)\\
& = & 
\end{eqnarray*}
}%
This completes the derivation of the blocked right-looking algorithm in  Figure~\ref{fig:LTLt_blk_right}.
 It casts the bulk of the computation in terms of the 
``sandwiched'' (skew-)symmetric rank-k update in (\ref{eqn:blk-right-2}).

What is somewhat surprising about this blocked right-looking algorithm for this operation is that its cost, when the blocking size $ b $ is reasonably large, is essentially 
$ m^3 / 3 $ flops which equals half the cost of the unblocked right-looking algorithm. 

 \begin{figure}[p]
    
\input LTLt_blk_right_left

\centering
    \footnotesize
    \FlaAlgorithm
    
\caption{Blocked right- and left-looking algorithms. In the blocked right-looking algorithm, the suffix ``\_0'' indicates that $l_{21}$, $\lambda_{31}$, and $l_{41}$ are implicitly equal to zero during the factorization sub-problem.}
    \label{fig:LTLt_blk_right}
\end{figure}

\NoShow{
\subsection{Lazy right-looking algorithm}

\begin{figure}[p]

\centering
    \FlaAlgorithm    
    \caption{Blocked lazy  right-looking algorithm.}
    \label{fig:LTLt_blk_lazy_right}
\end{figure}

Next, let us derive a blocked algorithm from the
invariant
\begin{equation}
    \label{PME-unb-right}
    \begin{array}{l}
\left( \begin{array}{c I c | c}
X_{TL} & \star & \star \\ \whline
x_{ML}^T & \chi_{MM} & \star \\ \hline
X_{BL} & x_{BM} & X_{BR}
\end{array} \right)
=
\left( \begin{array}{c I c | c}
T_{TL} & ~~~~~~~\star ~~~~~~~& \star \\ \whline
 \tau_{ML} e_l^T & 
 \multicolumn{2}{c}{
 \multirow{2}{*}{
 $
 \left( \begin{array}{c | c}
 1 &  0 \\ \hline
l_{BM} & L_{BR} 
\end{array}
\right)
 \left( \begin{array}{c | c}
 0 &  \star \\ \hline
\tau_{BM} 
e_f
& T_{BR} 
\end{array}
\right)
\left( \begin{array}{c | c}
 1 &  0 \\ \hline
l_{BM} & L_{BR} 
\end{array}
\right)^T
$
 }
 }
  \\ \cline{1-1}
0  &  \multicolumn{2}{c}{}
\end{array} \right) \wedge \\
~~ 
\color{gray}
\left( \begin{array}{c I c | c}
\widehat X_{TL} & \star & \star \\ \whline
\widehat x_{ML}^T & 0 & \star \\ \hline
\widehat X_{BL} & \widehat x_{BM} & \widehat X_{BR}
\end{array} \right) \wedge
\color{gray}
\left( \begin{array}{c I c | c}
\widehat X_{TL} & \star & \star \\ \whline
\widehat x_{ML}^T & 0 & \star \\ \hline
\widehat X_{BL} & \widehat x_{BM} & \widehat X_{BR}
\end{array} \right) \\
\color{gray}
~~ =
\left( \begin{array}{c I c | c}
\color{blue} L_{TL} & 0 & 0 \\ \whline
\color{blue} l_{ML}^T & 1 & 0 \\ \hline
\color{blue} L_{BL} & \color{blue} l_{BM} &  I
\end{array}\right)
\left( \begin{array}{c I c | c}
T_{TL} & - \tau_{ML} e_l  & 0 \\ \whline
\tau_{ML} e_l^T  & 0 & -\tau_{BM} (L_{BR} e_f)^T \\ \hline
0 & \tau_{BM} L_{BR} e_f & L_{BR} T_{BR} L_{BR}^T
\end{array} \right)
\left( \begin{array}{c I c | c}
\color{blue} L_{TL}^T & \color{blue} l_{ML} & \color{blue} L_{BL}^T \\ \whline
0 & 1 & \color{blue} l_{BM}^T \\ \hline
0 & 0 &  I
\end{array}\right).
\end{array}
\end{equation}
    
After repartitioning, we get
\begin{equation}
    \label{Before-blk-lazy_right}
\begin{array}{l}
\left( \begin{array}{c I c | c | c | c }
X_{00} & \star & \star & \star & \star \\ \whline
x_{10}^T & \chi_{11} & 
\star & \star& \star \\ \hline
X_{20} & 
x _{21} & X_{22} &  \star & \star \\ \hline
x_{30}^T & \chi_{31}
& x_{32}^T & \chi_{33}
& \star \\ \hline
X_{40} & x_{41}
& X_{42} & x_{43} & X_{44}
\end{array} \right) \\
\setlength{\arraycolsep}{2pt}
~~
=
\left( \begin{array}{c I c | c | c | c }
T_{00} & \hspace{0.5in}
\star
\hspace{0.5in}~ & 
\hspace{0.5in}
\star
\hspace{0.5in}~ & \hspace{0.5in}\star \hspace{0.5in}
~ & \hspace{0.5in}\star \hspace{0.5in}~ \\ \whline
\tau_{10} e_l^T & \multicolumn{4}{c}{
\multirow{4}{*}{
$
\left( \begin{array}{c | c | c | c}
1 & 0 & 0 & 0 \\ \hline
\color{blue} l_{21} & L_{22} & 0 & 0 \\ \hline
\color{blue} \lambda_{31} & l_{32}^T & 1 & 0 \\ \hline 
\color{blue} l_{41} & L_{42} & l_{43} & L_{44}
\end{array} 
\right)
\left( \begin{array}{c | c | c | c}
0 & - \tau_{21} e_f^T & 0 & 0 \\ \hline
\tau_{21} e_f & T_{22} & - \tau_{32} e_l & 0 \\ \hline
0 & \tau_{32} e_l & 0 & - \tau_{43} e_f^T \\ \hline 
0 & 0 & \tau_{43} e_f & T_{44}
\end{array} 
\right)
\left( \begin{array}{c | c | c | c}
1 &\color{blue} l_{21}^T & \color{blue} \lambda_{31} & \color{blue} l_{41}^T \\ \hline
0 & L_{22} & l_{32} & L_{42}^T \\ \hline
0 & 0 & 1 & l_{43}^T \\ \hline 
0 & 0 & 0 & L_{44}^T
\end{array} 
\right)
$
}
}
\\ \cline{1-1}
0 & 
\multicolumn{4}{c}
{~}
\\ \cline{1-1}
0 & 
\multicolumn{4}{c}
{~}
\\ \cline{1-1}
0 & 
\multicolumn{4}{c}
{~}
\end{array} \right)
\end{array}
\end{equation}
and at the bottom of the loop we find
\begin{equation}
    \label{After-unb-right}
\begin{array}{l}
\left( \begin{array}{c | c | c I c | c }
X_{00}^{\rm +} & \star & \star & \star & \star \\ \hline
x_{10}^{\rm +\!T} & \chi_{11}^{\rm +} & 
\star & \star& \star \\ \hline
X_{20}^{\rm +} & 
x _{21}^{\rm +} & X_{22}^{\rm +} &  \star & \star \\ \whline
x_{30}^{\rm +\!T} & \chi_{31}^{\rm +}
& x_{32}^{\rm +\!T} & \chi_{33}^{\rm +}
& \star \\ \hline
X_{40}^{\rm +} & x_{41}
& X_{42}^{\rm +} & x_{43}^{\rm +} & X_{44}^{\rm +}
\end{array} \right)
=
\setlength{\arraycolsep}{2pt}
\left( \begin{array}{c | c | c I c | c}
T_{00} & \star & \star & \star 
& \star \\ \hline
\tau_{10} e_l^T & 0 & 
~~~\star~~~ & ~~~~~~~\star ~~~~~~~& \star\\ \hline
0 &
\tau_{21} e_f
&
T_{22}
&
\star & \star \\ \whline
0 & 
0 &
\tau_{32} e_l^T
& 
\multicolumn{2}{c}{
\multirow{2}{*}{
$
\left( \begin{array}{c|c}
1 & 0 \\ \hline
\color{blue} l_{43} & L_{44}
\end{array} \right)
\left( \begin{array}{c|c}
0 & - \tau_{43} e_f^T \\ \hline
\tau_{43} e_f & T_{44}
\end{array} \right)
\left( \begin{array}{c|c}
1 & \color{blue} l_{43}^T \\ \hline
0 & L_{44}^T
\end{array} \right)
$
}
}
\\
\cline{1-3}
0 & 
0 & 
0 & 
\multicolumn{2}{c}
{}
\end{array}
\right)
.
\end{array}
\end{equation}
We note from (\ref{Before-blk-lazy_right}) that
\[
\begin{array}{rcl}
\left( \begin{array}{c | c | c | c }
 \chi_{11} & 
\star & \star& \star \\ \hline 
x _{21} & X_{22} &  \star & \star \\ \hline
 \chi_{31}
& x_{32}^T & \chi_{33}
& \star \\ \hline
 x_{41}
& X_{42} & x_{43} & X_{44}
\end{array} \right)
 & = &
 \left( \begin{array}{c | c | c | c}
1 & 0 & 0 & 0 \\ \hline
\color{blue} l_{21} & L_{22} & 0 & 0 \\ \hline
\color{blue} \lambda_{31} & l_{32}^T & 1 & 0 \\ \hline 
\color{blue} l_{41} & L_{42} & l_{43} & L_{44}
\end{array} 
\right)
\left( \begin{array}{c | c | c | c}
0 & - \tau_{21} e_f^T & 0 & 0 \\ \hline
\tau_{21} e_f & T_{22} & - \tau_{32} e_l & 0 \\ \hline
0 & \tau_{32} e_l & 0 & - \tau_{43} e_f^T \\ \hline 
0 & 0 & \tau_{43} e_f & T_{44}
\end{array} 
\right)
\left( \begin{array}{c | c | c | c}
1 &\color{blue} l_{21}^T & \color{blue} \lambda_{31} & \color{blue} l_{41}^T \\ \hline
0 & L_{22} & l_{32} & L_{42}^T \\ \hline
0 & 0 & 1 & l_{43}^T \\ \hline 
0 & 0 & 0 & L_{44}^T
\end{array} 
\right)
\end{array}
\]
and hence
\[
\begin{array}{rcl}
\left( \begin{array}{c | c }
 \chi_{11} & 
\star \\ \hline 
x _{21} & X_{22}  \\ \hline
 \chi_{31}
& x_{32}^T  \\ \hline
 x_{41}
& X_{42} 
\end{array} \right)
 & = &
 \left( \begin{array}{c | c | c }
1 & 0 & 0 \\ \hline
\color{blue} l_{21} & L_{22} & 0  \\ \hline
\color{blue} \lambda_{31} & l_{32}^T & 1 \\ \hline 
\color{blue} l_{41} & L_{42} & l_{43} 
\end{array} 
\right)
\left( \begin{array}{c | c }
0 & - \tau_{21} e_f^T \\ \hline
\tau_{21} e_f & T_{22} \\ \hline
0 & \tau_{32} e_l  
\end{array} 
\right)
\left( \begin{array}{c | c }
1 &\color{blue} l_{21}^T  \\ \hline
0 & L_{22}  
\end{array} 
\right)
.
\end{array}
\]
Also, 
\[
\begin{array}{l}
\left( \begin{array}{c | c }
 \chi_{33}
& \star \\ \hline
 x_{43} & X_{44}
\end{array} \right)
=
 \left( \begin{array}{c | c | c | c}
 \color{blue} \lambda_{31} & l_{32}^T & 1 & 0 \\ \hline 
\color{blue} l_{41} & L_{42} & l_{43} & L_{44}
\end{array} 
\right)
\left( \begin{array}{c | c | c | c}
0 & - \tau_{21} e_f^T & 0 & 0 \\ \hline
\tau_{21} e_f & T_{22} & - \tau_{32} e_l & 0 \\ \hline
0 & \tau_{32} e_l^T & 0 & - \tau_{43} e_f^T \\ \hline 
0 & 0 & \tau_{43} e_f & T_{44}
\end{array} 
\right)
\left( \begin{array}{ c | c}
 \color{blue} \lambda_{31} & \color{blue} l_{41}^T \\ \hline
 l_{32} & L_{42}^T \\ \hline
 1 & l_{43}^T \\ \hline 
 0 & L_{44}^T
\end{array} 
\right) \\
~~= 
\left( \begin{array}{c | c | c | c}
 \color{blue} \lambda_{31} & l_{32}^T & 1 & 0 \\ \hline 
\color{blue} l_{41} & L_{42} & l_{43} & L_{44}
\end{array} 
\right)
\left(
\left( \begin{array}{c | c | c | c}
0 & - \tau_{21} e_f^T & 0 & 0 \\ \hline
\tau_{21} e_f & T_{22} & - \tau_{32} e_l & 0 \\ \hline
0 & \tau_{32} e_l^T & 0 & 0 \\ \hline 
0 & 0 & 0 & 0
\end{array} 
\right)
+
\left( \begin{array}{c | c | c | c}
0 & 0 & 0 & 0 \\ \hline
0 & 0 & 0 & 0 \\ \hline
0 & 0 & 0 & - \tau_{43} e_f^T \\ \hline 
0 & 0 & \tau_{43} e_f & T_{44}
\end{array} 
\right)
\right)
\left( \begin{array}{ c | c}
 \color{blue} \lambda_{31} & \color{blue} l_{41}^T \\ \hline
 l_{32} & L_{42}^T \\ \hline
 1 & l_{43}^T \\ \hline 
 0 & L_{44}^T
\end{array} 
\right) \\
~~ =
\left( \begin{array}{c | c}
\lambda_{31} & l_{32}^T \\ \hline
l_{41} & L_{42}
\end{array}
\right)
\\
~~~~
\setlength{\arraycolsep}{2pt}
\left( 
\begin{array}{c | c}
\left( \begin{array}{c | c}
0 & - \tau_{21} e_f^T 
\\ \hline
\tau_{21} e_f & T_{22}
\end{array}
\right)
+
\left( \begin{array}{c | c}
1 & 0 \\ \hline
l_{43} & L_{43}
\end{array}
\right)
\left( \begin{array}{c | c}
0 & 0
\\ \hline
- \tau_{32} e_l^T  & 0
\end{array}
\right)
&
\left( \begin{array}{c | c}
0 & - \tau_{21} e_f^T 
\\ \hline
\tau_{21} e_f & T_{22}
\end{array}
\right)
+
\left( \begin{array}{c | c}
1 & 0 \\ \hline
l_{43} & L_{43}
\end{array}
\right)
\left( \begin{array}{c | c}
0 & \tau_{32} e_l^T 
\\ \hline
0 & 0
\end{array}
\right)
\end{array}
\right)
\left( \begin{array}{ c | c}
 \color{blue} \lambda_{31} & \color{blue} l_{41}^T \\ \hline
 l_{32} & L_{42}^T \\ \hline
 1 & l_{43}^T \\ \hline 
 0 & L_{44}^T
\end{array} 
\right)
\end{array}
\]
This allows us to employ Theorem~\ref{thm:right-update} by noting that in that theorem
\begin{itemize}
    \item 
    $ C_{TL}^{\rm +} = 
    \left( \begin{array}{c I c}
    T_{22} & \star \\ \whline
    \tau_{32} e_l^T & 0 
    \end{array} \right) $.
    \item 
    $ C_{BL}^{\rm +} = 
    \left( \begin{array}{c I c}
    0 & x_{43}^{\rm +} 
    \end{array} \right)
    = 
    - \left( 
    \begin{array}{c I c }
    L_{42} & l_{43}
    \end{array} \right)
    \left( \begin{array}{c I c}
    T_{22} & - \tau_{32} e_l \\ \whline
    \tau_{32} e_l^T & 0 
    \end{array} \right) 
    + 
    \left( \begin{array}{c I c}
    X_{42} & x_{43}
    \end{array} \right)
    \left( \begin{array}{c I c}
    L_{22} & 0 \\ \whline
    l_{32}^T& 1 
    \end{array} \right)^{-T} 
    $.

    \item
    $
    C_{BR}^{\rm +} = -
    \left( \begin{array}{c I c}
    0 & x_{43}^{\rm +}
    \end{array} \right)
    \left( \begin{array}{c }
    L_{42}^T \\ \whline
    l_{43}^T
    \end{array} \right)
    +
    \left( \begin{array}{c I c }
    L_{42} &  l_{43}
    \end{array} \right)
        \left(
    \left( \begin{array}{c I c}
    X_{42} & x_{43}
    \end{array} \right) 
    \left( \begin{array}{c I c }
     L_{22} & 0 \\ \whline
    l_{32}^T & 1
    \end{array} \right)^{-T}
    \right)^T
    + X_{44}
    $.
\end{itemize}
This suggest the following steps for the blocked right-looking algorithm:
\begin{itemize}
    \item 
    Create a copy:
    $ \left( \begin{array}{c I c}
    X_{42}^c & x_{43}^c
    \end{array} \right)
    :=
    \left( \begin{array}{c I c}
    X_{42} & x_{43}
    \end{array} \right)
    $. 
    \item 
    With a left- or right-looking algorithm, modified to only update the first $ b $ columns,  compute 
    \[
    \left( \begin{array}{c | c}
    \chi_{11} & x_{12}^T \\ \hline
    x_{21} & X_{22} \\ \hline
    \chi_{31} & x_{32}^T \\ \hline
    x_{41} & X_{42}
    \end{array} \right)
    :=
    \left( \begin{array}{c | c}
    0 & \star \\ \hline
    \tau_{21} e_f & T_{22} \\ \hline
    0 & \tau_{32} e_l^T \\ \hline
    0 & 0
    \end{array} \right)
    \quad
    \mbox{and}
    \quad
    \left( \begin{array}{c I c}
    L_{22} & 0 \\ \hline
    l_{32}^T & 1 \\ \hline
    L_{42} & l_{43}
    \end{array} \right)
    \]
    \item 
    $ \left( \begin{array}{c I c}
    X_{42}^c & x_{43}^c
    \end{array} \right)
    :=
    \left( \begin{array}{c I c}
    X_{42}^c & x_{43}^c
    \end{array} \right)
    \left( \begin{array}{c I c }
     L_{22} & 0 \\ \whline
    l_{32}^T & 1
    \end{array} \right)^{-T}
    $. 
    \item 
    $ x_{43} := \tau_{32} L_{42} e_l + x_{43}^c $.
    \item 
    $ X_{44} := X_{44} -
    x_{43} l_{43}^T + 
    \left( \begin{array}{c I c }
    L_{42} &  l_{43}
    \end{array} \right)
    \left( \begin{array}{c I c}
    X_{42}^c & x_{43}^c
    \end{array} \right)^T
    $.
\end{itemize}
This is summarized in the algorithm in Figure~\ref{fig:LTLt_blk_right}.
}

If we choose the block size  in the  algorithm equal to one ($ X_{22} $ is $ 0 \times 0  $), then this becomes the 
unblocked right-looking (Parlett-Reid) algorithm.  
Our blocked algorithm has some resemblance  to the blocked algorithm for computing the $ L T L^T $ factorization of a  symmetric matrix  given in~\cite{Miroslav2011} and can be modified to perform that operation. 
In their algorithm, blocks of the same matrix $ H $ that Aasen introduced are computed, which is what we avoid.  {\em If} a sandwiched (skew-)symmetric rank-k update were available, {\em then} our algorithm  avoids the workspace required  by their algorithm for parts of $ H $. Our companion paper \cite{LTLt_SISC_ArXiv} explores this possibility in practice.

A problem with this algorithm is the separate rank-2 update $X_{44} := X_{44} + ( l_{43} x_{43}^T - x_{43} l_{43}^T) $, since it requires an extra pass over  $ X_{44} $ and hence additional memory accesses.  In Sections~\ref{sec:2a} and~\ref{sec:2b} we show how blocked algorithms corresponding to Invariants~2a and~2b overcome this.

\subsection{Factoring the panel}

The factoring of the panel in the various blocked algorithms is accomplished by calling any of the unblocked algorithms, provided they are modified to not update any part of the matrix $ X $ outside the current panel. The blocked right- and left-looking algorithms require different assumptions about the first column of the $L$ panel, as seen in Figure~\ref{fig:LTLt_blk_right}. As discussed in Section~\ref{sec:specification}, the assumption of a non-zero first column of $L$ may be easily handled by pre-processing, or in the case of the unblocked left-looking algorithm (Section~\ref{sec:unb-left}), naturally included in the update steps.

\subsection{Left-looking algorithm}

 Next, let us again adopt the
invariant for the left-looking algorithm in~(\ref{eqn:inv_unb_left-1})--(\ref{eqn:inv_unb_left-2}).
\NoShow{
\begin{eqnarray*}
\left( \begin{array}{c I c | c}
X_{TL} & \star & \star \\ \whline
x_{ML}^T & \chi_{MM} & \star \\ \hline
X_{BL} & x_{BM} & X_{BR}
\end{array} \right)
=
\left( \begin{array}{c I c | c}
\color{blue} T_{TL} & \star & \star \\ \whline
\color{blue} \tau_{ML} e_l^T & 0 &  \star \\ \hline
0   & \widehat x_{BM}  & \widehat X_{BR}
\end{array} \right)  \wedge
\left( \begin{array}{c I c | c}
\widehat X_{TL} & \star & \star \\ \whline
\widehat x_{ML}^T & 0 & \star \\ \hline
\widehat X_{BL} & \widehat x_{BM} & \widehat X_{BR}
\end{array} \right) \\
~~ =
\left( \begin{array}{c I c | c}
\color{blue} L_{TL} & 0 & 0 \\ \whline
\color{blue} l_{ML}^T & 1 & 0 \\ \hline
\color{blue} L_{BL} & \color{blue} l_{BM} &  L_{BR}
\end{array}\right)
\left( \begin{array}{c I c | c}
\color{blue} T_{TL} & - \color{blue} \tau_{ML} e_l  & 0 \\ \whline
\color{blue} \tau_{ML} e_l^T  & 0 & -\tau_{BM} e_f^T \\ \hline
0 & \tau_{BM} e_f & T_{BR}
\end{array} \right)
\left( \begin{array}{c I c | c}
\color{blue} L_{TL}^T & \color{blue} l_{ML} & \color{blue} L_{BL}^T \\ \whline
0 & 1 & \color{blue} l_{BM}^T \\ \hline
0 & 0 &  L_{BR}^T
\end{array}\right).
\end{eqnarray*}
}
After repartitioning, in Step~6 we get
{
\setlength{\arraycolsep}{2pt}
\begin{equation}
    \label{Before-blk-left}
\begin{array}{l}
\left( \begin{array}{c I c | c | c | c }
X_{00} & \cellcolor{lightgray!35} \star & \cellcolor{lightgray!35} \star & \star & \star \\ \whline
\rowcolor{lightgray!35} x_{10}^T & \chi_{11} & 
\star & \star& \star \\ \hline
\rowcolor{lightgray!35} X_{20} & 
x _{21} & X_{22} &  \star & \star \\ \hline
x_{30}^T & \cellcolor{lightgray!35} \chi_{31}
& \cellcolor{lightgray!35} x_{32}^T & \chi_{33}
& \star \\ \hline
X_{40} & \cellcolor{lightgray!35} x_{41}
& \cellcolor{lightgray!35} X_{42} & x_{43} & X_{44}
\end{array} \right)
=
\left( \begin{array}{c I c | c | c | c }
\color{blue} T_{00} & \star & \star & \star & \star \\ \whline
\color{blue} \tau_{10} e_L^T & 0 & 
\star & \star& \star \\ \hline
0 & 
\widehat x _{21} & \widehat X_{22} &  \star & \star \\ \hline
0 & \widehat \chi_{31}
& \widehat x_{32}^T & 0
& \star \\ \hline
0 & \widehat x_{41}
& \widehat X_{42} & \widehat x_{43} & \widehat X_{44}
\end{array} \right) \wedge
\left( \begin{array}{c I c | c | c | c }
\widehat X_{00} & - \widehat x_{10} & -\widehat X_{20}^T & - \widehat x_{30} & \widehat X_{40}^T \\ \whline
\widehat x_{10}^T & 0 & 
- \widehat x_{21}^T & - \widehat \chi_{31} & - \widehat X_{42} \\ \hline
\widehat X_{20} & 
\widehat x _{21} & \widehat X_{22} &  - \widehat x_{32} & - \widehat X_{42}^T \\ \hline
\widehat x_{30}^T & \widehat \chi_{31}
& \widehat x_{32}^T & 0
& - \widehat x_{43}^T \\ \hline
\widehat X_{40} & \widehat x_{41}
& \widehat X_{42} & \widehat x_{43} & \widehat X_{44}
\end{array} \right)
 \\
\mbox{\hspace{0.1in}}
=\left( \begin{array}{c I c | c | c | c }
\color{blue} L_{00} & 0 & 0 & 0 & 0 \\ \whline
\color{blue} l_{10}^T & 1 & 
0 & 0 & 0 \\ \hline
\color{blue} L_{20} & 
\color{blue} l _{21} & L_{22} &  0 & 0 \\ \hline
\color{blue} l_{30}^T & \color{blue} \lambda_{31}
& l_{32}^T & 1
& 0 \\ \hline
\color{blue} L_{40} & \color{blue} l_{41}
& L_{42} & l_{43} & L_{44}
\end{array} \right)
\left( \begin{array}{c I c | c | c | c }
\color{blue} T_{00} & \color{blue} - \tau_{10} e_l & 0 & 0 & 0 \\ \whline
\color{blue} \tau_{10} e_l^T & 0 & 
- \tau_{21} e_f^T & 0 & 0 \\ \hline
0 & 
\tau_{21} e_F & T_{22} &  - \tau_{32} e_l & 0 \\ \hline
0 & 0
& \tau_{32} e_l^T & 0
& - \tau_{43} e_f^T \\ \hline
0 & 0
& 0 & \tau_{43} e_f & \widehat T_{44}
\end{array} \right)
\left( \begin{array}{c I c | c | c | c }
\color{blue} L_{00}^T & \color{blue} l_{10}  & \color{blue} L_{20}^T & \color{blue} l_{30} & \color{blue} L_{40}^T \\ \whline
0 & 1 & 
\color{blue} l_{21}^T & \color{blue} \lambda_{31} & \color{blue} l_{41}^T \\ \hline
0 & 
0  & L_{22} &  l_{32} & L_{42}^T \\ \hline
0 & 0
& 0 & 1
& l_{43}^T \\ \hline
0 & 0 
& 0 & 0 & L_{44}^T
\end{array} \right)
\end{array}
\end{equation}
}%
and at the bottom of the loop body, in Step~7 it must be that
{
\setlength{\arraycolsep}{2pt}
\begin{equation}
    \label{After-blk-right}
\begin{array}{l}
\left( \begin{array}{c | c | c I c | c }
X_{00}^{\rm +} & \star & \star & \star & \star \\ \hline
x_{10}^T & \chi_{11}^{\rm +} & 
\star & \star& \star \\ \hline
X_{20}^{\rm +} & 
x _{21}^{\rm +} & X_{22}^{\rm +} &  \star & \star \\ \whline
x_{30}^{\rm +\! T} & \chi_{31}^{\rm +}
& x_{32}^{\rm +\! T} & \chi_{33}^{\rm +}
& \star \\ \hline
X_{40}^{\rm +} & x_{41}^{\rm +}
& X_{42}^{\rm +} & x_{43}^{\rm +} & X_{44}^{\rm +}
\end{array} \right)
=
\left( \begin{array}{c | c | c I c | c }
\color{blue} T_{00} & \star & \star & \star & \star \\ \hline
\color{blue} \tau_{10} e_L^T & 0 & 
\star & \star& \star \\ \hline
0 & 
\color{blue}  \tau_{21} e_F & \color{blue}   T_{22} &  \star & \star \\ \whline
0 & 0
& \color{blue}  \tau_{32} e_l^T & 0
& \star \\ \hline
0 & 0
& 0 & \widehat x_{43} & \widehat X_{44}
\end{array} \right)
 \wedge
\left( \begin{array}{c | c | c I c | c }
\widehat X_{00} & - \widehat x_{10} & -\widehat X_{20}^T & - \widehat x_{30} & \widehat X_{40}^T \\ \hline
\widehat x_{10}^T & 0 & 
- \widehat x_{21}^T & - \widehat \chi_{31} & - \widehat X_{42} \\ \hline
\widehat X_{20} & 
\widehat x _{21} & \widehat X_{22} &  - \widehat x_{32} & - \widehat X_{42}^T \\ \whline
\widehat x_{30}^T & \widehat \chi_{31}
& \widehat x_{32}^T & 0
& - \widehat x_{43}^T \\ \hline
\widehat X_{40} & \widehat x_{41}
& \widehat X_{42} & \widehat x_{43} & \widehat X_{44}
\end{array} \right)
 \\
\mbox{\hspace{0.1in}}
=\left( \begin{array}{c | c | c I c | c }
\color{blue} L_{00} & 0 & 0 & 0 & 0 \\ \hline
\color{blue} l_{10}^T & 1 & 
0 & 0 & 0 \\ \hline
\color{blue} L_{20} & 
\color{blue} l _{21} & \color{blue} L_{22} &  0 & 0 \\ \whline
\color{blue} l_{30}^T & \color{blue} \lambda_{31}
& \color{blue} l_{32}^T & 1
& 0 \\ \hline
\color{blue} L_{40} & \color{blue} l_{41}
& \color{blue}  L_{42} & \color{blue}  l_{43} & L_{44}
\end{array} \right)
\left( \begin{array}{c | c | c I c | c }
\color{blue} T_{00} & \color{blue} - \tau_{10} e_l & 0 & 0 & 0 \\ \hline
\color{blue} \tau_{10} e_l^T & 0 & 
\color{blue}  - \tau_{21} e_f^T & 0 & 0 \\ \hline
0 & 
\color{blue}  \tau_{21} e_F & \color{blue}  T_{22} & \color{blue}  - \tau_{32} e_l & 0 \\ \whline
0 & 0
& \color{blue}  \tau_{32} e_l^T & 0
& - \tau_{43} e_f^T \\ \hline
0 & 0
& 0 & \tau_{43} e_f & \widehat T_{44}
\end{array} \right)
\left( \begin{array}{c | c | c I c | c }
\color{blue} L_{00}^T & \color{blue} l_{10}  & \color{blue} L_{20}^T & \color{blue} l_{30} & \color{blue} L_{40}^T \\ \hline
0 & 1 & 
\color{blue} l_{21}^T & \color{blue} \lambda_{31} & \color{blue} l_{41}^T \\ \hline
0 & 
0  & \color{blue} L_{22} &  \color{blue} l_{32} & \color{blue} L_{42}^T \\ \whline
0 & 0
& 0 & 1
& \color{blue} l_{43}^T \\ \hline
0 & 0 
& 0 & 0 & L_{44}^T
\end{array} \right).
\end{array}
\end{equation}
}%
Here we also highlight the ``known'' parts of $T$ in blue at each stage of the algorithm. Again separating what is known we find that
{
\begin{eqnarray*}
\setlength{\arraycolsep}{2pt}
\lefteqn{
\left( \begin{array}{ c | c } 
x _{21} & X_{22} \\ \hline
\chi_{31}
& x_{32}^T \\ \hline
x_{41}
& X_{42} 
\end{array} \right)
~=~
\left( \begin{array}{ c | c } 
\widehat x _{21} & \widehat X_{22} \\ \hline
\widehat \chi_{31}
& \widehat x_{32}^T \\ \hline
\widehat x_{41}
& \widehat X_{42} 
\end{array} \right)
=
\left( \begin{array}{c I c | c | c  }
\color{blue} L_{20} & 
\color{blue} l _{21} & L_{22} &  0  \\ \hline
\color{blue} l_{30}^T & \color{blue} \lambda_{31}
& l_{32}^T & 1 \\ \hline
\color{blue} L_{40} & \color{blue} l_{41}
& L_{42} & l_{43} 
\end{array} \right)
\left( \begin{array}{c I c | c }
\color{blue} T_{00} & \color{blue}-  \tau_{10} e_l & 0  \\ \whline
\color{blue} \tau_{10} e_l^T & 0 & 
  - \tau_{21} e_f^T  \\ \hline
0 & 
 \tau_{21} e_f &   T_{22}   \\ \hline
0 & 0
&  \tau_{32} e_l^T   
\end{array} \right)
\left( \begin{array}{c | c }
 \color{blue} l_{10}  & \color{blue} L_{20}^T  \\ \whline
1 & 
\color{blue} l_{21}^T \\ \hline 
0  & L_{22}^T
\end{array} \right)
}
\\
\setlength{\arraycolsep}{2pt}
&=&
\left( \begin{array}{c I c | c | c  }
\color{blue} L_{20} & 
\color{blue} l _{21} & L_{22} &  0  \\ \hline
\color{blue} l_{30}^T & \color{blue} \lambda_{31}
& l_{32}^T & 1 \\ \hline
\color{blue} L_{40} & \color{blue} l_{41}
& L_{42} & l_{43} 
\end{array} \right)
\left(
\left( \begin{array}{c I c | c }
\color{blue} T_{00} & \color{blue} - \tau_{10} e_l & 0  \\ \whline
\color{blue} \tau_{10} e_l^T & 0 & 
  0  \\ \hline
0 & 
0 &   0   \\ \hline
0 & 0
&  0  
\end{array} \right)
+
\left( \begin{array}{c I c | c }
0 & 0 & 0  \\ \whline
0  & 0 & 
  - \tau_{21} e_f^T  \\ \hline
0 & 
 \tau_{21} e_f &   T_{22}   \\ \hline
0 & 0
&  \tau_{32} e_l^T   
\end{array} \right)
\right)
\left( \begin{array}{c | c }
 \color{blue} l_{10}  & \color{blue} L_{20}^T  \\ \whline
1 & 
\color{blue} l_{21}^T \\ \hline 
0  & L_{22}^T
\end{array} \right)
\\
&=&
\setlength{\arraycolsep}{2pt}
\left( \begin{array}{c I c }
\color{blue} L_{20} & 
\color{blue} l _{21}  \\ \hline
\color{blue} l_{30}^T & \color{blue} \lambda_{31}
\\ \hline
\color{blue} L_{40} & \color{blue} l_{41}
\end{array} \right)
\left( \begin{array}{c I c  }
\color{blue} T_{00} & \color{blue}-  \tau_{10} e_l   \\ \whline
\color{blue} \tau_{10} e_l^T & 0 
\end{array} \right)
\left( \begin{array}{c | c }
 \color{blue} l_{10}  & \color{blue} L_{20}^T  \\ \whline
1 & 
\color{blue} l_{21}^T 
\end{array} \right) +
\left( \begin{array}{c | c | c  } 
\color{blue} l _{21} & L_{22} &  0  \\ \hline
\color{blue} \lambda_{31}
& l_{32}^T & 1 \\ \hline
\color{blue} l_{41}
& L_{42} & l_{43} 
\end{array} \right)
\left( \begin{array}{c | c }
0 & 
-  \tau_{21} e_f^T  \\ \hline
 \tau_{21} e_f &   T_{22}   \\ \hline
0
&  \tau_{32} e_l^T   
\end{array} \right)
\left( \begin{array}{c | c }
1 & 
\color{blue} l_{21}^T \\ \hline 
0  & L_{22}^T
\end{array} \right).
\end{eqnarray*}
}
From this we conclude that we must first update
\begin{eqnarray*}
\left( \begin{array}{ c | c } 
x _{21} & X_{22} \\ \hline
\chi_{31}
&  x_{32}^T \\ \hline
x_{41}
&  X_{42} 
\end{array} \right) & := &
\left( \begin{array}{ c | c } 
x _{21} & X_{22} \\ \hline
\chi_{31}
&  x_{32}^T \\ \hline
x_{41}
&  X_{42} 
\end{array} \right) 
-
\left( \begin{array}{c I c }
\color{blue} L_{20} & 
\color{blue} l _{21}  \\ \hline
\color{blue} l_{30}^T & \color{blue} \lambda_{31}
\\ \hline
\color{blue} L_{40} & \color{blue} l_{41}
\end{array} \right)
\left( \begin{array}{c I c  }
\color{blue} T_{00} & \color{blue}- \tau_{10} e_l   \\ \whline
\color{blue} \tau_{10} e_l^T & 0 
\end{array} \right)
\left( \begin{array}{c | c }
 \color{blue} l_{10}  & \color{blue} L_{20}^T  \\ \whline
1 & 
\color{blue} l_{21}^T 
\end{array} \right) \\
& = &
\left( \begin{array}{ c | c } 
x _{21} & X_{22} \\ \hline
\chi_{31}
&  x_{32}^T \\ \hline
x_{41}
&  X_{42} 
\end{array} \right) 
-
\left( \begin{array}{c I c }
\color{blue} L_{20} & 
\color{blue} l _{21}  \\ \hline
\color{blue} l_{30}^T & \color{blue} \lambda_{31}
\\ \hline
\color{blue} L_{40} & \color{blue} l_{41}
\end{array} \right)
\left( \begin{array}{c I c  }
X_{00} & \star   \\ \whline
x_{10}^T & 0 
\end{array} \right)
\left( \begin{array}{c | c }
 \color{blue} l_{10}  & \color{blue} L_{20}^T  \\ \whline
1 & 
\color{blue} l_{21}^T 
\end{array} \right).
\end{eqnarray*}
After this, the relevant parts of $ X $ satisfy
\[
\left( \begin{array}{ c | c } 
\chi_{11} & \star \\  \hline
x_{21} & X_{22} \\ \hline
\chi_{31}
&  x_{32}^T \\ \hline
x_{41}
&  X_{42} 
\end{array} \right)
=
\left( \begin{array}{c | c | c  } 
1 & 0  \\  \hline
\color{blue} l _{21} & L_{22} &  0  \\ \hline
\color{blue} \lambda_{31}
& l_{32}^T & 1 \\ \hline
\color{blue} l_{41}
& L_{42} & l_{43} 
\end{array} \right)
\left( \begin{array}{c | c }
0 & 
-  \tau_{21} e_f^T  \\ \hline
 \tau_{21} e_f &   T_{22}   \\ \hline
0
&  \tau_{32} e_l^T   
\end{array} \right)
\left( \begin{array}{c | c }
1 & 
\color{blue} l_{21}^T \\ \hline 
0  & L_{22}^T
\end{array} \right),
\]
which we recognize as a partial $ L T L^T $ factorization   that can be used to update the required parts of $ X $ and $ L $ via, for example, an unblocked left-looking algorithm that returns when the relevant columns have computed.  
Note that when the algorithm starts with the full matrix, the
first column of matrix  $ L $ has zeroes below the  diagonal while the last column that was computed 
in an earlier block iteration becomes that first column of $ L $ for the partial factorization.  This completes the derivation of the blocked left-looking algorithm in  Figure~\ref{fig:LTLt_blk_right}.

\subsection{Fused blocked right-looking algorithm: Variant 2a}

\label{sec:2a}

We now justify Invariant~2a given by~(\ref{eqn:inv_unb_right_2a-1})-(\ref{eqn:inv_unb_right_2a-2}).
Invariant~1 is given by
{
\setlength{\arraycolsep}{2pt}
  \begin{eqnarray}
  \nonumber 
  \lefteqn{
\left( \begin{array}{c I c | c}
X_{TL} & \star & \star \\ \whline
x_{ML}^T & \chi_{MM} & \star \\ \hline
X_{BL} & x_{BM} & X_{BR}
\end{array} \right)
=
\left( \begin{array}{c I c | c}
T_{TL} & \star & \star \\ \whline
 \tau_{ML} e_l^T & 0 &  \star \\ \hline
0  &  \tau_{BM} 
L_{BR} e_f
& L_{BR} T_{BR} L_{BR}^T
\end{array} \right)  \wedge 
\left( \begin{array}{c I c | c}
\widehat X_{TL} & 
- \widehat x_{ML}
& 
- \widehat X_{BL}^T
\\ \whline
\widehat x_{ML}^T & 0 & 
- \widehat x_{BM}^T \\ \hline
\widehat X_{BL} & \widehat x_{BM} & \widehat X_{BR}
\end{array} \right) } \\
\nonumber
&& ~~~~~ =
\left( \begin{array}{c I c | c}
\color{blue} L_{TL} & 0 & 0 \\ \whline
\color{blue} l_{ML}^T & 1 & 0 \\ \hline
\color{blue} L_{BL} & \color{blue} l_{BM} &  I
\end{array}\right)
\left( \begin{array}{c I c | c}
T_{TL} & - \tau_{ML} e_l  & 0 \\ \whline
\tau_{ML} e_l^T  & 0 & -\tau_{BM} (L_{BR} e_f)^T \\ \hline
0 & \tau_{BM} L_{BR} e_f & L_{BR} T_{BR} L_{BR}^T
\end{array} \right)
\left( \begin{array}{c I c | c}
\color{blue} L_{TL}^T & \color{blue} l_{ML} & \color{blue} L_{BL}^T \\ \whline
0 & 1 & \color{blue} l_{BM}^T \\ \hline
0 & 0 &  I
\end{array}\right).
\end{eqnarray}
    }%
In order to avoid the separate skew-symmetric rank-2 update in Variant~1, it may be beneficial to delay the application of the last Gauss transform to the remainder of the matrix.
This means that instead of 
\[
\left( \begin{array}{c | c}
\chi_{MM} & \star \\ \hline
x_{BM}  & X_{BR}
\end{array} \right)
=
\left( \begin{array}{c | c}
0 & \star \\ \hline
\tau_{BM} L_{BR} e_f  & L_{BR} T_{BR} L_{BBR}^T
\end{array} \right)
\]
we want to maintain
\begin{eqnarray*}
\left( \begin{array}{c | c}
\chi_{MM} & \star \\ \hline
x_{BM}  & X_{BR}
\end{array} \right)
&=&
\left( \begin{array}{c | c}
1 & 0 \\ \hline
l_{BM}  & I
\end{array} \right)
\left( \begin{array}{c | c}
0 & \star \\ \hline
\tau_{BM} L_{BR} e_f  & L_{BR} T_{BR} L_{BR}^T
\end{array} \right)
\left( \begin{array}{c | c}
1 & l_{BM}^T \\ \hline
0  & I
\end{array} \right) \\
&=&
\left( \begin{array}{c | c}
1 & 0 \\ \hline
l_{BM}  &  L_{BR}
\end{array} \right)
\left( \begin{array}{c | c}
0 & \star \\ \hline
\tau_{BM} e_f  &  T_{BR} 
\end{array} \right)
\left( \begin{array}{c | c}
1 & l_{BM}^T \\ \hline
0  & L_{BR}^T
\end{array} \right) 
.
\end{eqnarray*}%
Thus we arrive at Invariant~2a:
{
\setlength{\arraycolsep}{2pt}
\[
\begin{array}{l}
\left( \begin{array}{c I c | c}
X_{TL} & \star & \star \\ \whline
x_{ML}^T & \chi_{MM} & \star \\ \hline
X_{BL} & x_{BM} & X_{BR}
\end{array} \right)
=
\left( \begin{array}{c I c | c}
T_{TL} & ~~~~~~~~~~~~\star ~~~~~~~~~~~~& \star \\ \whline
 \tau_{ML} e_l^T & 
 \multicolumn{2}{c}{
 \multirow{2}{*}{
 $
 \left( \begin{array}{c | c}
 1 &  0 \\ \hline
\color{blue} l_{BM} & L_{BR} 
\end{array}
\right)
 \left( \begin{array}{c | c}
 0 &  \star \\ \hline
\tau_{BM} 
e_f
& T_{BR} 
\end{array}
\right)
\left( \begin{array}{c | c}
 1 &  \color{blue} l_{BM}^T \\ \hline
0 & L_{BR}^T
\end{array}
\right)
$
 }
 }
  \\ \cline{1-1}
0  &  \multicolumn{2}{c}{}
\end{array} \right)
\wedge \\
\\[-0.15in]
~~ 
\left( \begin{array}{c I c | c}
\widehat X_{TL} & \star & \star \\ \whline
\widehat x_{ML}^T & 0 & \star \\ \hline
\widehat X_{BL} & \widehat x_{BM} & \widehat X_{BR}
\end{array} \right)  =
\left( \begin{array}{c I c | c}
\color{blue} L_{TL} & 0 & 0 \\ \whline
\color{blue} l_{ML}^T & 1 & 0 \\ \hline
\color{blue} L_{BL} & \color{blue} l_{BM} &  L_{BR}
\end{array}\right)
\left( \begin{array}{c I c | c}
T_{TL} & - \tau_{ML} e_l  & 0 \\ \whline
\tau_{ML} e_l^T  & 0 & -\tau_{BM} e_f^T \\ \hline
0 & \tau_{BM} e_f & T_{BR}
\end{array} \right)
\left( \begin{array}{c I c | c}
\color{blue} L_{TL}^T & \color{blue} l_{ML} & \color{blue} L_{BL}^T \\ \whline
0 & 1 & \color{blue} l_{BM}^T \\ \hline
0 & 0 &  L_{BR}^T
\end{array}\right).
\end{array}
\]
}%

Now we  derive the corresponding  blocked algorithm from the
invariant.
In the body of the loop, the repartitioning now exposes a new block of columns and rows in each iteration.
After repartitioning in  Step~5a, we get for Step~6 that
{
\setlength{\unitlength}{2pt}
\begin{eqnarray*}
\label{Before-blk-right-2b}
\lefteqn{
\left( \begin{array}{c I c | c | c | c }
X_{00} & \cellcolor{lightgray!35}  \star & \cellcolor{lightgray!35}  \star & \star & \star \\ \whline
\cellcolor{lightgray!35} x_{10}^T & \cellcolor{lightgray!35} \chi_{11} & 
\cellcolor{lightgray!35} \star & \cellcolor{lightgray!35} \star& \cellcolor{lightgray!35} \star \\ \hline
\cellcolor{lightgray!35} X_{20} & 
\cellcolor{lightgray!35} x _{21} & \cellcolor{lightgray!35} X_{22} &  \cellcolor{lightgray!35} \star & \cellcolor{lightgray!35} \star \\ \hline
x_{30}^T & \cellcolor{lightgray!35} \chi_{31}
& \cellcolor{lightgray!35} x_{32}^T & \chi_{33}
& \star \\ \hline
X_{40} & \cellcolor{lightgray!35} x_{41}
& \cellcolor{lightgray!35} X_{42} & x_{43} & X_{44}
\end{array} \right) }\\
& 
= &
\setlength{\arraycolsep}{2pt}
\left( \begin{array}{c I c | c | c | c }
T_{00} & \mbox{\hspace{0.45in}}
\star
\mbox{\hspace{0.45in}} & 
\mbox{\hspace{0.45in}}
\star
\mbox{\hspace{0.45in}} & \mbox{\hspace{0.45in}}\star \mbox{\hspace{0.45in}}
 & \mbox{\hspace{0.4in}} \star \mbox{\hspace{0.4in}} \\ \whline
\tau_{10} e_l^T & 
\multicolumn{4}{c}{
\multirow{4}{*}{$
\left( \begin{array}{c | c | c | c}
1 & 0 & 0 & 0 \\\hline 
\color{blue} l_{21} & L_{22} & 0 &  0 \\ \hline
\color{blue} \lambda_{31} & l_{32}^T &
1 & 0 \\ \hline
\color{blue} l_{41} & L_{42} &
l_{43} &
L_{44}
\end{array}
\right) 
\left( \begin{array}{c | c | c | c}
0 & -\tau_{21} & 0 & 0 \\ \hline
\tau_{21} & T_{22} & -\tau_{32} e_l &  0 \\ \hline
0 & \tau_{32} e_l^T &
0 & - \tau_{43} e_f^T \\ \hline
0 & 0 &
\tau_{43} e_f &
T_{44}
\end{array}
\right)
\left( \begin{array}{c | c | c | c}
1 & \color{blue} l_{21}^T & \color{blue} \lambda_{31} & \color{blue} l_{41}^T \\ \hline
0 & L_{22}^T & l_{32}  &  L_{42}^T \\ \hline
0 & 0 & 
1 & l_{43}^T \\ \hline
0 & 0 & 0 &
L_{44}^T
\end{array}
\right)
$ } }
\\
\cline{1-1} 
0 & 
  \multicolumn{4}{c}{}
\\
\cline{1-1} 
0 & 
  \multicolumn{4}{c}{}
\\
\cline{1-1} 
0 & 
  \multicolumn{4}{c}{}
\end{array} \right),
\end{eqnarray*}%
}%
where the gray highlighting captures the block of rows and columns being exposed in this iteration.
At the bottom of the loop we find
for Step~7 that 
{
\begin{eqnarray}
    \label{eqn:blk:1-2b}
\lefteqn{
\left( \begin{array}{c | c | c I c | c }
X_{00}^{\rm +} & \cellcolor{lightgray!35} \star & \cellcolor{lightgray!35} \star & \star & \star \\ \hline
\rowcolor{lightgray!35} x_{10}^{\rm +\!T} & \chi_{11}^{\rm +} & 
\star & \star& \star \\ \hline
\rowcolor{lightgray!35} X_{20}^{\rm +} & 
\cellcolor{lightgray!35} x _{21}^{\rm +} & \cellcolor{lightgray!35} X_{22}^{\rm +} &  \star & \star \\ \whline
x_{30}^{\rm +\!T} & \cellcolor{lightgray!35} \chi_{31}^{\rm +}
& \cellcolor{lightgray!35} x_{32}^{\rm +\!T} & \chi_{33}^{\rm +}
& \star \\ \hline
X_{40}^{\rm +} & \cellcolor{lightgray!35} x_{41}^{\rm +} 
& \cellcolor{lightgray!35} X_{42}^{\rm +} & x_{43}^{\rm +} & X_{44}^{\rm +}
\end{array} \right)} \nonumber \\
&&=
\setlength{\arraycolsep}{2pt}
\left( \begin{array}{c | c | c I c | c}
T_{00} & \star & \star & \mbox{\hspace{0.55in}}\star\mbox{\hspace{0.55in}}
& \mbox{\hspace{0.4in}}\star\mbox{\hspace{0.4in}}\\ \hline
\tau_{10} e_l^T & 0 & 
~~~\star~~~ & \mbox{\hspace{0.4in}}\star\mbox{\hspace{0.4in}}& \mbox{\hspace{0.4in}}\star\mbox{\hspace{0.4in}} \\ \hline
0 &
\tau_{21} e_f
&
T_{22}
&
\star & \star \\ \whline
0 & 
0 &
\tau_{32} e_l^T
& 
\multicolumn{2}{c}{
 \multirow{2}{*}{
 $
 \left( \begin{array}{c | c}
 1 & 0 \\ \hline
 l_{43} & L_{44}
 \end{array} \right)
 \left( \begin{array}{c | c}
 0 & -\tau_{43} e_f^T \\ \hline
\tau_{43} e_f & T_{44}
 \end{array} \right)
 \left( \begin{array}{c | c}
 1 & l_{43}^T \\ \hline
 0 & L_{44}^T
 \end{array} \right)
 $
 }}
\\ \cline{1-3} 
0 & 
0 & 
0 & 
\multicolumn{2}{c}{}
\end{array} \right)
\NoShow{
\wedge \\
~~
\left( \begin{array}{c I c | c | c}
\widehat X_{00} & \star & \star & \star \\ \whline
\widehat x_{10}^T & 0 & 
\star & \star \\ \hline
\widehat x_{20}^T & 
\widehat \chi _{21} &  0 &  \star \\ \hline
\widehat X_{30} & \widehat x_{31}
& \widehat x_{32} & \widehat X_{33}
\end{array} \right)
=
\left( \begin{array}{c I c | c | c}
L_{00} & 0 & 0 & 0 \\ \whline
l_{10}^T & 1 & 
0 & 0 \\ \hline
l_{20}^T & 
\lambda _{21} & 
1
&  0 \\ \hline
L_{30} & l_{31}
& l_{32} & L_{33}
\end{array} \right)
\left( \begin{array}{c I c | c | c}
T_{00} & 
\tau_{10} e_l & 0 &  \\ \whline
\tau_{10} e_l^T & 0 & 
\tau_{21}  & 0 \\ \hline
0 & 
\tau_{21} & 
0
&  \tau_{32} e_f^T \\ \hline
0 & 
0 &
\tau_{32} e_f & T_{33}
\end{array} \right)
\left( \begin{array}{c I c | c | c}
L_{00}^T & l_{10} & l_{20} & L_{30}^T \\ \whline
0 & 1 & 
\lambda_{21} & l_{31}^T \\ \hline
0
& 
0 & 
1
&  l_{32}^T \\ \hline
0 & 
& 0 & L_{33}^T
\end{array} \right).
}
.
\end{eqnarray}
}
We observe that
{
\begin{eqnarray*}
  \lefteqn{
 \left( \begin{array}{c | c | c | c }
 \chi_{11} & 
\star & \star& \star \\ \hline 
x _{21} & X_{22} &  \star & \star \\ \hline
 \chi_{31}
& x_{32}^T & \chi_{33}
& \star \\ \hline
 x_{41}
& X_{42} & x_{43} & X_{44}
\end{array} \right)
} \\
&=&
\left( \begin{array}{c | c | c | c}
1 & 0 & 0 & 0 \\\hline 
\color{blue} l_{21} & L_{22} & 0 &  0 \\ \hline
\color{blue} \lambda_{31} & l_{32}^T &
1 & 0 \\ \hline
\color{blue} l_{41} & L_{42} &
l_{43} &
L_{44}
\end{array}
\right) 
\left( \begin{array}{c | c | c | c}
0 & -\tau_{21} & 0 & 0 \\ \hline
\tau_{21} & T_{22} & -\tau_{32} e_l &  0 \\ \hline
0 & \tau_{32} e_l^T &
0 & - \tau_{43} e_f^T \\ \hline
0 & 0 &
\tau_{43} e_f &
T_{44}
\end{array}
\right)
\left( \begin{array}{c | c | c | c}
1 & \color{blue} l_{21}^T & \color{blue} \lambda_{31} & \color{blue} l_{41}^T \\ \hline
0 & L_{22}^T & l_{32}  &  L_{42}^T \\ \hline
0 & 0 & 
1 & l_{43}^T \\ \hline
0 & 0 & 0 &
L_{44}^T
\end{array}
\right)
,
\end{eqnarray*}
}%
which implies that
\[
 \left( \begin{array}{c | c}
 \chi_{11} & 
\star  \\ \hline 
x _{21} & X_{22}  \\ \hline
 \chi_{31}
& x_{32}^T \\ \hline
 x_{41}
& X_{42} 
\end{array} \right) 
=
\left( \begin{array}{c | c | c }
1 & 0 & 0  \\ \hline
\color{blue} l_{21} & L_{22} & 0  \\ \hline
\color{blue} \lambda_{31} & l_{32}^T &
1  \\ \hline
\color{blue} l_{41} & L_{42} & l_{43} 
\end{array}
\right)
\left( \begin{array}{c | c }
0 & - \tau_{21} e_f^T  \\
\hline
\tau_{21} e_f^T & T_{22}   \\ \hline
0 & \tau_{32} e_l^T  
\end{array}
\right)
\left( \begin{array}{c | c}
1 & \color{blue} l_{21}^T \\ \hline
0 & L_{22}^T 
\end{array}
\right).
\]
Examining~\mbox{(\ref{eqn:blk:1-2b})} tells us that  
$
\left( \begin{array}{c | c}
\chi_{11}^{\rm +} & 
\star  \\ \hline 
x _{21}^{\rm +} & X_{22}^{\rm +}  \\ \hline
 \chi_{31}{\rm +}
& x_{32}^{\rm +\!T} \\ \hline
 x_{41}^{\rm +}
& X_{42}^{\rm +} 
\end{array} \right) 
~
\mbox{and}
~
\left( \begin{array}{c | c | c }
1 & 0 & 0  \\ \hline
\color{blue} l_{21} & L_{22} & 0  \\ \hline
\color{blue} \lambda_{31} & l_{32}^T &
1  \\ \hline
\color{blue} l_{41} & L_{42} &
l_{43} 
\end{array}
\right)
~
\mbox{are computed from}
~
\left( \begin{array}{c | c}
 \chi_{11} & 
\star  \\ \hline 
x _{21} & X_{22}  \\ \hline
 \chi_{31}
& x_{32}^T \\ \hline
 x_{41}
& X_{42} 
\end{array} \right) 
$
by factoring that panel.

The purpose of the game now becomes to update the remaining part of $ X $ by separating what is known from what is yet to be computed.  Notice that
{
\begin{eqnarray}
\nonumber
    \lefteqn{\hspace{-3em}
    \left( \begin{array}{c | c}
    \chi_{33} & \star \\ \hline
    x_{43}& X_{44}
    \end{array}
    \right) =
    \left( \begin{array}{c | c | c | c}
\lambda_{31} & l_{32}^T &
1 & 0 \\ \hline
l_{41} & L_{42} &
l_{43} &
L_{44}
\end{array}
\right)
\left( \begin{array}{c | c | c | c}
0 & -\tau_{21} e_f^T & 0 & 0 \\
\hline
\tau_{21} e_f & T_{22} & -\tau_{32} e_l &  0 \\ \hline
0 & \tau_{32} e_l^T &
0 & - \tau_{43} e_f^T \\ \hline
0 & 0 &
\tau_{43} e_f &
T_{44}
\end{array}
\right)
\left( \begin{array}{c | c}
\lambda_{42} & l_{41}^T \\ \hline
l_{32}  &  L_{42}^T \\ \hline
1 & l_{43}^T \\ \hline
0 &
L_{44}^T
\end{array}
\right)
    } \\
    \label{eqn:bk-right-2a-1}
    & = &
        \left( \begin{array}{c | c | c | c}
\lambda_{31} & l_{32}^T &
1 & 0 \\ \hline
l_{41} & L_{42} &
l_{43} &
L_{44}
\end{array}
\right)
\left[
\left( \begin{array}{c | c | c | c}
0 & -\tau_{21} e_f^T & 0 & 0 \\
\hline
\tau_{21} e_f & T_{22} & -\tau_{32} e_l &  0 \\ \hline
0 & \tau_{32} e_l^T &
0 & 0 \\ \hline
0 & 0 &
0 &
0
\end{array}
\right)
+ \right.
\\
& &  \hspace{1.5in} \left.
\left( \begin{array}{c | c | c | c}
0 & 0 & 0 & 0 \\
\hline
0 & 0 & 0 &  0 \\ \hline
0 & 0 &
0 & - \tau_{43} e_f^T \\ \hline
0 & 0 &
\tau_{43} e_f &
T_{44}
\end{array}
\right)
\right]
\left( \begin{array}{c | c}
\lambda_{42} & l_{41}^T \\ \hline
l_{32}  &  L_{42}^T \\ \hline
1 & l_{43}^T \\ \hline
0 &
L_{44}^T
\end{array}
\right)  \\
\nonumber
& = & 
    \left( \begin{array}{ c | c | c }
\lambda_{31} & l_{32}^T &
1 \\ \hline
l_{41} & L_{42} &
l_{43} 
\end{array}
\right)
\left( \begin{array}{c | c | c }
 0 & - \tau_{21} & 0 \\ \hline
 \tau_{21} &  T_{22} & -\tau_{32} e_l  \\ \hline
0 & \tau_{32} e_l^T &
0 
\end{array}
\right)
\left( \begin{array}{c | c}
\lambda_{31} & l_{41}^T \\ \hline
l_{32}  &  L_{42}^T \\ \hline
1 & l_{43}^T 
\end{array}
\right)
+
\left( \begin{array}{c | c}
    \chi_{33}^{\rm +} & \star \\ \hline
    x_{43}^{\rm +}& X_{44}^{\rm +}
    \end{array}
    \right)
.
\end{eqnarray}
}%
This prescribes  the update
\begin{eqnarray}
\setlength{\arraycolsep}{2pt}
\nonumber
\left( \begin{array}{c | c}
    \chi_{33} & \star \\ \hline
    x_{43}& X_{44}
    \end{array}
    \right)
    &:=&
    \left( \begin{array}{c | c}
    \chi_{33} & \star \\ \hline
    x_{43}& X_{44}
    \end{array}
    \right)
    -
 \left( \begin{array}{ c | c | c }
\lambda_{31} & l_{32}^T &
1 \\ \hline
l_{41} & L_{42} &
l_{43} 
\end{array}
\right)
\left( \begin{array}{c | c | c }
 0 & - \tau_{21} & 0 \\ \hline
 \tau_{21} &  T_{22} & -\tau_{32} e_l  \\ \hline
0 & \tau_{32} e_l^T &
0 
\end{array}
\right)
\left( \begin{array}{c | c}
\lambda_{31} & l_{41}^T \\ \hline
l_{32}  &  L_{42}^T \\ \hline
1 & l_{43}^T 
\end{array}
\right)  \\
\nonumber
&=&
    \left( \begin{array}{c | c}
    \chi_{33} & \star \\ \hline
    x_{43}& X_{44}
    \end{array}
    \right)
    -
 \left( \begin{array}{ c | c | c }
\lambda_{31} & l_{32}^T &
1 \\ \hline
l_{41} & L_{42} &
l_{43} 
\end{array}
\right)
\left( \begin{array}{c | c | c }

\chi_{11} & \star & \star\\ \hline
 x_{21} &  X_{22} & \star  \\ \hline
\chi_{31}  & x_{32}^T &
0 
\end{array}
\right)
\left( \begin{array}{c | c}
\lambda_{31} & l_{41}^T \\ \hline
l_{32}  &  L_{42}^T \\ \hline
1 & l_{43}^T 
\end{array}
\right)
.
    \end{eqnarray}
\NoShow{
Invoking Corollary~\ref{cor:inv} reveals
\begin{eqnarray*}
\lefteqn{
\left( \begin{array}{c I c | c}
T_{22} & \star & 0 \\ \whline 
\tau_{32} e_l^T &
\chi_{33}^{\rm +} & \star \\ \hline
0 &
x_{43}^{\rm +} & X_{44}^{\rm +}
\end{array} \right)
 = 
\left( \begin{array}{c I c | c}
T_{22} & \star & 0 \\ \whline 
\tau_{32} e_l^T & 
0 & \star \\ \hline
0 & 
\tau_{43} L_{44} e_f & L_{44}
 T_{44} L_{44}^T 
 \end{array} \right) }
 \\
& & =
 \left( \begin{array}{c I c | c}
 I & 0 & 0 \\ \whline
0 & 1 &  0 \\ \hline
0 & 0 &
L_{44}
\end{array}
\right)
\left( \begin{array}{c I c | c}
T_{22} & \star & 0 \\ \whline 
\tau_{32} e_l^T & 
0 & \star \\ \hline
0 & 
\tau_{43}  e_f & 
 T_{44}  
 \end{array} \right)
\left( \begin{array}{c I c | c}
 I & 0 & 0 \\ \whline
0 & 1 &  0 \\ \hline
0 & 0 &
L_{44}
\end{array}
\right)^T
\\
& & =
 \left( \begin{array}{c I c | c}
 I & 0 & 0 \\ \whline
0 & 1 &  0 \\ \hline
0 & 0 &
L_{44}
\end{array}
\right)
\left( \begin{array}{c | c | c}
L_{22} & 0 &  0 \\ \hline
l_{32}^T &
1 & 0 \\ \hline
L_{42} &
l_{43} &
L_{44}
\end{array}
\right)^{-1} \\
& & ~~~~~~
\left( \begin{array}{c | c | c}
L_{22} & 0 &  0 \\ \hline
l_{32}^T &
1 & 0 \\ \hline
L_{42} &
l_{43} &
L_{44}
\end{array}
\right) 
\left( \begin{array}{c I c | c}
T_{22} & \star & 0 \\ \whline 
\tau_{32} e_l^T & 
0 & \star \\ \hline
0 & 
\tau_{43}  e_f & 
 T_{44}  
 \end{array} \right)
\left( \begin{array}{c | c | c}
L_{22} & 0 &  0 \\ \hline
l_{32}^T &
1 & 0 \\ \hline
L_{42} &
l_{43} &
L_{44}
\end{array}
\right)^{T}
 \\
& &  ~~~~~~ ~~~~~~
\left( \begin{array}{c | c | c}
L_{22} & 0 &  0 \\ \hline
l_{32}^T &
1 & 0 \\ \hline
L_{42} &
l_{43} &
L_{44}
\end{array}
\right)^{-T}
\left( \begin{array}{c I c | c}
 I & 0 & 0 \\ \whline
0 & 1 &  0 \\ \hline
0 & 0 &
L_{44}
\end{array}
\right)^T
\\
& &  =
 \left( \begin{array}{c I c | c}
 L_{22} & 0 & 0 \\ \whline
l_{32}^T & 1 &  0 \\ \hline
L_{42} & l_{43} &
I
\end{array}
\right)^{-1}
\left( \begin{array}{c I c | c}
X_{22} & \star & \star \\ \whline 
x_{32}^T & 
0 & \star \\ \hline
X_{42} & 
x_{43} & 
 X_{44}  
 \end{array} \right) 
 \left( \begin{array}{c I c | c}
 L_{22} & 0 & 0 \\ \whline
l_{32}^T & 1 &  0 \\ \hline
L_{42} & l_{43} &
I
\end{array}
\right)^{-T}.
\end{eqnarray*}
This allows us to employ Theorem~\ref{thm:right-update} by noting that in that theorem
\begin{itemize}
    \item 
    $ C_{TL}^{\rm +} = 
    \left( \begin{array}{c I c}
    T_{22} & \star \\ \whline
    \tau_{32} e_l^T & 0 
    \end{array} \right) $.
    \item 
    $ 
    \begin{array}[t]{@{}r @{~} c @{~} l}
    \begin{array}[t]{c}
    \underbrace{
    \left( \begin{array}{c I c}
    0 & x_{43}^{\rm +} 
    \end{array} \right) 
    } \\
    C_{BL}^{\rm +}
    \end{array}
    & = &
    - 
    \begin{array}[t]{c}
    \underbrace{
    \left( 
    \begin{array}{c I c }
    L_{42} & l_{43}
    \end{array} \right)
    } \\
    B_{BL}
    \end{array}
    \begin{array}[t]{c}
    \underbrace{
    \left( \begin{array}{c I c}
    T_{22} & - \tau_{32} e_l \\ \whline
    \tau_{32} e_l^T & 0 
    \end{array} \right) 
    } \\
    C_{TL}^{\rm +}
    \end{array}
    + 
    \begin{array}[t]{c}
    \underbrace{
    \left( \begin{array}{c I c}
    X_{42} & x_{43}
    \end{array} \right) 
    } \\
    C_{BL}
    \end{array}
    \begin{array}[t]{c}
    \underbrace{
    \left( \begin{array}{c I c}
    L_{22} & 0 \\ \whline
    l_{32}^T& 1 
    \end{array} \right)^{-T} 
    }
    \\
    B_{TL}^{-T}
    \end{array} \\
    & = & 
    \left( \begin{array}{ c I c}
    \star\star &
    \tau_{32} L_{42} e_l + 
    \end{array}
    \right)
    \end{array}
    $.
    \item
    $
    \begin{array}[t]{r@{~}c@{~}l@{~}c@{~}l}
    \begin{array}[t]{c}
    \underbrace{
    X_{44}^{\rm +}
    } \\
    C_{BR}^{\rm +}
    \end{array}
    &= &
    \begin{array}[t]{c}
    \underbrace{
    X_{44}
    } \\
    C_{BR}
    \end{array}
    &-& 
    \begin{array}[t]{c}
    \underbrace{
    \left( \begin{array}{c | c}
    L_{42} &
    l_{43}
    \end{array} \right)
    } \\
    B_{BL}
    \end{array}
    \begin{array}[t]{c}
    \underbrace{
    \left( \begin{array}{c I c}
    T_{22} & \star \\ \whline
    \tau_{32} e_l^T & 0 
    \end{array} \right) 
    }
    \\
    C_{TL}^{\rm +}
    \end{array}
    \begin{array}[t]{c}
    \underbrace{
    \left( \begin{array}{c | c}
    L_{42} &
    l_{43}
    \end{array} \right)^T}
    \\
    B_{BL}^T
    \end{array}
    \\
    && & 
    + &
    \begin{array}[t]{c}
    \underbrace{    \left( \begin{array}{c | c }
    L_{42} & 
    l_{43}
    \end{array} \right)
    } \\
    B_{BL}
    \end{array}
    \begin{array}[t]{c}
    \underbrace{
     \left( \begin{array}{c I c}
    0 & x_{43}^{\rm +} 
    \end{array} \right)^T
    } \\
    C_{BL}^T
    \end{array}
    -
    \begin{array}[t]{c}
    \underbrace{
    \left( \begin{array}{c I c}
    0 & x_{43}^{\rm +} 
    \end{array} \right)
    } \\
    C_{BL}^{\rm +}
    \end{array}
    \begin{array}[t]{c}
    \underbrace{
    \left( \begin{array}{c | c }
    L_{42} & 
    l_{43}
\end{array} \right)^T
} \\
B_{BL}^T
\end{array} \\
    &= &
    X_{44} &-& 
    \left( \begin{array}{c | c}
    L_{42} &
    l_{43}
    \end{array} \right)
    \left( \begin{array}{c I c}
    T_{22} & \star \\ \whline
    \tau_{32} e_l^T & 0 
    \end{array} \right)
    \left( \begin{array}{c | c}
    L_{42} &
    l_{43}
    \end{array} \right)^T +
    ( l_{43} x_{43}^{\rm +\!T} -
    x_{43}^{\rm +} l_{43}^T ).
    \end{array}
    $.
\end{itemize}
}
\NoShow{
We can do slightly better, in  the process linking the blocked right-looking algorithm to Wimmer's unblocked algorithm.
\begin{eqnarray*}
\lefteqn{(\ref{eqn:bk-right-1}) = 
\left( \begin{array}{ c | c | c}
l_{32}^T &
1 & 0 \\ \hline
L_{42} &
l_{43} &
L_{44}
\end{array}
\right)
\left[
\left( \begin{array}{c | c | c}
 T_{22} & 0 &  0 \\ \hline
 0 &
0 & 0 \\ \hline
0 &
0 &
0
\end{array}
\right)
+
\left( \begin{array}{c | c | c}
 0 & -\tau_{32} e_l &  0 \\ \hline
 \tau_{32} e_l^T &
0 & 0 \\ \hline
0 &
0 &
0
\end{array}
\right)
\right.
}
\\
&&
\hspace{2in}
\left.
+
\left( \begin{array}{c | c | c}
 0 & 0 &  0 \\ \hline
0 &
0 & - \tau_{43} e_f^T \\ \hline
0 &
\tau_{43} e_f &
T_{44}
\end{array}
\right)
\right]
\left( \begin{array}{c | c}
l_{32}  &  L_{42}^T \\ \hline
1 & l_{34}^T \\ \hline
0 &
L_{44}^T
\end{array}
\right)   \\
& = & 
\left( \begin{array}{ c }
l_{32}^T  \\ \hline
L_{42}
\end{array}
\right)
T_{22}
\left( \begin{array}{c | c}
l_{32}  &  L_{42}^T 
\end{array}
\right)
+
\left( \begin{array}{ c | c }
l_{32}^T &
1  \\ \hline
L_{42} &
l_{43} 
\end{array}
\right)
\left[
\left( \begin{array}{c | c }
 0 & -\tau_{32} e_l \\ \hline
 \tau_{32} e_l^T &
0 
\end{array}
\right)
\right]
\left( \begin{array}{c | c}
l_{32}  &  L_{42}^T \\ \hline
1 & l_{34}^T 
\end{array}
\right) \\
& & \hspace{2in}
+
\left( \begin{array}{c | c}
1 & 0 \\ \hline
l_{43} &
I 
\end{array}
\right)
\left( \begin{array}{c | c}
    \chi_{33}^{\rm +} & \star \\ \hline
    x_{43}^{\rm +}& X_{44}^{\rm +}
    \end{array}
    \right)
\left( \begin{array}{c | c}
1 & l_{43}^T \\ \hline
0 &
I 
\end{array}
\right)\\
& = & 
\end{eqnarray*}
}%
This completes the derivation of algorithm Variant~2a in  Figure~\ref{fig:LTLt_blk_right2}.
Importantly, the separate skew-symmetric rank-2 update of Variant~1 does not appear in Variant~2a, having been fused  with the computation of what in Variant~1 was the previous iteration, reducing the number of times $ X_{44} $ needs to be brought in from memory.  
 
\begin{figure}[p]

\resetsteps      


\renewcommand{\routinename}{ \left[ X, L \right] := \mbox{\sc LTLt\_blk\_right\_2a/b}( X ) }


\renewcommand{\guard}{
  m( X_{TL} ) < m( X )-1
}


\renewcommand{\partitionings}{
$ L  = I $\\
  $
  X \rightarrow
  \left( \begin{array}{c I c | c}
  X_{TL} & \star & \star \\ \whline
  x_{ML}^T & \chi_{MM} & \star \\ \hline
  X_{BL} & x_{BM} & X_{BR}
  \end{array}
  \right)
  $
,
  $
  L \rightarrow
\left( \begin{array}{c I c | c}
  L_{TL} & 0 & 0 \\ \whline
  l_{ML}^T & \lambda_{MM} & 0 \\ \hline
  L_{BL} & l_{BM} & L_{BR}
  \end{array}
  \right)
  $ 
}

\renewcommand{\moreinitialize}{
 \\
 \mbox{{\color{blue} if} Variant 2b Compute first Gauss transform:} \\
  \quad
  \begin{tabular}{@{}l @{\quad} l}
  $\chi = $ first element of $x_{BM}$ \\
  $ L_{BR} e_f := x_{BM} / \chi $ &
  \mbox{(compute first column of $L_{BR}$)} \\
  $ x_{BM} := \chi e_f $ \\
  \end{tabular} \\
  \mbox{\color{blue} endif}}

\renewcommand{\partitionsizes}{
$ X_{TL} $ and $ L_{TL} $ are $ 0 \times 0 $
}


\renewcommand{\repartitionings}{
\setlength{\arraycolsep}{4pt}
\footnotesize
$  \left( \begin{array}{c I c | c}
  X_{TL} & \star & \star \\ \whline
  x_{ML}^T & \chi_{MM} & \star \\ \hline
  X_{BL} & x_{BM} & X_{BR}
  \end{array}
  \right)
  \rightarrow
  \left( \begin{array}{c I c | c | c | c}
  X_{00} & \star & \star & \star & \star \\ \whline
   x_{10}^T & \chi_{11} & \star & \star & \star\\ \hline
   X_{20} & x_{21} & X_{22} & \star & \star \\ \hline
   x_{30}^T & \chi_{31} & x_{32}^T & \chi_{33} & \star \\ \hline
   X_{40} & x_{41} & X_{42} & x_{43} & X_{44}
   \end{array} \right)
   $,
   $
   \left( \begin{array}{c I c | c}
  L_{TL} & 0 & 0 \\ \whline
  l_{ML}^T & \lambda_{MM}& 0 \\ \hline
  L_{BL} & l_{BM} & L_{BR}
  \end{array}
  \right)
  \rightarrow
  \cdots
   $
   }

\renewcommand{\repartitionsizes}{
  $ \chi_{ij} $ and $ \lambda_{ij}$ are scalars ...
  }


\renewcommand{\moveboundaries}{
\setlength{\arraycolsep}{4pt}
\footnotesize
$  \left( \begin{array}{c I c | c}
  X_{TL} & \star & \star \\ \whline
  x_{ML}^T & \chi_{MM} & \star \\ \hline
  X_{BL} & x_{BM} & X_{BR}
  \end{array}
  \right)
  \leftarrow
  \left( \begin{array}{c | c | c I c | c}
  X_{00} & \star & \star & \star & \star \\ \hline
   x_{10}^T & \chi_{11} & \star & \star & \star \\ \hline
   X_{20} & x_{21} & X_{22} & \star & \star \\ \whline
   x_{30}^T & \chi_{31} & x_{32}^T & \chi_{33} & \star \\ \hline
   X_{40} & x_{41} & X_{42} & x_{43} & X_{44}
   \end{array} \right)
   $,
   $
   \left( \begin{array}{c I c | c}
  L_{TL} & 0 & 0 \\ \whline
  l_{ML}^T & \lambda_{MM} & 0 \\ \hline
  L_{BL} & l_{BM} & L_{BR}
  \end{array}
  \right)
  \leftarrow
  \cdots
   $}


\renewcommand{\update}{
\setlength{\arraycolsep}{2pt}
\\
$
  \begin{array}{| l |} 
   \hline
      \mbox{\underline{\bf Fused right-looking (Variant~2a):}} \\
  \begin{array}[t]{@{}l@{}}
    \begin{array}{@{}l@{}}
\left[ \left( \begin{array}{c | c}
    0 & \star \\ \hline
    x_{21} & X_{22} \\ \hline
    \chi_{31} & x_{32}^T \\ \hline
    x_{41} & X_{42}
    \end{array} \right),
    \left( \begin{array}{c | c | c }
1 & 0 & 0 \\ \hline
l_{21} & L_{22} & 0   \\ \hline
\lambda_{31} & l_{32}^T &
1 \\ \hline
l_{41} & L_{42} &
l_{43} 
\end{array}
\right)
    \right] :=
    \mbox{\sc LTLt\_unb}( 
    \left( \begin{array}{c | c}
    0 & \star \\ \hline
    x_{21} & X_{22} \\ \hline
    \chi_{31} & x_{32}^T \\ \hline
    x_{41} & X_{42}
    \end{array} \right),
     \left( \begin{array}{c | c | c }
1 & 0 & 0 \\ \hline
l_{21} & L_{22} & 0   \\ \hline
\lambda_{31} & l_{32}^T &
1 \\ \hline
l_{41} & L_{42} &
l_{43} 
\end{array}
\right)
    )   
    \end{array}
    \\  
    \left( \begin{array}{c | c}
    0 & \star \\ \hline
    x_{43}& X_{44}
    \end{array}
    \right)
    :=
\left( \begin{array}{c | c}
    0 & \star \\ \hline
    x_{43}& X_{44}
    \end{array}
    \right)
    - 
 \left( \begin{array}{ c | c | c }
\lambda_{31} & l_{32}^T &
1 \\ \hline
l_{41} & L_{42} &
l_{43} 
\end{array}
\right)
\left( \begin{array}{c | c | c }

0 & \star & \star\\ \hline
 x_{21} &  X_{22} & \star  \\ \hline
\chi_{31}  & x_{32}^T &
0 
\end{array}
\right)
\left( \begin{array}{c | c}
\lambda_{31} & l_{41}^T \\ \hline
l_{32}  &  L_{42}^T \\ \hline
1 & l_{43}^T 
\end{array}
\right)
    \end{array}
\\ \hline
    \mbox{\underline{\bf Fused right-looking (Variant~2b):}}
\\
  \begin{array}[t]{@{}l}
    \begin{array}{@{}l}
    \left[ \left( \begin{array}{c | c}
 X_{22} & \star\\ \hline
 x_{32}^T & \chi_{33} \\ \hline
 X_{42} & x_{43}
    \end{array} \right),
    \left( \begin{array}{c | c | c}
L_{22} & 0  & 0 \\ \hline
l_{32}^T &
1 & 0 \\ \hline
L_{42} &
l_{43} & L_{44} e_f
\end{array}
\right)
    \right]   
 :=
    \mbox{\sc LTLt\_unb}( 
    \left( \begin{array}{c | c}
 X_{22} & \star\\ \hline
 x_{32}^T & \chi_{33} \\ \hline
 X_{42} & x_{43}
    \end{array} \right),
    \left( \begin{array}{c | c | c}
L_{22} & 0  & 0 \\ \hline
l_{32}^T &
1 & 0 \\ \hline
L_{42} &
l_{43} & L_{44} e_f
\end{array}
\right)
    )   
    \end{array}
    \\  
 X_{44} := 
X_{44} - 
\left( \begin{array}{c | c | c}
L_{42} &
l_{43} &
L_{44} e_f
\end{array}
\right)
\left( \begin{array}{c | c | c}
X_{22} & \star &  \star \\ \hline
x_{32}^T &
0 & \star  \\ \hline
0 &
e_f^T x_{43}  &
0
\end{array}
\right)
\left( \begin{array}{c}
L_{42}^T \\ \hline
l_{43}^T \\ \hline
(L_{44} e_f)^T
\end{array}
\right)
    \end{array}
    \\ \hline
\end{array}
$
\\
}

\centering
    \small
    \FlaAlgorithm
    
\caption{Fused blocked right-looking algorithms corresponding to Invariants~2a and~2b. For Variant~2a.}
    \label{fig:LTLt_blk_right2}
\end{figure}

\subsection{Fused blocked right-looking algorithm: Variant 2b}

\label{sec:2b}

Let us derive an alternative blocked right-looking algorithm corresponding to Invariant~2b given by~(\ref{eqn:inv_unb_right_2b-1})-(\ref{eqn:inv_unb_right_2b-2})
\NoShow{
\begin{eqnarray*}
\lefteqn{
\left( \begin{array}{c I c | c}
X_{TL} & \star & \star \\ \whline
x_{ML}^T & \chi_{MM} & \star \\ \hline
X_{BL} & x_{BM} & X_{BR}
\end{array} \right)
=
\left( \begin{array}{c I c | c}
T_{TL} & \star & \star \\ \whline
 \tau_{ML} e_l^T & 0 &  \star \\ \hline
0  &  \tau_{BM} 
L_{BR} e_f
& L_{BR} T_{BR} L_{BR}^T
\end{array} \right)  \wedge
\color{gray}
\left( \begin{array}{c I c | c}
\widehat X_{TL} & \star & \star \\ \whline
\widehat x_{ML}^T & 0 & \star \\ \hline
\widehat X_{BL} & \widehat x_{BM} & \widehat X_{BR}
\end{array} \right)} \\
\color{gray}
&& =
\left( \begin{array}{c I c | c}
\color{blue} L_{TL} & 0 & 0 \\ \whline
\color{blue} l_{ML}^T & 1 & 0 \\ \hline
\color{blue} L_{BL} & \color{blue} l_{BM} &  I
\end{array}\right)
\left( \begin{array}{c I c | c}
T_{TL} & - \tau_{ML} e_l  & 0 \\ \whline
\tau_{ML} e_l^T  & 0 & -\tau_{BM} (L_{BR} e_f)^T \\ \hline
0 & \tau_{BM} L_{BR} e_f & L_{BR} T_{BR} L_{BR}^T
\end{array} \right)
\left( \begin{array}{c I c | c}
\color{blue} L_{TL}^T & \color{blue} l_{ML} & \color{blue} L_{BL}^T \\ \whline
0 & 1 & \color{blue} l_{BM}^T \\ \hline
0 & 0 &  I
\end{array}\right).
\end{eqnarray*}
}  
This  invariant differs from Invariant~1 in that it also computes one more Gauss transform (but does not yet apply it).  The repartitioning now exposes a new block of columns and rows in each iteration.
After repartitioning, we get for Step~6 that
{
\setlength{\unitlength}{2pt}
\begin{eqnarray*}
\label{Before-blk-right-2a}
\lefteqn{
\left( \begin{array}{c I c | c | c | c }
X_{00} & \star & \cellcolor{lightgray!35}  \star & \cellcolor{lightgray!35}  \star & \star \\ \whline
x_{10}^T &  \chi_{11} & 
 \cellcolor{lightgray!35} \star & \cellcolor{lightgray!35} \star&  \star \\ \hline
\cellcolor{lightgray!35} X_{20} & 
\cellcolor{lightgray!35} x _{21} & \cellcolor{lightgray!35} X_{22} &  \cellcolor{lightgray!35} \star & \cellcolor{lightgray!35} \star \\ \hline
\cellcolor{lightgray!35} x_{30}^T & \cellcolor{lightgray!35} \chi_{31}
& \cellcolor{lightgray!35} x_{32}^T & \cellcolor{lightgray!35} \chi_{33}
& \cellcolor{lightgray!35} \star \\ \hline
X_{40} &  x_{41}
& \cellcolor{lightgray!35} \cellcolor{lightgray!35} X_{42} & \cellcolor{lightgray!35} x_{43} & X_{44}
\end{array} \right) }\\
& 
= &
\setlength{\arraycolsep}{2pt}
\left( \begin{array}{c I c | c | c | c }
T_{00} & \star & 
\mbox{\hspace{0.5in}}
\star
\mbox{\hspace{0.5in}} & \mbox{\hspace{0.5in}}\star \mbox{\hspace{0.5in}}
 & \mbox{\hspace{0.5in}} \star \mbox{\hspace{0.5in}} \\ \whline
\tau_{10} e_l^T & 0 & 
\star   & \star  & \star \\ \hline
0 & 
\tau_{21}
 & 
\multicolumn{3}{c}
{
\multirow{2}{*}{
$
\left( \begin{array}{c | c | c}
L_{22} & 0 &  0 \\ \hline
l_{32}^T &
1 & 0 \\ \hline
L_{42} &
l_{43} &
L_{44}
\end{array}
\right)
\left( \begin{array}{c | c | c}
T_{22} & -\tau_{32} e_l &  0 \\ \hline
\tau_{32} e_l^T &
0 & - \tau_{43} e_f^T \\ \hline
0 &
\tau_{43} e_f &
T_{44}
\end{array}
\right)
\left( \begin{array}{c | c | c}
L_{22}^T & l_{32}  &  L_{42}^T \\ \hline
0 &
1 & l_{43}^T \\ \hline
0 & 0 &
L_{44}^T
\end{array}
\right)
$ } }
\\
\cline{1-2} 
0 & 
0
&  \multicolumn{3}{c}{}
\\
\cline{1-2} 
0 & 
0 &  \multicolumn{3}{c}{}
\end{array} \right),
\end{eqnarray*}%
}%
where the gray highlighting captures the block of rows and columns being exposed that are updated with their final values in this iteration.
At the bottom of the loop we find
for Step~7 that 
{
\begin{equation}
\begin{array}{l}
\left( \begin{array}{c | c | c I c | c }
X_{00}^{\rm +} &  \star & \cellcolor{lightgray!35} \star & \cellcolor{lightgray!35} \star & \star \\ \hline
 x_{10}^{\rm +\!T} & \chi_{11}^{\rm +} & 
\cellcolor{lightgray!35} \star & \cellcolor{lightgray!35} \star& \star \\ \hline
\rowcolor{lightgray!35} X_{20}^{\rm +} & 
\cellcolor{lightgray!35} x _{21}^{\rm +} & \cellcolor{lightgray!35} X_{22}^{\rm +} &  \star & \star \\ \whline
\rowcolor{lightgray!35} x_{30}^{\rm +\!T} & \cellcolor{lightgray!35} \chi_{31}^{\rm +}
& \cellcolor{lightgray!35} x_{32}^{\rm +\!T} & \chi_{33}^{\rm +}
& \star \\ \hline
X_{40}^{\rm +} &  x_{41}^{\rm +} 
& \cellcolor{lightgray!35} X_{42}^{\rm +} & \cellcolor{lightgray!35} x_{43}^{\rm +} & X_{44}^{\rm +}
\end{array} \right)
=
\setlength{\arraycolsep}{2pt}
\left( \begin{array}{c | c | c I c | c}
T_{00} & \star & \star & \star 
& \star \\ \hline
\tau_{10} e_l^T & 0 & \star & \star & \star\\ \hline
0 &
\tau_{21} e_f
&
T_{22}
&
\star & \star \\ \whline
0 & 
0 &
\tau_{32} e_l^T
& 
0 & \star
\\
\hline
0 & 
0 & 
0 & 
\tau_{43} e_f &
L_{44} T_{44} L_{44}^T
\end{array} \right)
\NoShow{
\wedge \\
~~
\left( \begin{array}{c I c | c | c}
\widehat X_{00} & \star & \star & \star \\ \whline
\widehat x_{10}^T & 0 & 
\star & \star \\ \hline
\widehat x_{20}^T & 
\widehat \chi _{21} &  0 &  \star \\ \hline
\widehat X_{30} & \widehat x_{31}
& \widehat x_{32} & \widehat X_{33}
\end{array} \right)
=
\left( \begin{array}{c I c | c | c}
L_{00} & 0 & 0 & 0 \\ \whline
l_{10}^T & 1 & 
0 & 0 \\ \hline
l_{20}^T & 
\lambda _{21} & 
1
&  0 \\ \hline
L_{30} & l_{31}
& l_{32} & L_{33}
\end{array} \right)
\left( \begin{array}{c I c | c | c}
T_{00} & 
\tau_{10} e_l & 0 &  \\ \whline
\tau_{10} e_l^T & 0 & 
\tau_{21}  & 0 \\ \hline
0 & 
\tau_{21} & 
0
&  \tau_{32} e_f^T \\ \hline
0 & 
0 &
\tau_{32} e_f & T_{33}
\end{array} \right)
\left( \begin{array}{c I c | c | c}
L_{00}^T & l_{10} & l_{20} & L_{30}^T \\ \whline
0 & 1 & 
\lambda_{21} & l_{31}^T \\ \hline
0
& 
0 & 
1
&  l_{32}^T \\ \hline
0 & 
& 0 & L_{33}^T
\end{array} \right).
}
.
\end{array}
\end{equation}
}
We observe that
\begin{eqnarray*}
 \left( \begin{array}{ c | c | c }
X_{22} &  \star & \star \\ \hline
 x_{32}^T & \chi_{33}
& \star \\ \hline
 X_{42} & x_{43} & X_{44}
\end{array} \right) 
=
\left( \begin{array}{c | c | c}
L_{22} & 0 &  0 \\ \hline
l_{32}^T &
1 & 0 \\ \hline
L_{42} &
l_{43} &
L_{44}
\end{array}
\right)
\left( \begin{array}{c | c | c}
T_{22} & -\tau_{32} e_l &  0 \\ \hline
\tau_{32} e_l^T &
0 & - \tau_{43} e_f^T \\ \hline
0 &
\tau_{43} e_f &
T_{44}
\end{array}
\right)
\left( \begin{array}{c | c | c}
L_{22}^T & l_{32}  &  L_{42}^T \\ \hline
0 &
1 & l_{43}^T \\ \hline
0 & 0 &
L_{44}^T
\end{array}
\right)
,
\end{eqnarray*}
which implies that
\begin{eqnarray*}
 \left( \begin{array}{ c | c }
X_{22} &  \star  \\ \hline
 x_{32}^T & \chi_{33}
 \\ \hline
 X_{42} & x_{43} 
\end{array} \right) 
&=&
\left( \begin{array}{c | c | c}
L_{22} & 0 &  0 \\ \hline
l_{32}^T &
1 & 0 \\ \hline
L_{42} &
l_{43} &
L_{44}
\end{array}
\right)
\left( \begin{array}{c | c }
T_{22} & -\tau_{32} e_l  \\ \hline
\tau_{32} e_l^T &
0 \\ \hline
0 &
\tau_{43} e_f 
\end{array}
\right)
\left( \begin{array}{c | c }
L_{22}^T & l_{32}  \\ \hline
0 &
1  
\end{array}
\right) \\
&=&
\left( \begin{array}{c | c | c}
L_{22} & 0 &  0 \\ \hline
l_{32}^T &
1 & 0 \\ \hline
L_{42} &
l_{43} &
L_{44} e_f
\end{array}
\right)
\left( \begin{array}{c | c }
T_{22} & -\tau_{32} e_l  \\ \hline
\tau_{32} e_l^T &
0 \\ \hline
0 &
\tau_{43}  
\end{array}
\right)
\left( \begin{array}{c | c }
L_{22}^T & l_{32}  \\ \hline
0 &
1  
\end{array}
\right).
\end{eqnarray*}
This tells us that  
$
\left( \begin{array}{c | c }
X_{22}^{\rm +} & \star \\ \hline
 x_{32}^{\rm +\!T} &  \chi_{33}{\rm +}
\\ \hline
 X_{42}^{\rm +} & x_{43}^{\rm +}

\end{array} \right) 
$
and
$
\left( \begin{array}{c | c | c}
L_{22} & 0  & 0 \\ \hline
l_{32}^T &
1 & 0 \\ \hline
L_{42} &
l_{43} & L_{44} e_f
\end{array}
\right)
$
are computed from
$
 \left( \begin{array}{ c | c }
X_{22} &  \star  \\ \hline
 x_{32}^T & \chi_{33}
 \\ \hline
 X_{42} & x_{43} 
\end{array} \right) 
$
by factoring that panel, where the first column of that part of $ L $ is already available from previous computation.

\NoShow{
 \begin{figure}[p]
    
\input LTLt_blk_right2

\centering
    \small
    \FlaAlgorithm
    
\caption{Blocked rightlooking algorithms corresponding to Invariants~1 and~2. For Variant~1, the first column of $L_{42}$
is implicitly equal to zero when used to update the trailing principle submatrix of $ X $.}
    \label{fig:LTLt_blk_right2}
\end{figure}
}

Now,
\begin{eqnarray*}
X_{44}
&=&
\left( \begin{array}{c | c | c}
L_{42} &
l_{43} &
L_{44}
\end{array}
\right)
\left( \begin{array}{c | c | c}
T_{22} & -\tau_{32} e_l &  0 \\ \hline
\tau_{32} e_l^T &
0 & - \tau_{43} e_f^T \\ \hline
0 &
\tau_{43} e_f &
T_{44}
\end{array}
\right)
\left( \begin{array}{c}
L_{42}^T \\ \hline
l_{43}^T \\ \hline
L_{44}^T
\end{array}
\right) \\
& = &
\left( \begin{array}{c | c | c}
L_{42} &
l_{43} &
L_{44}
\end{array}
\right)
\left[
\left( \begin{array}{c | c | c}
T_{22} & -\tau_{32} e_l &  0 \\ \hline
\tau_{32} e_l^T &
0 & - \tau_{43} e_f^T \\ \hline
0 &
\tau_{43} e_f &
0
\end{array}
\right)
+
\left( \begin{array}{c | c | c}
0 & 0 &  0 \\ \hline
0 &
0 & 0 \\ \hline
0 &
0 &
T_{44}
\end{array}
\right)
\right]
\left( \begin{array}{c}
L_{42}^T \\ \hline
l_{43}^T \\ \hline
L_{44}^T
\end{array}
\right) \\
& = &
\left( \begin{array}{c | c | c}
L_{42} &
l_{43} &
L_{44} e_f
\end{array}
\right)
\left( \begin{array}{c | c | c}
T_{22} & -\tau_{32} e_l &  0 \\ \hline
\tau_{32} e_l^T &
0 & - \tau_{43}  \\ \hline
0 &
\tau_{43}  &
0
\end{array}
\right)
\left( \begin{array}{c | c | c}
L_{42} &
l_{43} &
L_{44} e_f
\end{array}
\right)^T
 + 
\begin{array}[t]{c}
\underbrace{L_{44} T_{44} L_{44}^T} \\
X_{44}^{\rm +}
\end{array}
\end{eqnarray*}
suggests the update
\[
X_{44} := 
X_{44} -
\left( \begin{array}{c | c | c}
L_{42} &
l_{43} &
L_{44} e_f
\end{array}
\right)
\left( \begin{array}{c | c | c}
T_{22} & -\tau_{32} e_l &  0 \\ \hline
\tau_{32} e_l^T &
0 & - \tau_{43}  \\ \hline
0 &
\tau_{43}  &
0
\end{array}
\right)
\left( \begin{array}{c | c | c}
L_{42} &
l_{43} &
L_{44} e_f
\end{array}
\right)^T.
\]
The resulting algorithm can be found in Figure~\ref{fig:LTLt_blk_right2}.
Again, the separate skew-symmetric rank-2 update of Variant~1 does not appear in Variant~2b, having been fused  with the computation of what in Variant~1 was the next iteration.  This reduces the number of times $ X_{44} $ needs to be brought in from memory.  

Note that before the start of the loop $X_{TL}$ is $0 \times 0$, but Invariant~2b prescribes a condition which is not equivalent to the precondition
    {
\setlength{\arraycolsep}{2pt}
  \begin{eqnarray*}
  \lefteqn{
\left( \begin{array}{c | c}
\chi_{MM} & \star \\ \hline
x_{BM} & X_{BR}
\end{array} \right)
=
\left( \begin{array}{c | c}
0 &  \star \\ \hline
\tau_{BM} e_f
& L_{BR} T_{BR} L_{BR}^T
\end{array} \right)  \wedge } \\
&& \left( \begin{array}{c | c}
0 & - \widehat x_{BM}^T \\ \hline
\widehat x_{BM} & \widehat X_{BR}
\end{array} \right)
 =
\left( \begin{array}{c | c}
1 & 0 \\ \hline
 l_{BM} &  I
\end{array}\right)
\left( \begin{array}{c | c}
0 & -\tau_{BM} ({ L_{BR} e_f})^T \\ \hline
\tau_{BM}
 L_{BR} e_f & L_{BR} T_{BR} L_{BR}^T
\end{array} \right)
\left( \begin{array}{c | c}
1 &  l_{BM}^T \\ \hline
0 &  I
\end{array}\right).
\end{eqnarray*}
    }
Assuming $l_{BM}=0$, this implies
    {
\setlength{\arraycolsep}{2pt}
  \begin{eqnarray*}
&& \left( \begin{array}{c | c}
\chi_{MM} & \star \\ \hline
x_{BM} & X_{BR}
\end{array} \right)
=
\left( \begin{array}{c | c}
0 &  \star \\ \hline
\tau_{BM} e_f
& \widehat X_{BR}
\end{array} \right) \\
&& L_{BR} e_f = \widehat x_{BM} / \tau_{BM}
\end{eqnarray*}
    }
and hence the initialization steps
  \begin{eqnarray*}
  L_{BR} e_f &:=& x_{BM} / \chi \\
  x_{BM} &:=& \chi e_f
\end{eqnarray*}
where $\chi$ is the first element of $x_{BM}$.

In the algorithm for Invariant~2a, a similar issue arises in that at the end of the loop $X_{BR}$ is $0\times 0$ but the final update to $\chi_{MM}$ has not yet been applied. However, in the skew-symmetric case $\chi_{MM}=0$ and no actual update is needed.

\subsection{Wimmer's blocked algorithm}

\label{sec:blk-right-wimmer}

There are a number of ways to arrive at Wimmer's blocked  algorithm, which employs a skew-symmetric rank-2k update rather than the ``sandwiched'' rank-k updated encountered so far.  Here we discuss a few.

\subsubsection{Accumulating rank-2k updates}

\NoShow{
Much as we find the formal derivation of algorithms useful, doing so to derive a blocked version of the  2-step algorithm as proposed by Wimmer~\cite{Wimmer2012}
requires the introduction of much notation which may obscure more than it exposes.   For this reason, we here embrace a more informal description.
}

\NoShow{
The loop invariant for the algorithm is
\[
\begin{array}{l}
\left( \begin{array}{c I c | c}
X_{TL} & \star & \star \\ \whline
x_{ML}^T & \chi_{MM} & \star \\ \hline
X_{BL} & x_{BM} & X_{BR}
\end{array} \right)
=
\left( \begin{array}{c I c | c}
T_{TL} & \star & \star \\ \whline
 \tau_{ML} e_l^T & 0 &  \star \\ \hline
0  &  \tau_{BM} 
L_{BR} e_f
& L_{BR} T_{BR} L_{BR}^T
\end{array} \right)  \wedge
\color{gray}
\left( \begin{array}{c I c | c}
\widehat X_{TL} & \star & \star \\ \whline
\widehat x_{ML}^T & 0 & \star \\ \hline
\widehat X_{BL} & \widehat x_{BM} & \widehat X_{BR}
\end{array} \right) \\
\color{gray}
~~ =
\left( \begin{array}{c I c | c}
\color{blue} L_{TL} & 0 & 0 \\ \whline
\color{blue} l_{ML}^T & 1 & 0 \\ \hline
\color{blue} L_{BL} & \color{blue} l_{BM} &  I
\end{array}\right)
\left( \begin{array}{c I c | c}
T_{TL} & - \tau_{ML} e_l  & 0 \\ \whline
\tau_{ML} e_l^T  & 0 & -\tau_{BM} (L_{BR} e_f)^T \\ \hline
0 & \tau_{BM} L_{BR} e_f & L_{BR} T_{BR} L_{BR}^T
\end{array} \right)
\left( \begin{array}{c I c | c}
\color{blue} L_{TL}^T & \color{blue} l_{ML} & \color{blue} L_{BL}^T \\ \whline
0 & 1 & \color{blue} l_{BM}^T \\ \hline
0 & 0 &  I
\end{array}\right),
\end{array}
\]
where either all or ever other column of the highlighted parts of $ L $ have been computed.
}
Assuming the unblocked Wimmer's algorithm is employed during the panel factorization (by executing that algorithm, but not updating beyond the current panel, thus performing a partial factorization), 
we observe 
that the computation during that stage performs a sequence of skew-symmetric rank-2 updates
\begin{equation}
\NoShow{X_{44} + l_{43} 
\begin{array}[t]{c}
\underbrace{( x_{43} + \tau_{32} l_{42} )^T} \\
w_{42}^T
\end{array}
- \begin{array}[t]{c}
\underbrace{( x_{43} + \tau_{32} l_{42} )} \\
w_{42}
\end{array}
l_{43}^T}
X_{44} + \left( \begin{array}{c} \lambda_{32} \\ \hline l_{42} \end{array} \right)
\left( \begin{array}{c | c} \chi_{32} & x_{42}^T \end{array} \right)
- 
\left( \begin{array}{c} \chi_{32} \\ \hline x_{42} \end{array} \right)
\left( \begin{array}{c | c} \lambda_{32} & l_{42}^T \end{array} \right).
\end{equation}
It is the accumulated action of those rank-2 updates that must be applied to part of the matrix (the trailing  matrix) that is not  updated during that panel factorization.
\NoShow{
In a blocked algorithms, during the factorization of the current panel, these only update within that panel, delaying the part of the updates that apply to the trailing matrix outside the current panel.
}

Here the explanation gets a little complicated.  During the factorization of the panel, the indexing of various parts of $ X $ and $ L $ refer to how the partitioning happens in the unblocked algorithm.  As we turn to what is left to be updated, the same indexing refers to submatrices relative to the partitioning for the blocked algorithm.

Once the current panel has been factored, it remains to apply the updates from panel factorization to 
\[
\left(
\begin{array}{c | c }
\chi_{33} & \star \\ \hline
x_{43} & X_{44}
\end{array}
\right)
\]
where the subscripts now refer to the partitions in the blocked algorithm.
The key is to recognize that, if $ W $ and $ Y $ each have  $ 2k $ columns, then
\begin{eqnarray*}
\setlength{\arraycolsep}{2pt}
W Y^T - Y W^T &=&
\left( \begin{array}{c | c | c | c | c}
w_0 & 0 & \cdots & w_{k-1} & 0
\end{array}
\right)
\left(
\begin{array}{c}
y_0^T \\ \hline
0 \\ \hline
\vdots \\ \hline
y_{k-1}^T \\ \hline
0
\end{array}
\right)
-
\left( \begin{array}{c | c | c | c | c}
y_0 & 0 & \cdots & y_{k-1} & 0
\end{array}
\right)
\left(
\begin{array}{c}
w_0^T \\ \hline
0 \\ \hline
\vdots \\ \hline
w_{k-1}^T \\ \hline
0
\end{array}
\right) \\
&=&
( w_0 y_0^T - y_0 w_0^T ) + \cdots +
( w_{k-1} y_{k-1}^T - y_{k-1} w_{k-1}^T ).
\end{eqnarray*}
This tells us that during the panel factorization we need to store the appropriate parts of each column of $X$ as it is factored, as the columns of a matrix $ W $ (padded with zeroes as necessary), so that upon completion of the panel factorization the remainder of the matrix can be updated with
\[
\left(
\begin{array}{c | c }
\chi_{33} & \star \\ \hline
x_{43} & X_{33}
\end{array}
\right)
:=
\left(
\begin{array}{c | c }
\chi_{33} & \star \\ \hline
x_{43} & X_{33}
\end{array}
\right)
+
W
\left(
\begin{array}{c | c }
1 & 0 \\ \hline
L_{42} & 
l_{43}
\end{array}
\right)^T  - \left(
\begin{array}{c | c }
1 & 0 \\ \hline
L_{42} & 
l_{43}
\end{array}
\right) W^T,
\]
which becomes a skew-symmetric rank-2k update if only the non-zero columns of $W$ and corresponding parts of $L$ are used. In practice, ``compressed'' matrices $W$ and $Y$ (= selected columns of $L$) are used with only the $k$ non-zero columns stored. However, using the expanded form makes the connection to our blocked algorithms clearer.

Likely, Wimmer took a similar approach~\cite{Wimmer2012}.

\subsubsection{Form $ W$ after the panel factorization}

\label{sec:Wimmer_blk_2}

Let us examine the relationship between the $ W $ discussed above and the update~(\ref{eqn:blk-right-2})--(\ref{eqn:blk-right-3}) in the blocked right-looking algorithm.
The columns of $ W $ can  be derived from $ L $ and $ X $ after the completion of the panel factorization, exploiting Theorem~\ref{thm:TandS} to modify the update in the blocked right-looking algorithm from Section~\ref{sec:blk-right}. 

The update~(\ref{eqn:blk-right-2}) to the trailing matrix in the blocked right-looking algorithm has the form
$ X := X - L T L^T $, which (as observed in Theorem~\ref{thm:TandS}), can be rewritten as $ X := X - ( W L^T - L W^T ) $, where $  W = L S $ as in~(\ref{eqn:TvsS}).
This gets us close to the update in Wimmer's blocked algorithm, but 
it still leaves the separate skew-symmetric rank-2 update in~(\ref{eqn:blk-right-3}).

Let's refine this idea:
Equation~(\ref{eqn:blk-right-2}) gives us
\begin{equation}
    \label{eqn:w1}
\left( \begin{array}{c | c}
    \chi_{33} & \star \\ \hline
    x_{43}& X_{44}
    \end{array}
    \right)
    :=
    \left( \begin{array}{c | c}
    \chi_{33} & \star \\ \hline
    x_{43}& X_{44}
    \end{array}
    \right)
    -
 \left( \begin{array}{ c | c }
l_{32}^T &
1 \\ \hline
L_{42} &
l_{43} 
\end{array}
\right)
\begin{array}[t]{c}
\underbrace{
\left( \begin{array}{c | c }
 T_{22} & -\tau_{32} e_l  \\ \hline
 \tau_{32} e_l^T  &
0 
\end{array}
\right)
} \\
S - S^T
\end{array}
\left( \begin{array}{c | c}
l_{32}  &  L_{42}^T \\ \hline
1 & l_{43}^T 
\end{array}
\right).
\end{equation}
If we let
\[
\left( \begin{array}{c | c}
    w_{32}^T & \omega_{33} \\ \hline
    W_{42} & w_{43}
    \end{array}
    \right)
    :=
\left( \begin{array}{ c | c }
l_{32}^T &
1 \\ \hline
L_{42} &
l_{43} 
\end{array}
\right)
\left( \begin{array}{c | c }
 S_{22} & s_{23}  \\ \hline
 s_{32}^T  &
0 
\end{array}
\right)
\]
then~(\ref{eqn:w1}) becomes
\[
\left( \begin{array}{c | c}
    \chi_{33} & \star \\ \hline
    x_{43}& X_{44}
    \end{array}
    \right)
    :=
    \left( \begin{array}{c | c}
    \chi_{33} & \star \\ \hline
    x_{43}& X_{44}
    \end{array}
    \right)
    -
        \left[
    \left( \begin{array}{c | c}
        w_{32}^T & \omega_{33} \\ \hline
    W_{42} & w_{43}
    \end{array}
    \right)
 \left( \begin{array}{c | c}
l_{32}  &  L_{42}^T \\ \hline
1 & l_{43}^T 
\end{array}
\right)
-
\left( \begin{array}{ c | c }
l_{32}^T &
1 \\ \hline
L_{42} &
l_{43} 
\end{array}
\right)
\left( \begin{array}{c | c}
    w_{32} & W_{42}^T \\ \hline
   \omega_{33} & w_{43}^T
    \end{array}
    \right)
\right]
\]
from which we conclude that 
\[
x_{43} :=
x_{43} -
        \left[
    \left( \begin{array}{c | c}
    W_{42} & w_{43}
    \end{array}
    \right)
 \left( \begin{array}{c  }
l_{32}   \\ \hline
1 
\end{array}
\right)
-
\left( \begin{array}{ c | c }
L_{42} &
l_{43} 
\end{array}
\right)
\left( \begin{array}{c }
    w_{32}  \\ \hline
   \omega_{33}
    \end{array}
    \right)
\right].
\]
Next, the update of $ X_{44} $ in (\ref{eqn:w1}) and (\ref{eqn:blk-right-3}) can be combined 
as
\begin{eqnarray*}
X_{44}
    &:=&
 X_{44}
    -
        \left[
    \left( \begin{array}{c | c}
    W_{42} & w_{43}
    \end{array}
    \right)
 \left( \begin{array}{ c}
  L_{42}^T \\ \hline
 l_{43}^T 
\end{array}
\right)
-
\left( \begin{array}{ c | c }
L_{42} &
l_{43} 
\end{array}
\right)
\left( \begin{array}{ c}
W_{42}^T \\ \hline
w_{43}^T
    \end{array}
    \right)
\right]
+ 
( l_{43} x_{43}^T
- x_{43} l_{43}^T )
\\
& = &
 X_{44}
    -
        \left[
    \left( \begin{array}{c | c}
    W_{42} & w_{43}+x_{43}
    \end{array}
    \right)
 \left( \begin{array}{ c}
  L_{42}^T \\ \hline
 l_{43}^T 
\end{array}
\right)
-
\left( \begin{array}{ c | c }
L_{42} &
l_{43} 
\end{array}
\right)
\left( \begin{array}{ c}
W_{42}^T \\ \hline
( w_{43} + x_{43})^T
    \end{array}
    \right)
\right]
\end{eqnarray*}
or, equivalently, via the steps
\begin{eqnarray*}
w_{43} & := & w_{43} + x_{43}
\\
X_{44} & :=  &
 X_{44}
    -
        \left[
    \left( \begin{array}{c | c}
    W_{42} & w_{43}
    \end{array}
    \right)
 \left( \begin{array}{ c}
  L_{42}^T \\ \hline
 l_{43}^T 
\end{array}
\right)
-
\left( \begin{array}{ c | c }
L_{42} &
l_{43} 
\end{array}
\right)
\left( \begin{array}{ c}
W_{42}^T \\ \hline
w_{43}^T
    \end{array}
    \right)
\right].
\end{eqnarray*}
Recalling that every other column of 
$ \left( \begin{array}{c|c}
W_{42} & w_{43} 
\end{array} \right) $ equals zero, this demonstrates that Wimmer's blocked algorithm is the right-looking blocked algorithm (except for the definition of $W$) and that the difference is in the details of the implementation of the update.

Incorporating the previously separate skew-symmetric rank-2 update not only saves computation, but it also means that $ X_{44} $ is not brought into memory an extra time for the skew-symmetric rank-2k update.

We will see that forming $ W $ after the panel factorization has completed gives us the flexibility of using different panel factorizations when it comes to incorporating pivoting.

\subsubsection{Deriving  Wimmer's blocked algorithm}

\label{sec:Wimmer_blk_var3}

Finally, Wimmer's blocked algorithm can be derived by starting the derivation process with  the postcondition of the algorithm as $ X = T \wedge \widehat X = W L^T - L W^T \wedge W = L S \wedge T = S - S^T $, with implicit assumptions about the structure of the various operands and an understanding that in the end only $ X $ and $ L $ are returned as results.
It may appear that all of $ W $ then needs to be computed, but a simple analysis shows that only the part of $ W $ needed in the current iteration needs to be kept.
This approach may end up yielding more variants.

\section{Adding pivoting}
\label{sec:pivoting}

We now briefly discuss  how symmetric pivoting can be added to the derivations and algorithms.

\subsection{Preparation}

{\em This section gives relevant results from~\cite{vandegeijn2023formal}.}

\begin{definition}
Given vector $ x $, 
$
\mbox{\sc iamax}(x)
$
returns  the index of the element in $ x $ with largest magnitude.
(In our discussion, indexing starts at zero).
\end{definition}

\begin{definition}
\label{def:perm_vector}
Given nonnegative integer $ \pi $, the $m\times m$ matrix $ P( \pi ) $ is the permutation matrix of appropriate size (defining $m$) that, when applied to a vector, swaps the top element, $ \chi_0 $, with the element indexed by $ \pi $, $ \chi_{\pi} $:
\[
P( \pi ) = 
\left\{
\begin{array}{cl}
I & \mbox{if $ \pi = 0 $} \\
\left( \begin{array}{c |c | c | c}
0 & 0 & 1 & 0 \\ \hline
0 & I_{\pi-1} & 0 & 0 \\ \hline
1 & 0 & 0 & 0 \\ \hline
0 & 0 & 0 & I_{m-\pi-1}
\end{array}
\right) & \mbox{otherwise,}
\end{array}
\right.
\]
where $ I_k $ is a $ k \times k $ identity matrix and $ 0 $ equals a submatrix (or vector) of all zeroes of appropriate size.
\end{definition}
Applying $ P( \pi ) $ to  $ m \times n $ matrix $ A $ swaps the top row with the row indexed with $ \pi $. Some key results regarding permutations and their action on a matrix play an important role when pivoting is added.
First some more definitions.
\begin{definition}
We call a vector $ p = \left( \begin{array}{c}
\pi_0 \\
\vdots \\
\pi_{n-1} 
\end{array}
\right) $ a permutation vector if  each $ \pi_i \in \{ 0, \ldots, m-i-1 \} $.
Here $ n $ equals the number of permutations and $ m \geq n $ is the row size of the matrix to which the permutations are applied.
\end{definition}
Associated with a permutation vector is the permutation matrix $ P( p ) $ that applies the permutations encoded in the vector $ p $:
\begin{definition}
\label{def:perm_matrix}
Given permutation vector $ p $ of size $ n $, 
\[
P( p ) = 
\left( \begin{array}{c | c}
I_{m-1} & 0 \\ \hline
0 & P( \pi_{n-1} )
\end{array}
\right)
\cdots 
\left( \begin{array}{c | c}
1 & 0 \\ \hline
0 & P( \pi_1 )
\end{array}
\right)
P( \pi_0 )
\]
is an $m\times m$ permutation matrix, where $ I_k $ is a $ k \times k $ identity matrix.
\end{definition}

A classic result about permutation matrices is
\begin{theorem}
For any permutation matrix $ P $, its inverse  equals its inverse:
$
P^{-1} = P^T$.
\end{theorem}
An immediate consequence is
\begin{corollary}
Let $ P( \pi ) $ be as defined in \ref{def:perm_vector}.  Then
$ P( \pi )^{-1} = P( \pi )^T = P( \pi ) $.
\end{corollary}
This captures that undoing the swapping of two rows of a matrix is to swap them again.

To derive the PME, we'll need to be able to apply permutations defined with partitioned permutation vectors.  
The following theorem exposes that to apply all permutations that were encountered, one can apply the first batch (given by $ p_T $) and then the second batch (given by $ p_B $).  Undoing these permutations means first undoing the second batch and then the undoing the first batch.
\begin{theorem}
\label{thm:part_perm}
Partition permutation vector $ 
p = \left( \begin{array}{c}
p_T  \\ \hline
p_B 
\end{array}
\right)$.
Then
\[
P( p ) = 
\left( \begin{array}{c | c}
I & 0 \\ \hline
0 & P( p_B ) 
\end{array}
\right)
P( p_T ) 
\quad
\mbox{and}
\quad
P( p )^{-1} =
P( p_T )^{-1}
\left( \begin{array}{c | c}
I & 0 \\ \hline
0 & P( p_B )
\end{array}
\right)^{-1}.
\]
Here, $ I $ is the identity ``of appropriate size'' in the context in which it is used.
\end{theorem}
Its proof follows immediately from   Definition~\ref{def:perm_matrix}.

A final corollary will become instrumental as we relate the state of variables before the update (in Step~6) to the state after the update (in Step~7), in order to determine updates (in Step~8).
\begin{corollary}
\label{cor:P}
    Partition permutation vector $ 
p = \left( \begin{array}{c}
p_1  \\ \hline
p_2  
\end{array}
\right)$.
Then
\NoShow{
\[
P( \left( \begin{array}{c}
p_1  \\ \hline
p_2  
\end{array}
\right) )^{-1} =
P( \pi_1 )^{-1}
\left( \begin{array}{c | c}
1 & 0 \\ \hline
0 & P( p_2 ) 
\end{array}
\right)^{-1}
=
P( \pi_1 )
\left( \begin{array}{c | c}
1 & 0 \\ \hline
0 & P( p_2 ) 
\end{array}
\right)^{-1}
 .
\]
and
}
$
P( p_1 ) P( \left( \begin{array}{c}
p_1  \\ \hline
p_2  
\end{array}
\right) )^{-1} =
\left( \begin{array}{c | c}
I & 0 \\ \hline
0 & P( p_2 ) 
\end{array}
\right)^{-1}$.
\end{corollary}
A special case of this is when $ p_1 = \pi_1 $, a scalar:
$
P( \pi_1 ) P( \left( \begin{array}{c}
\pi_1  \\ \hline
p_2  
\end{array}
\right) )^{-1} =
\left( \begin{array}{c | c}
1 & 0 \\ \hline
0 & P( p_2 ) 
\end{array}
\right)^{-1}$.

It is well known that when computing the LU factorization with partial pivoting the net results satisfies $ P A = L U $, where $ P $ is the accumulation of the action of individual row swaps (permutations).  The resulting matrix has the property that $ \vert L \vert \leq J $, where $ \vert L \vert $ results from taking the element-wise absolute value and 
$ J $ is the matrix of all ones and appropriate size.  This guarantees that every entry in $ L $ is less than or equal to one in magnitude, which reduces the element growth that could cause numerical instability.

\subsection{Deriving the PME}

Taking insights from the LU factorization with pivoting into account, we expect pivoting to result in the computation of $ P $, $ L $, and $ T $ such that $ P X P^T = L T L^T $ or, equivalently,  
$ X = P^{-1} L T L^T P^{-T} $.  
Although $ P^{-1} = P^T $, the inverses are exposed to emphasize that $ P^{-1} $ undoes permutations that were encountered in the execution of the algorithm.

The PME now becomes
{
\setlength{\arraycolsep}{2pt}
\begin{eqnarray*}
\lefteqn{
\left( \begin{array}{c I c | c}
X_{TL} & \star & \star \\ \whline
x_{ML}^T & \chi_{MM} & \star \\ \hline
X_{BL} & x_{BM} & X_{BR}
\end{array} \right)
=
\left( \begin{array}{c I c | c}
T_{TL} & \star & \star \\ \whline
\tau_{ML} e_l^T & 0 &  \star \\ \hline
0  & \tau_{BM} e_f & T_{BR}
\end{array} \right) \wedge 
\left( \begin{array}{c I c | c}
\widehat X_{TL} & \widehat x_{ML} & \widehat X_{BL}^T \\ \whline
\widehat x_{ML}^T & 0 & \widehat x_{BM}^T \\ \hline
\widehat X_{BL} & \widehat x_{BM} & \widehat X_{BR}
\end{array} \right) }
\\
& = &
P( 
\left( \begin{array}{c}
p_T \\ \whline
\pi_M \\ \hline
p_B
\end{array}
\right)
)^{-1}
\left( \begin{array}{c I c | c}
\widetilde L_{TL} & 0 & 0 \\ \whline
\widetilde l_{ML}^T & 1 & 0 \\ \hline
\widetilde L_{BL} & \widetilde l_{BM} &  \widetilde L_{BR}
\end{array}\right)
\left( \begin{array}{c I c | c}
T_{TL} & - \tau_{ML} e_l  & 0 \\ \whline
\tau_{ML} e_l^T  & 0 & -\tau_{BM} e_f^T \\ \hline
0 & \tau_{BM} e_f & T_{BR}
\end{array} \right)
\left( \begin{array}{c I c | c}
\widetilde L_{TL}^T & \widetilde l_{ML} & L_{BL}^T \\ \whline
0 & 1 & \widetilde l_{BM}^T \\ \hline
0 & 0 &  \widetilde L_{BR}^T
\end{array}\right)
P( 
\left( \begin{array}{c}
p_T \\ \whline
\pi_M \\ \hline
p_B
\end{array}
\right)
)^{-T},
\end{eqnarray*}
}%
plus the condition that forces the elements of $ L $ to be less than one in magnitude.
Here, we use $ \widetilde L $ to denote the final $ L $ while in our later discussion $ L $ will be used for the matrix that contains the currently computed parts of $ L $.

This can be rewritten as
{
\setlength{\arraycolsep}{2pt}
\begin{eqnarray*}
\lefteqn{
\left( \begin{array}{c I c | c}
X_{TL} & \star & \star \\ \whline
x_{ML}^T & \chi_{MM} & \star \\ \hline
X_{BL} & x_{BM} & X_{BR}
\end{array} \right)
=
\left( \begin{array}{c I c | c}
T_{TL} & \star & \star \\ \whline
\tau_{ML} e_l^T & 0 &  \star \\ \hline
0  & \tau_{BM} e_f & T_{BR}
\end{array} \right) \wedge }
\\
&&
\left( \begin{array}{c I c | c}
\widehat X_{TL} & \widehat x_{ML} & \widehat X_{BL}^T \\ \whline
\widehat x_{ML}^T & 0 & \widehat x_{BM}^T \\ \hline
\widehat X_{BL} & \widehat x_{BM} & \widehat X_{BR}
\end{array} \right) =
P( 
\left( \begin{array}{c}
p_T \\ \whline
\pi_M 
\end{array}
\right)
)^{-1}
\left( \begin{array}{c I c | c}
\widetilde L_{TL} & 0 & 0 \\ \whline
\widetilde l_{ML}^T & 1 & 0 \\ \hline
P( p_B )^{-1} \widetilde L_{BL} & P( p_B )^{-1} \widetilde l_{BM} &  I
\end{array}\right) 
\\
&&\hspace{1.4in}
\left( \begin{array}{c I c | c}
T_{TL} & - \tau_{ML} e_l  & 0 \\ \whline
\tau_{ML} e_l^T  & 0 & -\tau_{BM} (P( p_B )^{-1} \widetilde L_{BR} e_f)^T \\ \hline
0 & \tau_{BM} P( p_B )^{-1} \widetilde L_{BR} e_f & P( p_B )^{-1} \widetilde L_{BR} T_{BR} \widetilde L_{BR}^T P( p_B )^{-T}
\end{array} \right) 
\\
&&\hspace{1.4in}
\left( \begin{array}{c I c | c}
\widetilde L_{TL}^T & \widetilde l_{ML} & ( P( p_B )^{-1}  \widetilde L_{BL})^T \\ \whline
0 & 1 & ( P( p_B )^{-1} \widetilde l_{BM} )^T \\ \hline
0 & 0 &  I
\end{array}\right)
P( 
\left( \begin{array}{c}
p_T \\ \whline
\pi_M 
\end{array}
\right)
)^{-T}.
\end{eqnarray*}
}
The important observation here is that $ P( p_B )^{-1} \widetilde L_{BR} $ equals the final (yet to be computed) $ \widetilde L_{BR} $ but with its rows not yet permuted with permutations yet to be computed.

\subsection{Adding pivoting to the unblocked right-looking algorithm}

\begin{figure}[tbp] 
\input LTLt_piv_unb_right
\centering
    \small 
    \FlaAlgorithm    
    \caption{Unblocked right- and left-looking algorithms with pivoting.}
    \label{fig:LTLt_piv_unb_right}
\end{figure}

The loop-invariant for the right-looking algorithm with pivoting becomes
{
\setlength{\arraycolsep}{2pt}
\begin{eqnarray*}
\nonumber
\lefteqn{
\left( \begin{array}{c I c | c}
X_{TL} & \star & \star \\ \whline
x_{ML}^T & \chi_{MM} & \star \\ \hline
X_{BL} & x_{BM} & X_{BR}
\end{array} \right)
=
\left( \begin{array}{c I c | c}
T_{TL} & \star & \star \\ \whline
\tau_{ML} e_l^T & 0 &  \star \\ \hline
0  & \tau_{BM} P( p_B )^{-1} \widetilde L_{BR} e_f & P( p_B )^{-1} \widetilde L_{BR} T_{BR} \widetilde L_{BR}^T P( p_B )^{-T}
\end{array} \right) \wedge } \\
& & 
\left( \begin{array}{c I c | c}
\widehat X_{TL} & \widehat x_{ML} & \widehat X_{BL}^T \\ \whline
\widehat x_{ML}^T & 0 & \widehat x_{BM}^T \\ \hline
\widehat X_{BL} & \widehat x_{BM} & \widehat X_{BR}
\end{array} \right) = 
P( 
\left( \begin{array}{c}
\color{blue}
p_T \\ \whline
\color{blue} \pi_M 
\end{array}
\right)
)^{-1}
\left( \begin{array}{c I c | c}
\color{blue} \widetilde L_{TL} & 0 & 0 \\ \whline
\color{blue} \widetilde l_{ML}^T & 1 & 0 \\ \hline
\color{blue} P( p_B )^{-1} \widetilde L_{BL} & \color{blue} P( p_B )^{-1} \widetilde l_{BM} &  I
\end{array}\right) \\
&&\hspace{1.4in}
\left( \begin{array}{c I c | c}
T_{TL} & - \tau_{ML} e_l  & 0 \\ \whline
\tau_{ML} e_l^T  & 0 & -\tau_{BM} (P( p_B )^{-1} \widetilde L_{BR} e_f)^T \\ \hline
0 & \tau_{BM} P( p_B )^{-1}\widetilde  L_{BR} e_f & P( p_B )^{-1} \widetilde L_{BR} T_{BR} \widetilde L_{BR}^T P( p_B )^{-T}
\end{array} \right) \\
&&\hspace{1.4in}
\left( \begin{array}{c I c | c}
\color{blue} \widetilde L_{TL}^T & \color{blue} \widetilde l_{ML} & \color{blue} ( P( p_B )^{-1}  \widetilde L_{BL})^T \\ \whline
0 & 1 & \color{blue} ( P( p_B )^{-1} \widetilde l_{BM} )^T \\ \hline
0 & 0 &  I
\end{array}\right)
P( 
\left( \begin{array}{c}
\color{blue} p_T \\ \whline
\color{blue} \pi_M 
\end{array}
\right)
)^{-T},
\end{eqnarray*}
}%
where the parts of $ L $ highlighted in blue have already been computed%
\footnote{$ p_B $ has not yet been computed but $ P( p_B )^{-1} \widetilde L_{BL} $ and
$ P( p_B )^{-1} \widetilde l_{BM} $ are available in the corresponding parts of $ L$.}.


At the top of the loop, in Step~6,
{
\setlength{\arraycolsep}{2pt}
\begin{eqnarray*}
\lefteqn{
\left( \begin{array}{c I c | c | c}
X_{00} & \star & \star & \star \\ \whline
x_{10}^T & \chi_{11} & 
\star & \star \\ \hline
x_{20}^T & 
\chi _{21} & \chi_{22} &  \star \\ \hline
X_{30} & x_{31}
& x_{32} & X_{33}
\end{array} \right)
= }\\
&&
\setlength{\arraycolsep}{2pt}
\left( 
\begin{array}{c I c | c | c }
 T_{00} & \star & ~~~~~~~~~~~~~~~~~~~\star~~~~~~~~~~~~~~~~~~~ & \star \\ \whline
\tau_{10} e_l^T  & 0 & 
\star & \star \\ \hline
0 & 
\multirow{2}{*}{$
\tau_{21}
p(  
\left( \begin{array}{c}
\pi_2 \\ \hline
p_3
\end{array}
\right) )^{-1}
\left( \begin{array}{c}
1   \\ \hline
\widetilde l_{32} 
\end{array}
\right)
$} & 
\multicolumn{2}{c}
{
\multirow{2}{*}{
$
p(  
\left( \begin{array}{c}
\pi_2 \\ \hline
p_3
\end{array}
\right) )^{-1}
\left( \begin{array}{c | c}
1 &  0 \\ \hline
\widetilde l_{32} &
\widetilde L_{33}
\end{array}
\right)
\left( \begin{array}{c | c}
0 &  \star \\ \hline
\tau_{32} e_f &
T_{33}
\end{array}
\right)
\left( \begin{array}{c | c}
1 & \widetilde  l_{32}^T \\ \hline
0 &
\widetilde L_{33}^T
\end{array}
\right)
p(  
\left( \begin{array}{c}
\pi_2 \\ \hline
p_3
\end{array}
\right) )^{-T}
$ } }
\\
\cline{1-1} 
 0 & 
&  \multicolumn{2}{c}{}
\end{array}
\right)
\NoShow{
\wedge \\
~~
\left( \begin{array}{c I c | c | c}
\widehat X_{00} & \star & \star & \star \\ \whline
\widehat x_{10}^T & 0 & 
\star & \star \\ \hline
\widehat x_{20}^T & 
\widehat \chi _{21} &  0 &  \star \\ \hline
\widehat X_{30} & \widehat x_{31}
& \widehat x_{32} & \widehat X_{33}
\end{array} \right) \\
~~~~
=
\left( \begin{array}{c I c | c | c}
\widetilde L_{00} & 0 & 0 & 0 \\ \whline
\widetilde l_{10}^T & 1 & 
0 & 0 \\ \hline
\widetilde l_{20}^T & 
\widetilde \lambda _{21} & 
1
&  0 \\ \hline
\widetilde L_{30} & \widetilde l_{31}
& \widetilde l_{32} & \widetilde L_{33}
\end{array} \right)
\left( \begin{array}{c I c | c | c}
T_{00} & 
\tau_{10} e_l & 0 &  \\ \whline
\tau_{10} e_l^T & 0 & 
\tau_{21}  & 0 \\ \hline
0 & 
\tau_{21} & 
0
&  \tau_{32} e_f^T \\ \hline
0 & 
0 &
\tau_{32} e_f & T_{33}
\end{array} \right)
\left( \begin{array}{c I c | c | c}
\widetilde L_{00}^T & \widetilde l_{10} & \widetilde l_{20} & \widetilde L_{30}^T \\ \whline
0 & 1 & 
\widetilde  \lambda{21} & \widetilde l_{31}^T \\ \hline
0
& 
0 & 
1
&  \widetilde l_{32}^T \\ \hline
0 & 
& 0 & \widetilde L_{33}^T
\end{array} \right)
}
\end{eqnarray*}
}%
and  $ L $ contains
{
\[
\left( \begin{array}{c | c I c | c}
L_{00} & \star & \star & \star \\ \hline
l_{10}^{\rm T} & \lambda_{11} & 
\star & \star \\ \whline
l_{20}^{\rm T} & 
\lambda_{21} & \lambda_{22} &  \star \\ \hline
L_{30} & l_{31}
& l_{32} & L_{33}
\end{array} \right)
=
\left( \begin{array}{c I c | c | c }
\widetilde L_{00} & 0 & 0 & 0 \\ \whline
\widetilde l_{10}^T & 1 & 0 & 0 \\ \hline
\multirow{2}{*}{
$
P( 
\left( \begin{array}{c}
\pi_2 \\ \hline
p_3
\end{array} \right))^{-1}
\left( \begin{array}{c}
\widetilde l_{20}^T \\ \hline
\widetilde L_{30}
\end{array}
\right)
$
}
& 
\multirow{2}{*}{
$
P( 
\left( \begin{array}{c}
\pi_2 \\ \hline
p_3
\end{array} \right))^{-1}
\left( \begin{array}{c}
\widetilde \lambda_{21} \\ \hline
\widetilde l_{31}
\end{array}
\right)
$
}
& 1 & 0 \\ \cline{3-4}
&  & 0 & I
\end{array} \right).
\]
}%
At the bottom of the loop, in Step~7,  $ X $ must contain 
{
\[
\begin{array}{l}
\left( \begin{array}{c | c I c | c}
X_{00}^{\rm +} & \star & \star & \star \\ \hline
x_{10}^{\rm +\!T} & \chi_{11} & 
\star & \star \\ \whline
x_{20}^{\rm +\!T} & 
\chi_{21}^{\rm +} & \chi_{22}^{\rm +} &  \star \\ \hline
X_{30}^{\rm +} & x_{31}^{\rm +}
& x_{32}^{\rm +} & X_{33}^{\rm +}
\end{array} \right)
=
\left( \begin{array}{c | c I c | c}
T_{00} & \star & \star & \star \\ \hline
 \tau_{10} e_l^T & 0 & 
~~~\star~~~ & \star \\ \whline
0 & 
\tau_{21} & 0 & \star
\\
\hline
 0 & 
 0 & 
\tau_{32}
P( p_3 )^{-1}
 L_{33} e_f
 &
P( p_3 )^{-1}
\widetilde L_{33}
T_{33} 
\widetilde L_{33}^T
P( p_3 )^{-T}
\end{array} \right)
\NoShow{
\wedge \\
~~
\left( \begin{array}{c I c | c | c}
\widehat X_{00} & \star & \star & \star \\ \whline
\widehat x_{10}^T & 0 & 
\star & \star \\ \hline
\widehat x_{20}^T & 
\widehat \chi _{21} &  0 &  \star \\ \hline
\widehat X_{30} & \widehat x_{31}
& \widehat x_{32} & \widehat X_{33}
\end{array} \right)
=
\left( \begin{array}{c I c | c | c}
\widetilde L_{00} & 0 & 0 & 0 \\ \whline
\widetilde l_{10}^T & 1 & 
0 & 0 \\ \hline
\widetilde l_{20}^T & 
\lambda _{21} & 
1
&  0 \\ \hline
\widetilde L_{30} &  l_{31}
& \widetilde l_{32} &  L_{33}
\end{array} \right)
\left( \begin{array}{c I c | c | c}
T_{00} & 
\tau_{10} e_l & 0 &  \\ \whline
\tau_{10} e_l^T & 0 & 
\tau_{21}  & 0 \\ \hline
0 & 
\tau_{21} & 
0
&  \tau_{32} e_f^T \\ \hline
0 & 
0 &
\tau_{32} e_f & T_{33}
\end{array} \right)
\left( \begin{array}{c I c | c | c}
\widetilde L_{00}^T & \widetilde l_{10} & \widetilde \widetilde l_{20} & \widetilde L_{30}^T \\ \whline
0 & 1 & 
  \lambda{21} & \widetilde\widetilde l_{31}^T \\ \hline
0
& 
0 & 
1
& \widetilde l_{32}^T \\ \hline
0 & 
& 0 & \widetilde L_{33}^T
\end{array} \right)
}
\end{array}
\]
}%
and $ L $ must contain
{
\[
\left( \begin{array}{c | c I c | c}
L_{00}^{\rm +} & \star & \star & \star \\ \hline
l_{10}^{\rm +\!T} & \lambda_{11}^{\rm +} & 
\star & \star \\ \whline
l_{20}^{\rm +\!T} & 
\lambda_{21}^{\rm +} & \lambda_{22}^{\rm +} &  \star \\ \hline
L_{30}^{\rm +} & l_{31}^{\rm +}
& l_{32}^{\rm +} & L_{33}^{\rm +}
\end{array} \right)
=
\left( \begin{array}{c | c I c | c}
\widetilde L_{00} & 0 & 0 & \star \\ \hline
\widetilde l_{10} & 1 &
0 & - \\ \whline
\widetilde l_{20}^{T} & 
\widetilde  \lambda_{21}& 1 &  0 \\ \hline
P( p_3 )^{-1} \widetilde L_{30} & P( p_3 )^{-1} \widetilde l_{31}
& P( p_3 )^{-1} \widetilde l_{32}& I
\end{array} \right).
\]
}

Now,
{
\setlength{\arraycolsep}{2pt}
\[
\tau_{21}
\left( \begin{array}{c}
1 \\ \hline
P( p_3 )^{-1} \widetilde l_{32}
\end{array} \right)
=
\tau_{21}
\left( \begin{array}{c | c}
1 & 0 \\ \hline
0 & P( p_3  )^{-1}
\end{array} \right)
\left( \begin{array}{c}
1 \\ \hline
\widetilde l_{32}
\end{array} \right)
=
\tau_{21}
P( \pi_2 )
P( 
\left( \begin{array}{c}
\pi_2 \\  \hline
p_3 
\end{array} \right)
)^{-1} 
\left( \begin{array}{c}
1 \\ \hline
\widetilde l_{32} 
\end{array} \right)
=
P( \pi_2 )
\left( \begin{array}{c}
\chi_{21} \\  \hline
x_{31} 
\end{array} \right).
\]
}%
Since $P(p_3)^{-1} \widetilde l_{32} = l_{32}$, this prescribes the updates
\begin{itemize}
    \item 
    $ \pi_2 := \mbox{\sc iamax}( 
    \left( \begin{array}{c}
\chi_{21} \\  \hline
x_{31} 
\end{array} \right) ) $:  Determine the index, relative to the first element of the input, of the element with largest absolute value. 
\item 
$
\left( \begin{array}{c}
\chi_{21} \\  \hline
x_{31} 
\end{array} \right)
:=
P( \pi_2 )
\left( \begin{array}{c}
\chi_{21} \\  \hline
x_{31} 
\end{array} \right)
$.
\item 
$ l_{32} := x_{31} / \chi_{21} $.
\end{itemize}
Also
\[
\left( \begin{array}{c | c }
\widetilde l_{20}^{T} & 
\widetilde  \lambda_{21} \\ \hline
P( p_3 )^{-1} \widetilde  L_{30} & P( p_3 )^{-1} \widetilde l_{31}
\end{array} \right)
=
P( \pi_2 )
P( \left( \begin{array}{c}
\pi_2 \\ \hline
p_3 
\end{array} \right) )^{-1}
\left( \begin{array}{c | c }
\widetilde l_{20}^{T} & 
\widetilde  \lambda_{21} \\ \hline
\widetilde L_{30} & \widetilde  l_{31}
\end{array} \right),
\]
which tells us to update 
\begin{itemize}
    \item 
    $
    \left( \begin{array}{c | c }
l_{20}^{T} & 
\lambda_{21} \\ \hline
L_{30} &  l_{31}
\end{array} \right) :=
P( \pi_2 )
\left( \begin{array}{c | c }
l_{20}^{T} & 
 \lambda_{21} \\ \hline
L_{30} &   l_{31}
\end{array} \right)$.
\end{itemize}

The final question is how to compute 
$ \left( \begin{array}{c | c}
\chi_{22}^{\rm +} & \star \\ \hline
x_{32}^{\rm +} & X_{33}^{\rm +}
\end{array} \right) $
from 
$ \left( \begin{array}{c | c}
\chi_{22} & \star \\ \hline
x_{32} & X_{33}
\end{array} \right) $.  Notice \mbox{that%
\footnote{
We show all steps to illustrate that this is a matter of judiciously applying rules about how Gauss transforms and/or permutations interact.}}%
,
\allowdisplaybreaks
\begin{eqnarray*}
\lefteqn{
\left( \begin{array}{c | c}
\chi_{22}^{\rm +} & \star \\ \hline
x_{32}^{\rm +} & X_{33}^{\rm +}
\end{array} \right)
 =
\left( \begin{array}{c | c}
0 & \star \\ \hline
\tau_{32} P( p_3 )^{-1} \widetilde L_{33} e_f & P( p_3 )^{-1}  \widetilde L_{33}
 T_{33} \widetilde L_{33}^T P( p_3 )^{-T}
 \end{array} \right)}
 \\
&= &
 \left( \begin{array}{c | c}
1 &  0 \\ \hline
0 &
P( p_3 )^{-1}
\end{array}
\right)
 \left( \begin{array}{c | c}
1 &  0 \\ \hline
0 &
\widetilde L_{33}
\end{array}
\right)
\left( \begin{array}{c | c}
0 &  - \tau_{32} e_f^T \\ \hline
\tau_{32} e_f &
T_{33}
\end{array}
\right)
\left( \begin{array}{c | c}
1 &  0 \\ \hline
0 &
\widetilde L_{33}^T
\end{array}
\right)
 \left( \begin{array}{c | c}
1 &  0 \\ \hline
0 &
P( p_3 )^{-T}
\end{array}
\right)
\\
&= &
 \left( \begin{array}{c | c}
1 &  0 \\ \hline
0 &
P( p_3 )^{-1}
\end{array}
\right)
 \left( \begin{array}{c | c}
1 &  0 \\ \hline
- \widetilde l_{32} &
I
\end{array}
\right)
\left( \begin{array}{c | c}
1 &  0 \\ \hline
\widetilde l_{32} &
\widetilde L_{33}
\end{array}
\right)
\left( \begin{array}{c | c}
0 &  - \tau_{32} e_f^T \\ \hline
\tau_{32} e_f &
T_{33}
\end{array}
\right) 
\\
& &
\hspace{1in}
\left( \begin{array}{c | c}
1 &  \widetilde l_{32}^T  \\ \hline
0 &
\widetilde L_{33}^T
\end{array}
\right)
\left( \begin{array}{c | c}
1 &  - \widetilde l_{32}^T \\ \hline
0 &
I
\end{array}
\right)
 \left( \begin{array}{c | c}
1 &  0 \\ \hline
0 &
P( p_3 )^{-T}
\end{array}
\right)
\\
& = &
 \left( \begin{array}{c | c}
1 &  0 \\ \hline
- P( p_3 )^{-1} \widetilde l_{32} &
I
\end{array}
\right)
 \left( \begin{array}{c | c}
1 &  0 \\ \hline
0 &
P( p_3 )^{-1}
\end{array}
\right)
\left( \begin{array}{c | c}
1 &  0 \\ \hline
\widetilde l_{32} &
\widetilde L_{33}
\end{array}
\right)
\left( \begin{array}{c | c}
0 &  - \tau_{32} e_f^T \\ \hline
\tau_{32} e_f &
T_{33}
\end{array}
\right) \\
& &
\hspace{1in}
\left( \begin{array}{c | c}
1 &  \widetilde l_{32}^T  \\ \hline
0 &
\widetilde L_{33}^T
\end{array}
\right)
 \left( \begin{array}{c | c}
1 &  0 \\ \hline
0 &
P( p_3 )^{-T}
\end{array}
\right)
\left( \begin{array}{c | c}
1 &  - ( P( p_3 )^{-1} \widetilde l_{32} )^T \\ \hline
0 &
I
\end{array}
\right)
\\
&=& 
 \left( \begin{array}{c | c}
1 &  0 \\ \hline
- P( p_3 )^{-1} \widetilde l_{32} &
I
\end{array}
\right)
P( \pi_2 )
P( 
 \left( \begin{array}{c}
\pi_2 \\ \hline
p_3
\end{array}
\right)
)^{-1}
\left( \begin{array}{c | c}
1 &  0 \\ \hline
\widetilde l_{32} &
\widetilde L_{33}
\end{array}
\right)
\left( \begin{array}{c | c}
0 &  - \tau_{32} e_f^T \\ \hline
\tau_{32} e_f &
T_{33}
\end{array}
\right) \\
& & \hspace{1in}
\left( \begin{array}{c | c}
1 & \widetilde l_{32}^T  \\ \hline
0 &
\widetilde L_{33}^T
\end{array}
\right)
P( 
 \left( \begin{array}{c}
\pi_2 \\ \hline
p_3
\end{array}
\right)
)^{-T}
P( \pi_2 )
\left( \begin{array}{c | c}
1 &  - ( P( p_3 )^{-1} \widetilde l_{32} )^T \\ \hline
0 &
I
\end{array}
\right)
\\
&=& 
 \left( \begin{array}{c | c}
1 &  0 \\ \hline
- P( p_3 )^{-1} \widetilde l_{32} &
I
\end{array}
\right)
P( \pi_2 )
\left( \begin{array}{c | c}
0 &  \star \\ \hline
x_{32} &
X_{33}
\end{array}
\right)
P( \pi_2 )
\left( \begin{array}{c | c}
1 &  - ( P( p_3 )^{-1} \widetilde l_{32} )^T \\ \hline
0 &
I
\end{array}
\right) \\
&=&
 \left( \begin{array}{c | c}
1 &  0 \\ \hline
- l_{32} &
I
\end{array}
\right)
P( \pi_2 )
\left( \begin{array}{c | c}
0 &  \star \\ \hline
x_{32} &
X_{33}
\end{array}
\right)
P( \pi_2 )
\left( \begin{array}{c | c}
1 &  - l_{32}^T \\ \hline
0 &
I
\end{array}
\right)
.
\end{eqnarray*}
For the last step, recognize that by now $ P( p_3 )^{-1} \widetilde l_{32} $ has been computed and stored in $  l_{32} $.
This prescribes the updates
\begin{eqnarray*}
\left( \begin{array}{c | c}
0 &  \star \\ \hline
x_{32} &
X_{33}
\end{array}
\right)
& := &
P( \pi_2 )
\left( \begin{array}{c | c}
0 &  \star \\ \hline
x_{32} &
X_{33}
\end{array}
\right)
P( \pi_2 ) \\
X_{33} & := & X_{33} + ( l_{32} x_{32} - x_{32} l_{32}^T ) .
\end{eqnarray*}
This completes the formal derivation of the algorithm in Figure~\ref{fig:LTLt_piv_unb_right}.

\subsection{Adding pivoting to the other algorithms}

While it is possible to judiciously push the FLAME methodology through to add pivoting to the other algorithms, we do not do so in this paper.  At some point, it makes sense to take the lessons that have been learned, and add the pivoting in a way that is guided by the process, but does not go through all the steps. 

For the left-looking unblocked algorithm we  show the  result in Figure~\ref{fig:LTLt_piv_unb_right}.  Adding pivoting to Wimmer's right-looking algorithm is similarly straight forward.

When adding pivoting to blocked algorithms, one needs to consider not only what part of the pivoting happens during the blocked algorithm, but also what needs to happen within the panel factorization.
In particular:
\begin{itemize}
    \item 
    Can we add pivoting to an unblocked panel factorization based on a right-looking algorithm?
    To answer this, assume we have already processed one or more columns, updating only columns within the panel. 
    If, for the next column, pivoting dictates that a column outside the current panel must be swapped into the panel, then that column will not yet have been updated consistent with the columns within the panel that were already processed.  We conclude that pivoting cannot be added to a panel factorization based on an unblocked (or blocked) right-looking algorithm.
    \item 
    Can we add pivoting to an unblocked panel factorization based on a left-looking algorithm?
    To answer this, assume we have already processed one or more columns {\em and the remainder of the matrix has been pivoted consistently}.
    Then, we can process the next column, regardless of whether or not it is pivoted from outside of the current panel, as it is updated according to prior computation \emph{after} being pivoted in.  In other words, pivoting can be added to an unblocked left-looking algorithm that only factors a current panel.
\end{itemize}
We conclude that pivoting can only be added to the unblocked left-looking algorithm that only updates a current panel.

Next, we turn to the blocked algorithms themselves.
We note that 
    pivoting cannot be added to the blocked left-looking algorithm, since, when  starting the factorization with a new block, one cannot know what parts of the trailing matrix to swap into that block. 
    Pivoting can however be added to the blocked right-looking algorithms since a next panel can be factored with a modified left-looking algorithm, after which the remainder of the matrix has already been pivoted and the appropriate prior parts of $ L $ can be consistently pivoted.

The updates for the blocked right-looking algorithms are given in 
Figure~\ref{fig:LTLt_piv_blk}.
The easiest way to arrive at an algorithm similar to the  blocked Wimmer's algorithm with pivoting is to modify the update of the trailing matrix as discussed in Section~\ref{sec:Wimmer_blk_2}.  Alternatively, the accummulation of $ W $ can be added to the left-looking unblocked algorithm with pivoting.

The interplay between pivoting and the various algorithms also yields the observation that only one level of blocking can be employed when pivoting: a blocked right-looking algorithm that uses an unblocked left-looking algorithm within the panel.

\begin{figure}[tbp]

\centering
{
    \footnotesize
    \input LTLt_piv_blk.tex
    }
    \caption{Pivoted blocked right-looking algorithms. All of the remaining parts of $ X$ are passed into the panel factorization so that symmetric pivoting can be applied.  The matrix $X$ is only updated (other than pivoting) up to the double lines. The vector $p_2^\star$ omits the first element, while the scalar $p_4^f$ is the first element of $p_4$ only. The very first pivot, $\pi_0$, is not computed and is assumed to be zero. }
    \label{fig:LTLt_piv_blk}
\end{figure}

\NoShow{        
\subsection{Adding pivoting to the blocked right-looking algorithm}

I am just storing stuff that I started here.  I suspect we are NOT going to include this in the paper.

{\footnotesize
\begin{equation}
    \label{Before-piv-unb-right}
\begin{array}{l}
\left( \begin{array}{c I c | c | c | c }
X_{00} & \star & \star & \star & \star \\ \whline
x_{10}^T & \chi_{11} & 
\star & \star& \star \\ \hline
X_{20} & 
x _{21} & X_{22} &  \star & \star \\ \hline
x_{30}^T & \chi_{31}
& x_{32}^T & \chi_{33}
& \star \\ \hline
X_{40} & x_{41}
& X_{42} & x_{43} & X_{44}
\end{array} \right) \\
\setlength{\arraycolsep}{2pt}
~~
=
\left( \begin{array}{c I c | c | c | c }
T_{00} & \star & 
\hspace{0.5in}
\star
\hspace{0.5in}~ & \hspace{0.5in}\star \hspace{0.5in}
~ & \hspace{0.5in}\star \hspace{0.5in}~ \\ \whline
\tau_{10} e_l^T & 0 & 
\hspace{0.5in}\star \hspace{0.5in}~  & \hspace{0.5in} \star \hspace{0.5in}~ & \hspace{0.5in} \star \hspace{0.5in} ~ \\ \hline
0 & 
\multirow{3}{*}{$
\tau_{31}
P( \left( \begin{array}{c}
p_2 \\ \hline
\pi_3 \\ \hline
p_4
\end{array}
\right) )^{-1}
\left( \begin{array}{c | c | c}
\widetilde L_{22} & 0 &  0 \\ \hline
\widetilde l_{32}^T &
1 & 0 \\ \hline
\widetilde L_{42} &
\widetilde l_{43} &
\widetilde L_{44}
\end{array}
\right) e_f
$} & 
\multicolumn{3}{c}
{
\multirow{2}{*}{
$
P( \left( \begin{array}{c}
p_2 \\ \hline
\pi_3 \\ \hline
p_4
\end{array}
\right) )^{-1}
\left( \begin{array}{c | c | c}
\widetilde L_{22} & 0 &  0 \\ \hline
\widetilde l_{32}^T &
1 & 0 \\ \hline
\widetilde L_{42} &
\widetilde l_{43} &
\widetilde L_{44}
\end{array}
\right)
\left( \begin{array}{c | c | c}
T_{22} & -\tau_{32} e_l &  0 \\ \hline
\tau_{32} e_l^T &
0 & - \tau_{43} e_f^T \\ \hline
0 &
\tau_{43} e_f &
T_{44}
\end{array}
\right) 
\left( \begin{array}{c | c | c}
\widetilde L_{22}^T & \widetilde l_{32}  &  \widetilde L_{42}^T \\ \hline
0 &
1 & \widetilde l_{34}^T \\ \hline
0 & 0 &
\widetilde L_{44}^T
\end{array}
\right)
P( \left( \begin{array}{c}
p_2 \\ \hline
\pi_3 \\ \hline
p_4
\end{array}
\right) )^{-T}
$ } }
\\
\cline{1-1} 
0 & 
&  \multicolumn{3}{c}{}
\\
\cline{1-1} 
0 & 
&  \multicolumn{3}{c}{}
\end{array} \right)
\NoShow{
\wedge \\
~~
\left( \begin{array}{c I c | c | c}
\widehat X_{00} & \star & \star & \star \\ \whline
\widehat x_{10}^T & 0 & 
\star & \star \\ \hline
\widehat x_{20}^T & 
\widehat \chi _{21} &  0 &  \star \\ \hline
\widehat X_{30} & \widehat x_{31}
& \widehat x_{32} & \widehat X_{33}
\end{array} \right) \\
~~~~
=
\left( \begin{array}{c I c | c | c}
\widetilde L_{00} & 0 & 0 & 0 \\ \whline
\widetilde l_{10}^T & 1 & 
0 & 0 \\ \hline
\widetilde l_{20}^T & 
\lambda _{21} & 
1
&  0 \\ \hline
\widetilde L_{30} & \widetilde l_{31}
& \widetilde l_{32} & \widetilde L_{33}
\end{array} \right)
\left( \begin{array}{c I c | c | c}
T_{00} & 
\tau_{10} e_l & 0 &  \\ \whline
\tau_{10} e_l^T & 0 & 
\tau_{21}  & 0 \\ \hline
0 & 
\tau_{21} & 
0
&  \tau_{32} e_f^T \\ \hline
0 & 
0 &
\tau_{32} e_f & T_{33}
\end{array} \right)
\left( \begin{array}{c I c | c | c}
\widetilde L_{00}^T & \widetilde l_{10} & \widetilde l_{20} & \widetilde L_{30}^T \\ \whline
0 & 1 & 
\widetilde  \lambda{21} & \widetilde l_{31}^T \\ \hline
0
& 
0 & 
1
&  \widetilde l_{32}^T \\ \hline
0 & 
& 0 & \widetilde L_{33}^T
\end{array} \right)
}
\end{array}
\end{equation}
}%
and at the bottom of the loop we find
{\footnotesize
\begin{equation}
    \label{After-piv-unb-right}
\begin{array}{l}
\left( \begin{array}{c | c | c I c | c }
X_{00}^{\rm +} & \star & \star & \star & \star \\ \hline
x_{10}^{\rm +\!T} & \chi_{11}^{\rm +} & 
\star & \star& \star \\ \hline
X_{20}^{\rm +} & 
x _{21}^{\rm +} & X_{22}^{\rm +} &  \star & \star \\ \whline
x_{30}^{\rm +\!T} & \chi_{31}^{\rm +}
& x_{32}^{\rm +\!T} & \chi_{33}^{\rm +}
& \star \\ \hline
X_{40}^{\rm +} & x_{41}
& X_{42}^{\rm +} & x_{43}^{\rm +} & X_{44}^{\rm +}
\end{array} \right)
=
\setlength{\arraycolsep}{2pt}
\left( \begin{array}{c | c | c I c | c}
T_{00} & \star & \star & \star 
& \star \\ \hline
\tau_{10} e_l^T & 0 & 
~~~\star~~~ & ~~~~~~~\star ~~~~~~~& \star\\ \hline
0 &
\tau_{21} e_f
&
T_{22}
&
\star & \star \\ \whline
0 & 
0 &
\tau_{32} e_l^T
& 
\multicolumn{2}{c}{
\multirow{2}{*}{
$
\left( \begin{array}{c | c}
1 & 0 \\ \hline
\widetilde l_{43} &
\widetilde L_{44}
\end{array}
\right)
\left( \begin{array}{c | c}
0 & \star \\ \hline
\tau_{43} 
e_f &
T_{44} 
\end{array}
\right)
\left( \begin{array}{c | c}
1 & \widetilde l_{43}^T \\ \hline
0 &
\widetilde L_{44}^T
\end{array}
\right)
$
}
}
\\
\cline{1-3}
0 & 
0 & 
0 & 
\multicolumn{2}{c}{}
\end{array} \right)
\NoShow{
\wedge \\
~~
\left( \begin{array}{c I c | c | c}
\widehat X_{00} & \star & \star & \star \\ \whline
\widehat x_{10}^T & 0 & 
\star & \star \\ \hline
\widehat x_{20}^T & 
\widehat \chi _{21} &  0 &  \star \\ \hline
\widehat X_{30} & \widehat x_{31}
& \widehat x_{32} & \widehat X_{33}
\end{array} \right)
=
\left( \begin{array}{c I c | c | c}
\widetilde L_{00} & 0 & 0 & 0 \\ \whline
\widetilde l_{10}^T & 1 & 
0 & 0 \\ \hline
\widetilde l_{20}^T & 
\lambda _{21} & 
1
&  0 \\ \hline
\widetilde L_{30} & \widetilde l_{31}
& \widetilde l_{32} & \widetilde L_{33}
\end{array} \right)
\left( \begin{array}{c I c | c | c}
T_{00} & 
\tau_{10} e_l & 0 &  \\ \whline
\tau_{10} e_l^T & 0 & 
\tau_{21}  & 0 \\ \hline
0 & 
\tau_{21} & 
0
&  \tau_{32} e_f^T \\ \hline
0 & 
0 &
\tau_{32} e_f & T_{33}
\end{array} \right)
\left( \begin{array}{c I c | c | c}
\widetilde L_{00}^T & \widetilde l_{10} & \widetilde l_{20} & \widetilde L_{30}^T \\ \whline
0 & 1 & 
\widetilde  \lambda{21} & \widetilde l_{31}^T \\ \hline
0
& 
0 & 
1
&  \widetilde l_{32}^T \\ \hline
0 & 
& 0 & \widetilde L_{33}^T
\end{array} \right).
}
.
\end{array}
\end{equation}
}
}


\section{Conclusion}

We have systematically derived a number of algorithms for the triangular tridiagonalization of a skew-symmetric matrix  by applying the FLAME methodology.  These include classic algorithms like modifications of the Parlett-Reid and Aasen's algorithms (which  were originally proposed for the triangular tridiagonalization of a symmetric matrix) and Wimmer's algorithms.  A number of these algorithms appear to be new: the blocked left-looking algorithm, for which we have not been able to find an equivalent in the literature, and the three blocked right-looking algorithms.  A twist on the traditional FLAME approach that expresses the loop invariant in terms of the final result simplified some the derivations.  We exposed an important link between Wimmer's blocked algorithm and the blocked right-looking algorithms.

Here are a few additional key takeaways: 
\begin{itemize}
    \item 
    We believe that both the presentation of the algorithms with an extension of the FLAME notation and the derivations are easier to follow than traditional expositions of similar algorithms.
    
    \item
    We expose the steps to be systematic, making it perhaps possible to extend systems that mechanically derive linear algebra algorithms, like Cl1ck~\cite{Cl1ck,LoopInvariants,PME}, so that these kinds  of operations are within their scope. 
    \item 
    The derivations can be  adapted to yield algorithms for the symmetric problem.
    
    \item
    We have shown how to derive algorithms that include pivoting.  We have not yet discovered how to systematicaly determine from the PME that pivoting cannot be added to a specific algorithm. This remains an open question.
    \item 
    To elegantly represent these algorithms in code, the FLAME APIs will need to be modified much like the notation needed to be extended.  This is discussed in~\cite{LTLt_SISC_ArXiv}.
    \item 
    To attain high performance without requiring extra workspace, and the related performance degradation due to movement of data, interfaces to new BLAS-like operations will need to be defined and high-performance implementations instantiated. 
    High-performance implementations of various matrix-matrix multiplications  include strategic packing for data locality~\cite{Goto1}.  The multiplication by the skew-symmetric tridiagonal matrix in the ``sandwiched'' version of these operations can be incorporated into that packing, which reduces the number of times data moves between memory layers and avoids workspace.  Exploiting this is within the scope of the BLAS-like Library Instantiation Software (BLIS)~\cite{BLIS1}.  
    This is discussed in~\cite{LTLt_SISC_ArXiv}.
    \item 
Theorem~\ref{thm:TandS} is a key insight that may  apply to other operations involving skew-symmetric matrices.
\end{itemize}
Thus, this paper provides new insights for the development of next-generation linear algebra software libraries with broader functionality and more flexibility.

\NoShow{
based on the FLAME notation system for the $LTL^T$ factorization of a skew-symmetric matrix. The systematically derived variant is twice as fast as the one constructed via traditional approaches. This will boost multiple tasks in computational sciences, mostly related to the handling of Fermions in statistical physics [Cite:mVMC]. 

The objective is to take a skew-symmetric matrix $ X $ and compute its $ L T L^T $ factorization, where $ L $ is unit lower triangular and $ T $ is a skew-symmetric tridiagonal matrix. Eventually, (symmetric) pivoting needs to be added.  
A skew-symmetric matrix has zeroes on the diagonal and its upper triangular entries equal the negative of the corresponding lower triangular entries.
}

\subsection*{Acknowledgments}

The FLAME methodology is the result of collaborations with many of our colleagues at UT Austin and around the world, many of whom appear as authors of works cited in this paper.  We dedicate this paper to all of them.

We in particular would like to thank Ishna Satyarth and Chao Yin, who provided important feedback and are coauthors for the paper that discusses the practical implementations of the derived algorithsm~\cite{LTLt_SISC_ArXiv}.

Over the years, the FLAME methodology was supported by a number of National Science Foundation (NSF) grants and by gifts from industry.
This specific paper was sponsored in part by the NSF under Awards CSSI-2003921 and CSSI-2003931 as well as research gifts from AMD, Arm, and Oracle.

This work was funded at SMU in part by the National Science Foundation (grant
CHE-2143725) and the US Department of Energy (grant DE-SC0022893)

RuQing G. Xu acknowledges the Global Science Graduate Course program and the Computational Science Alliance of The University of Tokyo for supporting his research work. 

{\em Any opinions, findings and conclusions or recommendations expressed in this material are those of the author(s) and do not necessarily reflect the views of the National Science Foundation (NSF) or the Department of Energy (DOE).}

\bibliography{biblio}

\end{document}